\documentclass[useAMS,usenatbib]{mn2e}
\pdfoutput=1
\usepackage{graphicx}
\usepackage{amsmath}
\usepackage{color}

\def \etal {et~al.~}

\newcommand{\be}{\begin{equation}}
\newcommand{\ee}{\end{equation}}

\newcommand{\ifm}[1]{\relax\ifmmode#1\else$\mathsurround=0pt #1$\fi}
\newcommand{\kms}{\ifmmode\,{\rm km}\,{\rm s}^{-1}\else km$\,$s$^{-1}$\fi}

\newcommand{\ltsima}{$\; \buildrel < \over \sim \;$}
\newcommand{\lsim}{\lower.5ex\hbox{\ltsima}}
\newcommand{\gtsima}{$\; \buildrel > \over \sim \;$}
\newcommand{\gsim}{\lower.5ex\hbox{\gtsima}}

\definecolor{green}{rgb}{0,0.5,0}
\definecolor{darkred}{rgb}{0.6,0,0}
\definecolor{grey}{rgb}{0.4,0.5,0.7}

\def\gtsima{$\; \buildrel > \over \sim \;$}
\def\ltsima{$\; \buildrel < \over \sim \;$}
\def\gta{\lower.7ex\hbox{\gtsima}}
\def\lta{\lower.7ex\hbox{\ltsima}}

\def\M11{M_{11}}
\def\V100{V_{100}}
\def\R1{R_{Mpc}}
\def\T6{T_6}

\begin{document}

\title{On stellar mass loss from galaxies in groups and clusters}

\pagerange{\pageref{firstpage}--\pageref{lastpage}} \pubyear{2005}

\author[E. Tollet \etal]{\'{E}douard~Tollet$^{1,2}$\thanks{edouard.tollet@obspm.fr}, 
Andrea~Cattaneo$^{1,3}$,
Gary~A.~Mamon$^{3}$,
\newauthor
Thibaud~Moutard$^{4}$
and Frank~C.~van den Bosch$^{5}$
\\
$^1$ Observatoire de Paris, GEPI, CNRS \& PSL Research University, 61 Avenue de l'observatoire, 75014 Paris, France\\
$^2$ Universit{\'e} Paris Diderot, Sorbonne Paris Cit{\'e}, Paris, France\\
$^3$ Institut d'Astrophysique de Paris, CNRS \& UPMC, 98 bis Boulevard Arago, 75014 Paris, France\\
$^4$ Department of Astronomy \& Physics, Saint Mary’s University, 923 Robie Street, Halifax, Nova Scotia, B3H 3C3, Canada\\
$^5$ Astronomy Department, Yale University , Box 208101, New Haven, CT 06520-8101, USA}

\date{Accepted for publication in MNRAS 18 July 2017. Received 13 July 2017; in original form 10 April 2017}
\pagerange{\pageref{firstpage}--\pageref{lastpage}} \pubyear{2017}
\maketitle

\label{firstpage}


\begin{abstract}

We estimate the stellar mass that satellite galaxies
lose once they enter groups (and clusters) by 
identifying groups in a high-resolution cosmological N-body simulation,
assigning entry masses to satellite galaxies with halo abundance matching at the entry time,
and comparing the predicted conditional stellar mass function of satellite galaxies at $z\simeq0$ with observations.
Our results depend on the mass of the stars that form in satellite galaxies after the entry time.
A model in which star formation shuts down completely as soon a galaxy enters a group environment is ruled out
because it underpredicts the stellar masses of satellite galaxies even in the absence of tidal stripping.
The greater is the stellar mass that is allowed to form, the greater the fraction that needs to be tidally stripped.
The stellar mass fraction lost by satellite galaxies after entering a group or cluster environment is consistent
with any value in the range $0-40\%$.

To place stronger constraints, we consider a more refined model of tidal stripping of galaxies on elongated orbits (where stripping occurs at orbit pericentres). Our model predicts less tidal stripping: satellite galaxies lose $\sim 20-25\%$ of their stellar mass since their entry into the group.
This finding is consistent with a slow-starvation delayed-quenching picture, 
in which galaxies that enter a group or cluster environment keep forming stars until at least the first pericentric passage.

\end{abstract} 



\section{Introduction}

\subsection{Constraints from the intracluster light}

The notion of intracluster light (ICL) began to emerge after \citet{zwicky57} 
and 
\citet{welch_sastry71} detected a diffuse luminous background around NCG 4874 and NGC 4889,
the two supergiant elliptical galaxies that dominate the central region of the Coma cluster.
Since central dominant (cD) galaxies with extensive outer envelopes are a common occurrence in very rich clusters \citep*{matthews_etal64,morgan_lesh65,bautz_morgan70},
\citet{gallagher_ostriker72} interpreted the ICL at the centre of Coma as `a diffuse intergalactic cloud of stars evaporated from colliding galaxies', which 
`will contribute to the formation of a cD system's envelope or in fact may constitute the cD ``galaxy'' itself'. 
Analytical calculations and N-body simulations have confirmed that tidal stripping by companion galaxies and by the cluster's potential can explain the origin of the ICL 
(e.g.,
\citealt{merritt83,mamon87,ghigna_etal98,hayashi_etal03,mihos_etal05,willman_etal04};
\citealt*{abadi_etal06,purcell_etal07}).
 \citet{klimentowski_etal09}, \citet*{lokas_etal11} and \citet{kazantzidis_etal11} found a median stellar mass loss of $\sim 30-35\%$ at each pericentric passage.

Forty years after \citet{gallagher_ostriker72}, there is still no consensus whether the outer envelopes of cD galaxies belong to the ICL or to the galaxies themselves.
The issue could be dismissed as largely semantic but is also the reason why the galaxy stellar mass functions (SMFs) of
\citet{baldry_etal12} and \citet{bernardi_etal13} differ by $\sim 0.5\,$dex at high masses (but see \citealp{bernardi_etal17}).

The ICL contributes $\sim 10\% - 30\%$ of a cluster's total luminosity
(\citealt{zibetti_etal05};
\citealt*{gonzalez_etal05,krick_etal06}) and sometimes even more \citep{lin_mohr04}.
However, stellar haloes  formed out of disrupted satellites are also present in lower mass systems. In particular, it has become clear that the stellar halo of the Milky
Way contains considerable substructure in the form of stellar streams \citep{helmi_etal99,yanny_etal03,bell_etal08}. 
In some cases, the streams can be unambiguously associated
with the satellite galaxies from which they came (\citealt*{ibata_etal94}; \citealt{odenkirchen_etal02}). Similar streams have also been detected in our neighbour galaxy M31 (e.g.
\citealt{ferguson_etal02}). 
In this article, we assess the impact of tidal stripping on the stellar masses of galaxies not just in clusters but across a broad range of environments.

\subsection{Constraints from the growth of cD galaxies}

\citet*{conroy_etal07a} studied the mass growth of cD galaxies from $z\sim 1$ to $z\sim 0$ by using
the abundance matching (AM) technique
\citep{marinoni_hudson02,vale_ostriker04}, described in detail by
\citet*{behroozi_etal10},
which assumes that halo masses (or equivalent properties) are strongly
correlated with observable galaxy properties such as luminosity or stellar mass.

Let $n_*(m_*)$ be the galaxy SMF, which can be determined from observations, and let $n_h(M_h)$ be the halo mass function, which depends on the cosmology and can be determined either analytically 
(e.g. \citealt{press_schechter74}) or from N-body simulations.
If the stellar mass $m_*$ is a growing function of the halo mass $M_{\rm h}$
with negligible scatter\footnote{Throughout this article, we use lower-case
  letters for stellar masses and radii within subhaloes, and upper-case
  letters for dark matter masses and subhalo radial coordinates within haloes.} ,
then the number density of galaxies with stellar mass $>m_*$ will be equal to the number density of haloes with mass
$>M_{\rm h}$:
\begin{equation}
\int_{m_*}^\infty n_*(m_*'){\rm\,d}m_*'=\int_{M_{\rm h}}^\infty n_h(M_{\rm
  h}'){\rm\,d}M_{\rm h}'\ .
\label{am}
\end{equation}
Given the stellar masses (from observations), AM determines the halo masses 
by solving Eq.~(\ref{am}) for $M_{\rm h}$, and conversely, given the halo
masses  (from a simulation), AM provides the stellar masses by solving
  Eq.~(\ref{am}) for $m_*$.
In other words, the most massive halo is assigned the
largest stellar mass, the
second most massive halo is assigned the second largest stellar mass, and so on.
The power of AM was nicely illustrated by \cite{conroy_etal06}, who painted
galaxy luminosities on halo maximum circular velocities to predict very well
the two-point correlation function of galaxies in bins of luminosity. 

\citet{conroy_etal07a} used the results of AM at $z=1$ 
to populate haloes at $z=1$ with galaxies, and they followed these galaxies until $z=0$.
They asked what happens when a smaller halo disappears into a larger one due to hierarchical merging.
Three possibilities were considered: i) the galaxy in the smaller halo merges with the central galaxy of the larger halo;
ii) the galaxy in the smaller halo becomes a satellite galaxy in the larger halo;
iii) the galaxy in the smaller halo is disrupted and its stars become part of the ICL of the larger halo.
The first assumption lead to cD galaxies that were far too bright. 
The second assumption underestimated  the total luminosity of the central galaxy plus the ICL by about a magnitude.
The third assumption was found to be in reasonably good agreement with the observations.

\citet{khang_vandenbosch08} and \citet{cattaneo_etal11} provided independent arguments in support of \citet{conroy_etal07a}'s conclusion.
\citet{khang_vandenbosch08} argued that tidal disruption is necessary to avoid that mergers with bluer satellites spoil the colours of massive red galaxies.
In \citet{cattaneo_etal11}, we used a method intermediate between semianalytic and HOD modelling to quantify the relative importance of gas accretion and mergers in the mass growth of galaxies.
We calibrated our model to fit the galaxy SMF around and below the knee of the Schechter function, where most of the galaxies are, 
and we found that the SMF above a stellar mass of  $3\times 10^{11}\,{\rm M}_\odot$ was overestimated by $\sim 0.2\,$dex.
This discrepancy disappeared when we included a simple model of tidal
stripping, assuming a fixed relative loss of stellar mass at each orbit,
calibrated  on the simulations of  \citet{klimentowski_etal09}.

\subsection{Constraints from satellite galaxies}

While \citet{conroy_etal07a} and \citet{cattaneo_etal11} focussed on tides as a mechanism to prevent overgrowth of cD galaxies between $z\sim 1$ and $z\sim 0$,
\citet{liu_etal10} compared the predictions of three semi-analytical models
(SAMs) 
\citep{delucia_blaizot07,bower_etal06,kang_etal05} with
the conditional SMF of satellite galaxies in SDSS groups (the conditional SMF $\phi(m_{\rm \ast}|M_{\rm h})$ is defined so that ${\rm d}\phi$ is the number of satellite galaxies
with stellar mass between $m_{\rm \ast}$ and $m_{\rm \ast}+{\rm d}m_{\rm \ast}$ in a host system of halo mass $M_{\rm h}$).
In all three models, the number of satellite galaxies was systematically
over-predicted, particularly at low halo masses.
Tidal stripping was considered as a possible explanation.
A mechanism to reduce the number of satellites in massive haloes is also necessary to bring SAMs in agreement with the observed clustering
properties of red galaxies \citep{delatorre_etal11}.

The galaxy stellar mass - metallicity relation is another source of observational evidence.
Galaxies that are more massive have higher metal abundances (e.g. \citealt{gallazzi_etal05}).
\citet{pasquali_etal10} found that satellite galaxies have higher metallicity than central galaxies of the same mass. 
They interpreted this observation as a consequence of tidal stripping, which has reduced the stellar masses of satellite galaxies, while preserving
their stellar metallicities. 
\citet{henriques_thomas10} confirmed that incorporating tidal disruption improves the  agreement with the mass - metallicity relation but
their interpretation of this finding differs from that of  \citet{pasquali_etal10}. For  \citet{pasquali_etal10},
tidal stripping of stars changes the position of galaxies on the stellar mass - metallicity relation by reducing their masses at constant metallicity.
However, ram-pressure and tidal stripping of gas can produce a similar shift by increasing metallicity at constant stellar mass
(starvation of gas accretion shuts down star formation and causes galaxies to behave like close boxes; \citealt{peng_etal15}).
Collectively, these articles highlight the difficulty of disentangling the effects of stripping and starvation.

\subsection{This work}

In this article, we build on \citet{conroy_etal07a}'s method and extend its scope to the investigation of the conditional SMF of satellites.
We start by identifying group and cluster haloes in a cosmological N-body simulation of dissipationless hierarchical clustering.
Merger trees extracted from the simulation allow us to follow galaxies from the entry time to the present.
We use AM to assign stellar masses to satellite galaxies at the entry time. 
By comparing the distribution of entry stellar masses $m_{\rm entry}$ to the distribution of satellite masses $m_{\rm s}$ observed in the Universe today \citep{yang_etal09,yang_etal12},
we can derive a {\it lower} limit $m_{\rm entry}-m_{\rm s}$ to the stellar mass lost through tidal stripping (the lower limit is zero if statistically $m_{\rm entry}\lsim m_{\rm s}$).

For a better estimate of the stellar mass $\Delta m_{\rm strip}$ lost through tidal stripping , we must increment this lower limit by the mass $\Delta m_*$ of the stars formed in the satellite after the entry time:
\begin{equation}
\Delta m_{\rm strip} = m_{\rm entry}+\Delta m_* - m_{\rm s}.
\label{DeltaM_strip}
\end{equation}
While $m_{\rm entry}$ is determined from AM at the entry time, estimating $\Delta m_*$ requires additional assumptions.
We can however derive an {\it upper} limit for $\Delta m_*$ by assuming that the masses of satellite galaxies grow with those of their subhaloes following the same stellar mass - halo mass relation
established for central galaxies from AM (see Sect.~\ref{secAM_no_strip} for more details).
We can therefore determine a lower and and an upper limit for $\Delta m_{\rm strip}$, 
which we can compare to a theoretical estimate of $\Delta m_{\rm strip}$ based on a model described in Sect.~5.2.

Implementing this research programme requires an accurate determination of the subhalo mass function and of the orbits of subhaloes in groups and clusters
(the strengths of the tides depends on the pericentric radius).
In this article, we describe in detail how we solve
the technical problem of reconstructing the orbits of \emph{orphan galaxies}, the subhaloes of which are no longer resolved by the N-body simulation.
We have tested the convergence of our scheme by comparing our results when we use merger trees from a simulation with $512^3$ particles and from another with $1024^3$ particles, both of which where run for the same cosmology and the same initial conditions.

The plan of the article is thus as follows.
In Sect.~2, we describe the N-body simulation and the way we analyse it (identification of haloes and subhaloes, measurement of halo properties, construction of merger trees).
In Sect.~3, we present our scheme to handle orphan galaxies (how we compute their orbits and how we decide at which time they merge with the central galaxy).
In Sect.~4, we explain how we use AM to compute $m_*(M_{\rm h},z)$, the stellar mass of the central galaxy in a halo of mass $M_{\rm h}$ at redshift $z$.
In Sect.~5, we elaborate on our two different models to assign stellar masses to subhaloes: 
one in which $m_s$ grows following the same relation for central galaxies, the other in which there is a complete shutdown of star formation in groups and clusters.
We also present our model (following the simpler
model of \citealp{mamon00})
for computing the stellar mass stripped from
galaxies in a simple form of the impulsive stripping approximation, where stripping occurs instantaneously at the pericentric passage.
In Sect.~6, we compare our predicted distribution for $m_{\rm entry}+\Delta m_*$ and $m_{\rm entry}+\Delta m_*-\Delta m_{\rm strip}$ with observations of the conditional SMF 
(the distribution for $m_{\rm s}$ as a function of the global environment).
For each model, we also compare the results of our calculations with the masses of central galaxies.
Finally, in Sect.~7, we discuss the uncertainties that affect our results and summarize the conclusions of the article.

\section{N-body simulation and merger-tree extraction}
\label{sim}

We use a cosmological N-body simulation with $\Omega_{\rm m}=0.308$,
$\Omega_\Lambda=0.692$, $\Omega_{\rm b} = 0.0481$,  $\sigma_8=0.807$ and
$H_0=67.8\,{\rm\,km\,s}^{-1}\, Mpc^{-1}$
(\citealt{planck14}, {\it Planck} + WP + BAO). 
The simulation has a computational volume of $(100{\rm\,Mpc})^3$ and contains $1024^3$ particles. 
The same simulation, with the same initial conditions, has also been run with $512^3$ particles to test for convergence.

287 snapshots (regularly spaced in the logarithm of the expansion factor)
were saved to disc from $z=16.7$ to $z=0$. The corresponding output  times
are in steps of 145 Myr (at $z=0$) or smaller.
We have processed each snapshot with the halo finder HaloMaker \citep{tweed_etal09}, which is based on AdaptaHOP \citep{aubert_etal04}.
AdaptaHOP is an excursion and percolation algorithm. It selects all particles
above a density threshold and links each one to its 32 nearest neighbours.
If the density distribution within a halo has more more than one peak separated by saddle points, AdaptaHOP decomposes it into a main host halo and a hierarchy of subhaloes, sub-subhaloes, etc. For simplicity of language, we shall refer to all substructures as subhaloes independently of their rank in the hierarchy.
The halo masses that we measure from the N-body simulation are exclusive, i.e., they do not include those of subhaloes.
By construction, a host halo is always more massive than its most massive
subhalo.

We further assume that, to belong to a halo or subhalo, a particle must be gravitationally bound to it.

For each halo containing at least 100 bound particles,
we determine the inertia ellipsoid, which is centred on its centre of mass,
and we rescale it until the overdensity, defined as the mean density inside the inertia ellipsoid
divided by the critical density of the Universe, 
equals overdensity contrast given by the fitting formulae of
\citet{bryan_norman98} in the case a {\it Planck} cosmology ($\Delta_c=102$ at $z=0$)\footnote{The formulae of \citet{bryan_norman98} are
a fit to the predictions of the spherical top-hat collapse model but we have assumed that small deviations from sphericity do not change the virial density of DM haloes.}. 
The halo virial mass $M_{\rm h}$ is the mass of the gravitationally bound N-body particles contained within the virial ellipsoid (i.e., the rescaled inertia ellipsoid).
The virial radius $R_{\rm vir} = \left(a\,b\,c\right)^{1/3}$ is that of a sphere whose
volume equals that of the virial ellipsoid of semi-axes $a$, $b$, and $c$.
We fit the spherically averaged density distribution of each halo with the
NFW profile \citep*{navarro_etal96} to measure its concentration $c$. 

In the case of a subhalo, we use the same procedure to obtain a first estimate of its mass and radius. Then, we shrink the subhalo by peeling off its outer layers
until the density at the recomputed radius $R_{\rm t}$ is at least as large as the host density at the position where the subhalo is located.
The subhalo mass $M_{\rm s}$ and the concentration parameter $c$ is recomputed accordingly.
The particles peeled off the outer layers of a subhalo are reassigned to the host halo if they are gravitationally bound to it.

The TreeMaker algorithm \citep{tweed_etal09} is used  to link haloes/subhaloes identified at different redshifts and generate merger trees.
A halo is identified as the descendent of another when it inherits more than half of its progenitor's particles.
Because of this definition, a halo can have many progenitors but at most one descendent. The main progenitor is always the one with the largest virial mass.
A halo/subhalo is found to have no descendent if it loses more than half its mass but no single halo accretes enough mass from it to qualify as its descendent.
Typically, this happens when a smaller halo crosses a larger one at high speed.
In these cases, the subhalo may be no longer identified but its particles are not assigned to the larger one because they are not gravitationally bound to it.
However, subhaloes that disappear without leaving any descendents are rare and of scarce statistical significance.
Most of those subhaloes disappear close to the pericenter, where the contrast against the host is weakest, and a fraction of them is detected again by the halo finder after their passage. As these objects are a possible source of artefacts for our model, we decide to ignore all subhaloes that have never been detected as central or field halo in any previous snapshot.

\section{Orphan galaxies and ghost subhaloes}

When a halo enters a group or a cluster and becomes a subhalo,
it begins to lose mass owing to the tides exerted by the gravitational potential of the host (Fig.~\ref{fig_traj}).
Eventually, the mass loss may be so large that the subhalo
is no longer identified by the halo finder. 
In SAMs, galaxies associated with subhaloes that are no longer identified are called \emph{orphans}  \citep{springel_etal01,guo_etal10}.  
In reality, a subhalo still exists. 
We have simply lost our capacity to detect it. 
We therefore call it a \emph{ghost} subhalo.  
Orphan galaxies may also be
created by non-physical artifacts from the halo finder itself
\citep{knebe_etal11,srisawat_etal13,avila_etal14}.

Although our model is not a SAM, we face the same problem of deciding how long a galaxy will survive after it becomes an orphan.
Immediately merging the orphan galaxy with the central galaxy of the host system
may produce too few satellites and  too massive central galaxies.
The brute force solution is to increase the number of particles until the results converge above a specified mass.
The most commonly followed alternative is to
estimate the survival time analytically from the orbital decay time through dynamical friction, $t_{\rm df}$
(see \citealt{knebe_etal15} and \citealt{pujol_etal17}, for an overview of the
orphan problem in semianalytic models of galaxy formation).

In its simplest version, this assumption is coupled to that of progressive decay on circular orbits
\citep{somerville_primack99,hatton_etal03,cattaneo_etal06,cora_etal06,gargiulo_etal15}.
However, this approach neglects the high typical orbital elongations of satellites
\citep{ghigna_etal98}, which are important for our analysis because
the strength of the tides depends on the pericentric radius.  A
more sophisticated approach is to track the defunct subhalo's most bound
particle until a time $t_{\rm df}$ has elapsed
\citep{delucia_blaizot07,benson12,gonzalez_etal14}.  Here, we follow a third
approach, first applied by \citet{lee_yi13}, which consists of following the
orbits of ghost suhaloes semianalytically by integrating their equations of
motion in presence of two forces: the gravitational attraction of the host
halo (computed assuming an NFW profile) and dynamical friction (the physical
reason why satellites lose energy, spiral in and eventually fall onto the
central galaxy).

When a subhalo ceases to be identified by the halo finder (more precisely, when it is not the main progenitor of its descendant in the merger tree\footnote{The merger tree is constructed by linking
haloes that the halo finder has identified in output files at different timesteps. A halo can have more than one progenitor but at most one descendent. If the halo fragments, its descendent is the fragment that has inherited more
than half of its particle. If none of the fragments contain more than half of
the particles that the halo finder had assigned to the halo at the previous timestep,
the halo is considered to have disappeared. The halo or subhalo is also considered to have disappeared if tidal stripping has been so strong that it has lost more than half of its particles from one timestep to the next
because the algorithm that constructs the tree is not able to recognise that the stripped halo is the descendent of its progenitor.}),
we save its mass $M_{\rm s}$, position ${\bf R}$ and velocity ${\bf V}$ in the host halo's reference frame at the time of last detection.
If the system that disappears is a halo, we treat its descendant's main progenitor as if it were its host.
These values are the initial conditions from which we start integrating the equations of motions for the ghost subhalo
(the red curve in Fig.~\ref{fig_traj} shows the orbit of a ghost subhalo computed in this manner after the subhalo was no longer resolved\footnote{The rapid decrease in size when passing from the black circles to the red circles in Fig.~\ref{fig_traj} is an artifact of the halo finder.}).

We model the ghost subhalo as point particle of mass $M_{\rm s}$ that moves under the action of two forces: the gravitational attraction of the host and the dynamical friction drag.
However, we shall soon see that the dynamical friction drag depends on $M_{\rm s}$.
Therefore, we cannot integrate the equations of motions without computing the evolution of $M_{\rm s}$ due to tidal stripping at each timestep.
The structure of our calculation is thus as follows.
In Sect.~\ref{tidal_circ}, we describe our method for computing tidal stripping of ghost haloes, which is based on the instantaneous tide approximation
(the tidal radius of the ghost subhalo at a given time is entirely determined by the configuration of the subhalo - host system at that time,
although we never allow the tidal radius to grow again after a ghost halo has been stripped).
In Sect.~\ref{orbital_motion}, we present the equations of motions that we integrate to compute the orbits of ghost subhaloes.
Finally, in Sect.~\ref{surv_time}, we discuss the time at which we should stop their integration because the satellite galaxy associated with the ghost subhalo has merged with the central one.

\subsection{Tidal stripping of ghost subhaloes}
\label{tidal_circ}

Once the halo finder identifies a structure as a subhalo, the radius it returns is no longer the virial radius but the tidal radius $R_{\rm t}$,
computed with the equation:
\begin{equation} 
\label{tidal_radius}
\frac{M_{\rm s}(R_{\rm t})}{R_{\rm t}^3}= |\alpha| \frac{M_{\rm h}(R)}{R^3},
\end{equation}
where $M_{\rm s}$ is the subhalo mass within $R_{\rm t}$, $R$ is the distance of the subhalo from the centre of mass of the host,
$M_{\rm h}(R)$ is the host halo mass within $R$,
and $|\alpha|=1$.

Eq.~(\ref{tidal_radius}) has a theoretical justification because it is the prediction of tidal theory in the approximation of circular orbits and instantaneous tide
(Appendix~A).
However, according to this theory, $\alpha$  should be the local logarithmic slope of the DM mean-density profile
($\alpha<0$ because density decreases with radius).
For an NFW profile, $-2.6\lsim \alpha\lsim -2.2$ for $R=R_{\rm vir}$, but numerical experiments in Appendix~A  suggest that the appropriate $|\alpha|$
(the one that gives the correct value of $R_{\rm t}$ ) is even higher ($\alpha\lsim -3$).
For $R\rightarrow 0$, $\alpha\rightarrow -1$ for an NFW profile. However, the presence of a massive central galaxy could imply that $\alpha\ll -1$ even at relatively small radii (Fig.~\ref{fig_alpha}).
Therefore, the value $|\alpha|=1$ assumed by the halo finder is therefore likely to overestimate $R_{\rm t}$, at least within the circular-orbit approximation.
This is not a problem for the orbits of detected subhaloes, which are computed self-consistently by the N-body simulation,
but it is a point that we must consider when computing the tidal radii of ghost subhaloes.

In this article, we compute $R_{\rm t}$ using $\alpha =-3$ for all ghost subhaloes and we never allow its value to grow again
(although Eq.~\ref{tidal_radius} is instantaneous and thus gives growing values of $R_{\rm t}$ between the pericentre and the apocentre).
The implications of assuming $\alpha=-3$ will be discussed in Appendix~A, after we have explained all the elements that enter our analysis.

\subsection{Orbital motion}
\label{orbital_motion}

A ghost subhalo is assumed to move in the gravitational potential $\Phi({\bf r})$ of the halo directly above it in the hierarchy of substructures. 
$10\%$ of ghosts are sub-subhaloes. For these systems, $\Phi$ is the gravitational potential of the subhalo that contains them.
In $46\%$ of these cases (which correspond to $4.6\%$ of all ghost systems), the subhalo merges with its host before the ghost sub-subhalo merges with the subhalo.
When this happens, the ghost sub-subhalo is promoted to ghost subhalo and continues its orbital motion in the gravitational potential of the host halo.

The equation of motion for a ghost subhalo is:
\begin{equation}
\dot{\bf V}= -{\bf\nabla}\Phi+{\bf a}_{\rm df},
\label{eom}
\end{equation}
where $\Phi$ is computed
assuming an NFW profile for the density distribution $\rho_{\rm h}$ of the system directly above the subhalo in the hierarchy of substructures (heretofore, the host halo, even if it is a subhalo)
and where
\begin{equation}
{\bf a}_{\rm df}=-{4\pi\,G^2\rho_{\rm h}\, M_{\rm s}\,\ln\Lambda\over V^3}\,
f\left({V\over\sqrt{2}\,\sigma}\right)\,
{\bf V}
\label{adf}
\end{equation}
with
\begin{equation}
f(x) = {\rm erf}(x)-{2\over\sqrt\pi}\,x\,\exp(-x^2)
\end{equation}
is the acceleration due to the dynamical friction force \citep{chandrasekhar43}.
In Eq.~(\ref{adf}), $M_{\rm s}$ is the mass of the ghost subhalo,
$\rho_{\rm h}$ is the density of the host at the location of the ghost subhalo,
$\sigma$ is the radial velocity dispersion of the DM particles (assumed to be Maxwellian) and
$\ln\Lambda$ is the so-called Coulomb logarithm.

$M_{\rm s}$ is computed assuming that the subhalo is described by the same NFW profile it had when it was last detected
truncated at the radius $R_{\rm t}$ introduced in Sect.~\ref{tidal_circ}
The host density is computed from the NFW profile $\rho_{\rm h}(R)$ of the host halo, where $R$ is the distance of the ghost subhalo from the centre of the host.
The radial velocity dispersion $\sigma$ is taken from the Appendix~A of \citet{duarte_mamon15}\footnote{\citet{duarte_mamon15} solved the Jeans equation of local dynamical equilibrium 
for a velocity anisotropy $\beta =1-2\sigma_r^2/\sigma_{\rm t}^2$ with the radial dependence $\beta =(1/2)[1-R/(R+R_0)]$, which \citet{mamon_lokas05b} found to match well
the velocity anisotropies measured in cosmological simulations ($R_0$ is the scale radius of the NFW profile).}.
The Coulomb logarithm is given by:
\begin{equation}
\ln\Lambda=\ln\left(1+{M_{\rm h}\over M_{\rm s}}\right).
\label{lambda}
\end{equation}

\subsection{Survival time}
\label{surv_time}

The problem of computing the survival time $t_{\rm surv}$ is that of determining after how many pericentric passages we can stop integrating Eq.~(\ref{eom})
because we can consider that the satellite galaxy has merged with the central galaxy of the host halo.
Our calculation is based on a modified version of the standard dynamical friction in \citet{binney_tremaine08},
which we briefly rederive to clarify its assumption.

In the simplifying case of circular orbits, Eq.~(\ref{eom}) implies that the specific angular momentum loss due
to the dynamical friction force is:
\begin{equation}
{{\rm d}\over{\rm d}t}(R\,V_{\rm c}) = -R\,a_{\rm df}.
\label{decay}
\end{equation}
For a singular isothermal sphere, $V_{\rm c}=\sqrt{2}\sigma$ is independent of radius and $\rho_{\rm h}= V_{\rm c}^2/(4\pi{\rm G}R^2)$. Hence,
inserting Eq.~(\ref{adf}) into Eq.~(\ref{decay}) leads to:
\begin{equation}
{{\rm d}R\over{\rm dt}}=-f(1) \ln\Lambda{{\rm G}\,M_{\rm s}\over R\,V_{\rm c}} .
\label{decay2}
\end{equation}
The time the satellite takes to spiral in from $R=R_{\rm vir}$ to $R=0$ is thus \citep{binney_tremaine08}:
\begin{equation}
t_{\rm df}=\int_0^{R_{\rm vir}}{V_{\rm c}\over f(1)\ln\Lambda{\rm G}M_{\rm
    s}}R{\rm\,d}R 
= {A\over\ln\Lambda}{M_{\rm h}\over M_{\rm s}}{R_{\rm vir}\over V_{\rm c}}\ ,
\label{t_df}
\end{equation}
with $A=1/[2f(1)]\simeq 1.17$, since $V_{\rm c}^2={\rm G}M_{\rm h}/R_{\rm vir}$.

On eccentric orbits, $\rho_{\rm h}$ varies on a timescale $t_{\rm dyn}\ll t_{\rm df}$ invalidating Eq.~(\ref{t_df}) (see  \citealp{mamon96,chan_etal97,cora_etal97})\footnote{Even on circular
  orbits, the timescale for orbital decay can be up to four smaller than
  predicted by the Chandrasekhar formula, because of resonances between the halo
  and subhalo \citep{prugniel_combes92}.}.
Furthermore, in Eq.~(\ref{t_df}), we could take $M_{\rm s}$ out of the integral because we assumed it to be constant.
Real subhaloes are stripped by the tidal field of the host.
This reduces the dynamical friction force and slows down the orbital decay \citep{mamon87, yi_etal13}.

\citet{jiang_etal08} have investigated these effects with cosmological hydrodynamic simulations.
They have found that  Eq.~(\ref{t_df}) gives an accurate measure of the timescale on which a satellite
initially at $R=R_{\rm vir}$ merges with the central galaxy if the Coulomb logarithm is computed with Eq.~(\ref{lambda}) and if the coefficient $A=1.17$ is replaced by:
\begin{equation} 
\label{jiang}
A=1.17\,(0.94\,\epsilon^{0.6}+0.6) \ ,
\end{equation}
where $\epsilon$ is the orbital circularity, that is, the ratio of the angular momentum to that of a circular orbit with the same total energy ($\epsilon=1$ for circular orbit
and $\epsilon=0$ for radial orbit).

\citet{moster_etal13} modelled orphan galaxies/ghost subhaloes in a manner
similar to ours. They used Eq.~(\ref{t_df}) with $A=2.34$, 
this value being based on idealised simulations of orbital decay by
\citet{boylan_etal08}.
Given the mean circularity $\left\langle \epsilon\right\rangle \simeq 0.55$
found by \citeauthor{jiang_etal08}, the mean value of $A$ in
Eq.~(\ref{jiang}) is 1.47. 
Hence, our dynamical-friction times are shorter than those used by \citeauthor{moster_etal13} by 
40\% on average.

\begin{figure}
\includegraphics[width=1.0\hsize,angle=0]{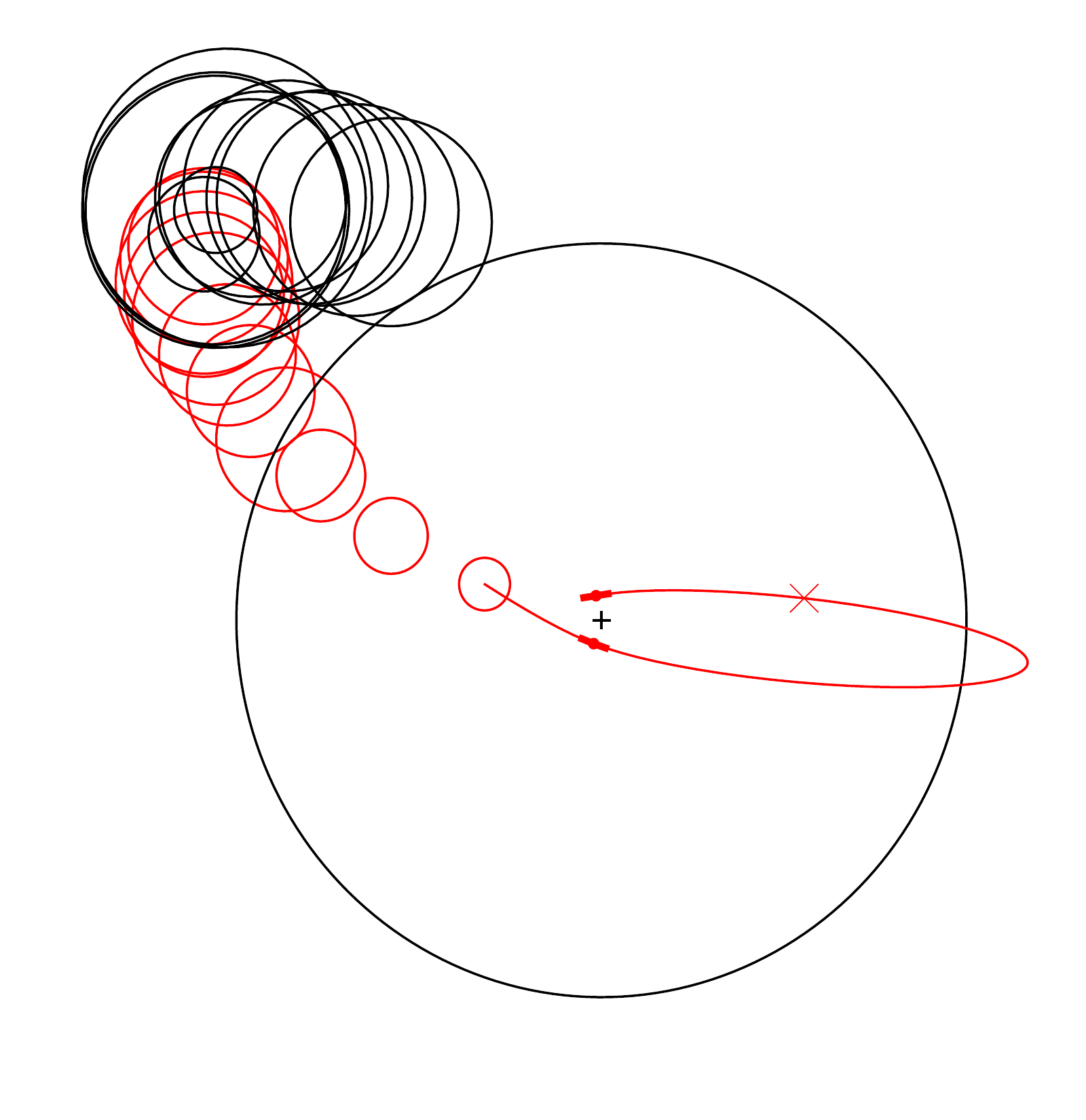} 
\caption{Trajectory and size of a small halo that becomes a subhalo of a larger  one and eventually merges with it.
Before the halo finder identifies the small halo as a subhalo,  
its virial radii are shown as \emph{small black circles}. Their overlapping demonstrates how good the time resolution of our merger trees is.
Once the halo finder identifies it as a subhalo (this occurs several timesteps before the subhalo enters the virial radius of the host halo, shown by the (\emph{large black circle}),
the sizes of the tidal radii are shown as  \emph{red circles},
the smallest of which denotes the subhalo's last detection in the N-body simulation. 
The subhalo then becomes a ghost subhalo and its orbit (\emph{solid red line}) is followed
analytically, solving Eq.~(\ref{eom}) in conjunction with 
Eqs.~(\ref{adf}) and~(\ref{tidal_radius}). 
The two \emph{small red filled circles} correspond to the first and the
second pericentric passage of the ghost subhalo since the time of last
detection. 
The \emph{red cross} indicates the position of the subhalo when the dynamical friction countdown timer comes to zero,
The \emph{black plus sign} denotes the centre of mass of the host system.
The \emph{thick} part of the solid line shows the part of the orbit around the pericentre along which the tides are supposed to act on the stars in the impulsive approximation.}
\label{fig_traj}
\end{figure}

\begin{figure}
\includegraphics[width=1.\hsize,angle=0]{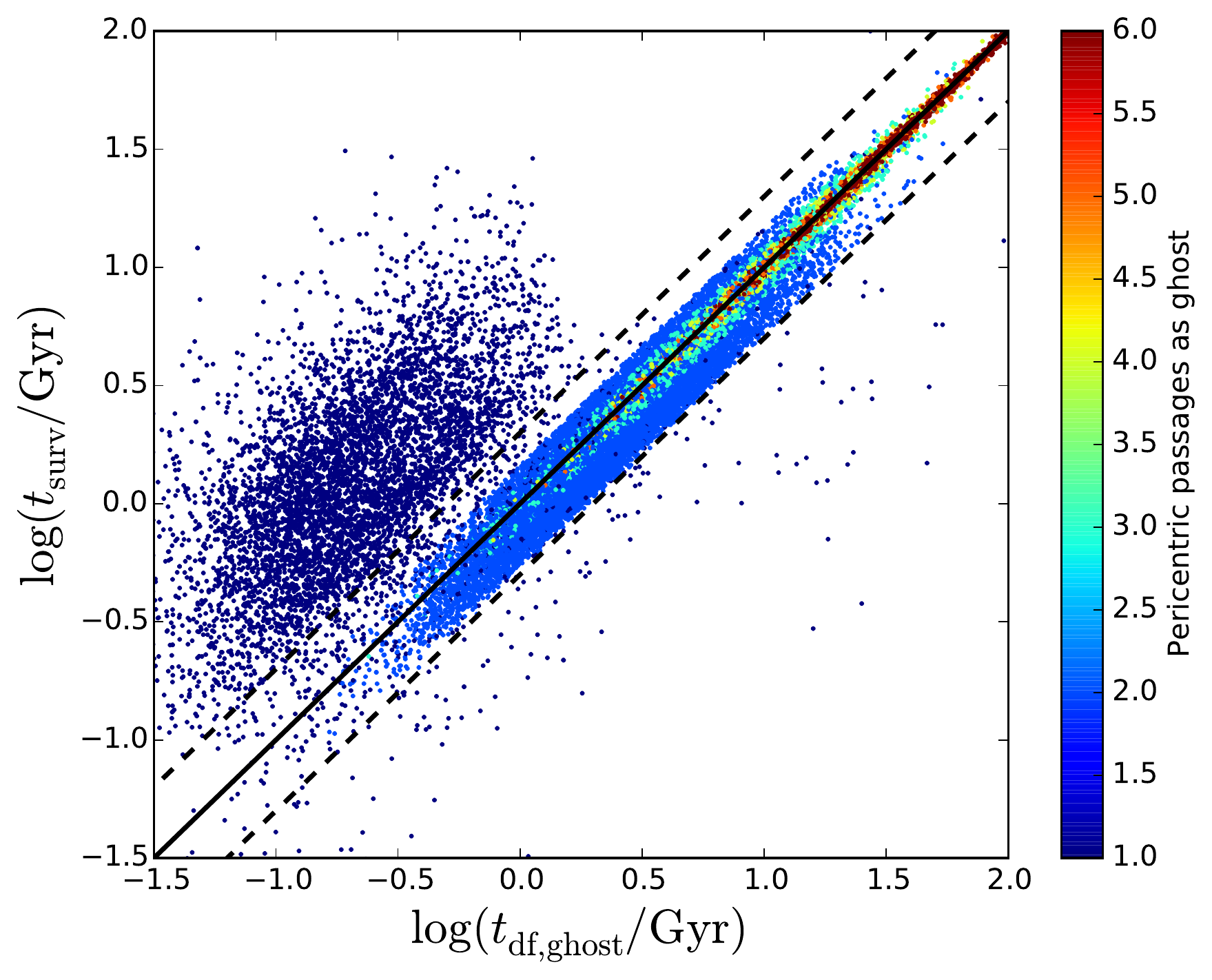} 
\caption{Comparison, for ghost subhaloes, 
between the expected survival time $t_{\rm df,ghost}$ based on Eqs.(\ref{t_df})-(\ref{jiang}) and the actual survival time
  $t_{\rm surv}=t_{\rm merg}-t_{\rm ghost}$ in our model, where mergers can
  occur only at a pericentric passage.
  For
  clarity, we show only 10\% of the ghosts of the
  simulation box.  The \emph{black solid line} represents $t_{\rm surv}=t_{\rm df,ghost}$.  The
  \emph{dashed lines} correspond to a scatter $\pm 0.3\,$dex and enclose $74\%$ of
  the points on the diagram.  Points are colour-coded according to the number
  of pericentric passages between the subhalo's last detection and the merger time in
  our model.  The fractions of ghost subhaloes that merge at the
  1st and the 2nd pericentric passage are $23\%$ and $40\%$, respectively.}
\label{fig_jiang}
\end{figure}

The dynamical friction time $t_{\rm df}$ computed 
with Eqs.~(\ref{t_df}) and~(\ref{jiang}) sets the initial value of the merging countdown time, which begins to tick 
for detected and ghost subhaloes at the time they first enter the virial radius.
When the time $t_{\rm df}$ elapses, a subhalo can be at any point of its orbit (orbits shrink by dynamical friction but, for highly elongated orbits, the
apocentre may still be far away from the centre of the host).
Physically, however,
the merger of a satellite galaxy with the central one is expected to occur at a pericentric passage.
We therefore assume that a ghost subhalo merges with its host at the pericentric passage that is closest in time to
when the merging countdown timer rings
(marked by a red cross in the example of Fig.~\ref{fig_traj}).

In formulae, let $t_{\rm entry}$ be the time at which the subhalo enters the virial radius  for the first time (the large black circle in Fig.~\ref{fig_traj} shows $R_{\rm vir}$ at $t_{\rm entry}$), let
$t_{\rm ghost}$ be the time at which the subhalo ceases to be detected (the last red empty circle in Fig.~\ref{fig_traj}) and
 let $t_{\rm merg}$ be the time of the pericentric passage  at which the galaxy merger takes place (the point where the red curve ends).
Then, $t_{\rm df,ghost}\equiv t_{\rm entry} + t_{\rm df} - t_{\rm ghost}$ is the remaining dynamical friction countdown timer when the subhalo turns into a ghost subhalo.
The survival time of the ghost subhalo from the time it becomes a ghost, $t_{\rm surv}\equiv t_{\rm merg}-t_{\rm ghost}$, can be both larger or smaller than $t_{\rm df,ghost}$
depending on whether the nearest pericentric passage occurs before or after the cosmic time $t_{\rm entry} + t_{\rm df}$.
However, Fig.~\ref{fig_jiang} shows that most ghost subhaloes ($74\%$) lie  on a tight
correlation $t_{\rm surv}\simeq t_{\rm df,ghost}$.
Nearly all the outliers merge at their first pericentric
passage.
They are subhaloes that ceased being detected short after a pericentric passage and for which the merging countdown timer rang while they were still detected.
As we assume that mergers can occur only at pericentric passages, these subhaloes were obliged to make another orbit even though their merging countdown timer had come to zero.

\section{The entry masses}
\label{secAM}

\begin{figure}
\includegraphics[width=1.\hsize,angle=0]{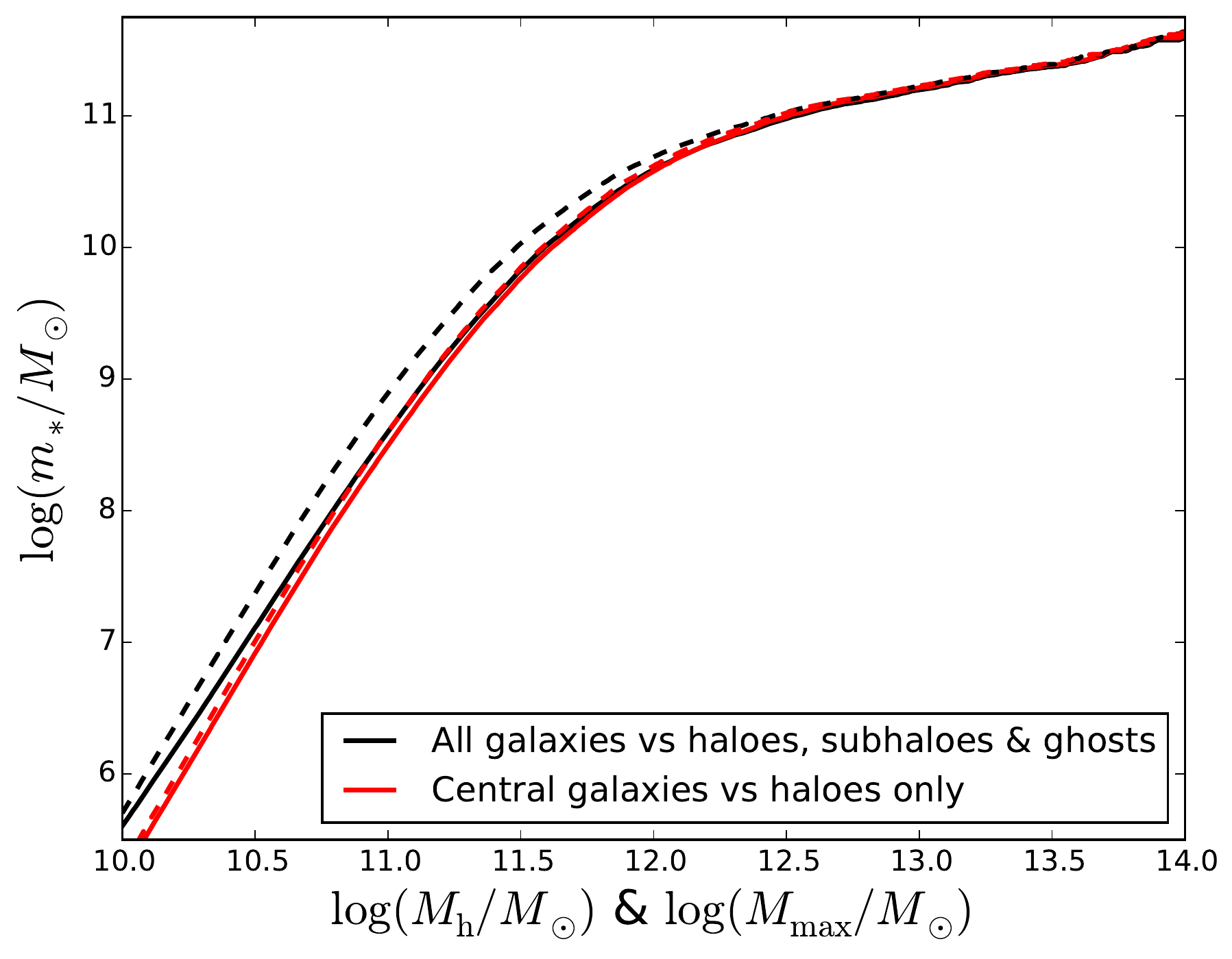} 
\caption{Stellar mass - halo mass relation computed by abundance matching (AM)
  (Eq.~\ref{am}) using the local ($z<0.2$) data of \citet{yang_etal12}.
The AM is performed between 
the stellar mass function (SMF) of central galaxies and the halo mass
function without subhaloes (\emph{red}) or between
the total SMF (central and satellite galaxies) 
to the total mass function of haloes and subhaloes, and ghosts
(\emph{black}).
The halo mass function is computed from the virial mass $M_{\rm h}$ measured
in the N-body simulation at $z\simeq 0.1$, using the procedure described in
Sect.~\ref{sim} (\emph{dashed lines}) or from 
the maximum virial mass $M_{\rm max}$ 
that a halo/subhalo and its main progenitor ever had over its entire history
at $z\gsim 0.1$ (\emph{solid lines}).
}
\label{fig_AM_2ways}
\end{figure}

\begin{figure} 
\includegraphics[width=1.\hsize,angle=0]{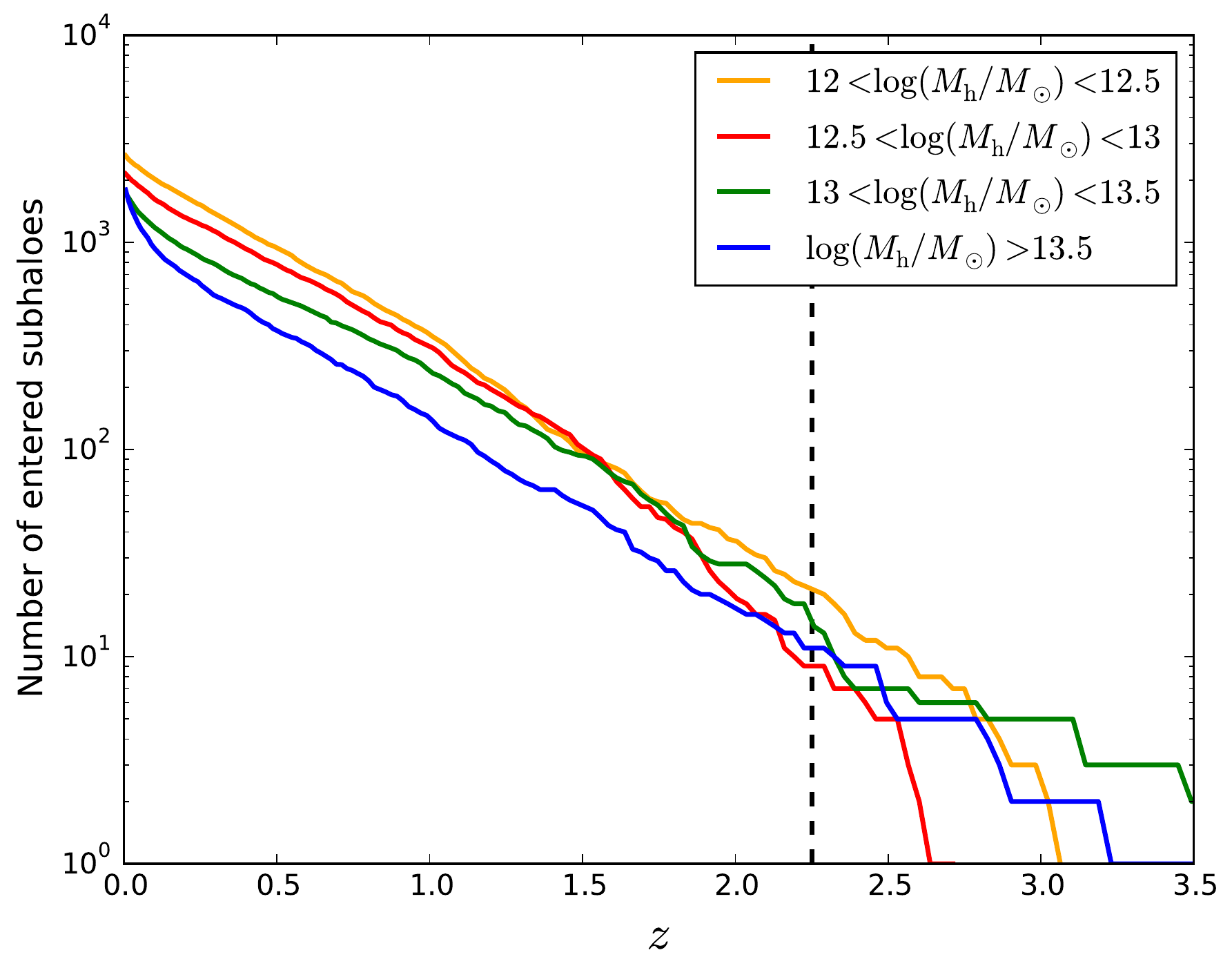} 
\caption{Cumulative distribution of halo entry redshifts
for different bins of host halo mass.
The yellow, red, green and blue curves show 
the mean distribution ${\rm d}N/{\rm d}z$ for the entry redshift $z=z_{\rm entry}$ in the bin of host halo mass
$12<{\rm log}(M_{\rm h}/M_\odot)\le 12.5$, $12.5<{\rm log}(M_{\rm h}/M_\odot)\le 13$, $13<{\rm log}(M_{\rm h}/M_\odot)\le 13.5$ and ${\rm log}(M_{\rm h}/M_\odot)>13.5$, respectively.
The vertical dashed line at $z=2.5$ marks the upper boundary of the redshift range probed by
\citet{muzzin_etal13}'s data.
At $z<0.2$, entry masses are computed from the SMF of \citet{yang_etal12},
who also provided a central/satellite decomposition. 
}
\label{fig_z_entry}
\end{figure}

This section explains our procedure to assign entry masses to galaxies that enter a group or cluster environment.
The entry redshift $z_{\rm entry}$ is the redshift at which the subhalo associated with a satellite galaxy is detected as subhalo of its host for the first time, 
and differs for each galaxy.
Since until $z_{\rm entry}$ all galaxies are central, we can assume that at
$z_{\rm entry}$ galaxies still obey the stellar mass - halo mass relation for
central galaxies, so that $m_{\rm entry}=m_{\rm c}(M_{\rm h},z_{\rm entry})$,  
where $m_{\rm c}$ is the stellar mass of the central galaxy for a halo of mass $M_{\rm h}$ at $z_{\rm entry}$.

In principle, we could derive $m_{\rm c}(M_{\rm h},z_{\rm entry})$ by solving Eq.~(\ref{am}),
where $n_*$ is the SMF of central galaxies at $z_{\rm entry}$ and $n_{\rm h}$ is the halo mass function (without subhaloes) at $z_{\rm entry}$.
In practice, while it is very easy for a theorist to measure $n_{\rm h}$ in an N-body simulation with or without subhaloes at any redshift,
separating central and satellite galaxies in the observations is much more
difficult: this requires large spectroscopic surveys and has only been done
so far for local data 
(the Main Galaxy Sample of the Sloan Digital Sky Survey, hereafter SDSS, where nearly all
galaxies lie at $z<0.2$; \citealt{yang_etal09,yang_etal12}).
The red dashed line in Fig.~\ref{fig_AM_2ways} shows the
stellar - halo mass  (SMHM)
relation that we obtain when we apply the procedure described in this paragraph
using \citet{yang_etal12}'s measurement of the local SMF of central galaxies
on the 7th data release of the SDSS.

The problem of this approach is that many of our satellite galaxies have $z_{\rm entry}$ outside the redshift range probed by \citet{yang_etal12}.
Fig.~\ref{fig_z_entry} shows the cumulative distribution of $z_{\rm entry}$ for different bins of host-halo (group) mass.
The median entry redshift is $z_{\rm entry}\simeq 0.1$ for $M_{\rm h}>10^{13.5}\,M_\odot$ but $z_{\rm entry}\simeq 0.3$ for  $M_{\rm h}<10^{13.5}\,M_\odot$,
while less than one satellite in a thousand has $z_{\rm entry}>2.5$.
At $z>0.2$, we only have the total SMF of galaxies, with no splitting between
centrals and satellites. 
To explore the consequences of approximating the SMF of central galaxies with the total SMF,
we start by making this approximation in the local Universe, where we know the correct answer $m_{\rm c}=m_{\rm c}(M_{\rm h})$ 
given by the red dashed curve in Fig.~\ref{fig_AM_2ways}.
If $n_*$ is the total SMF and $n_{\rm h}$ is the halo mass function including subhaloes and ghosts, then Eq.~(\ref{am}) gives the relation $m_*=m_*(M_{\rm h})$ 
shown by the black dashed curve in Fig.~\ref{fig_AM_2ways}.
The difference between the black dashed curve and the red dashed curve is
sufficiently 
large that approximating the latter with the former would compromise our analysis.

The black dashed curve lies above the red dashed curve because DM haloes are
stripped more easily than the compact luminous galaxies at their centres.
Subhaloes are stripped more heavily and have higher stellar-to-dark matter
mass ratios $m_*/M_{\rm h}$ than haloes.  To demonstrate that tidal stripping
of subhaloes is the physical origin for the difference between the black
dashed curve and the red dashed curve, we have recomputed the curves using
the maximum mass $M_{\rm max}$ that a halo ever had across its history rather
than the virial mass $M_{\rm h}$ at $z\simeq 0.1$ as an estimator of $M_{\rm
  h}$. In this definition, $M_{\rm max}$ cannot decrease. Thus, this procedure removes the
effects of mass loss from haloes/subhaloes in our analysis.  The relation
$m_{\rm c}=m_{\rm c}(M_{\rm max})$ for central galaxies (the red solid line) is
very similar to $m_{\rm c}=m_{\rm c}(M_{\rm h})$ (the red dashed line)
because, for haloes, mass loss is usually negligible.  However, the relation
$m_*=m_*(M_{\rm max})$ for all galaxies (the black solid line) is
significantly different from $m_*=m_*(M_{\rm h})$ (the black dashed line).

The main conclusion of Fig.~\ref{fig_AM_2ways} is that
 $m_*(M_{\rm max})\simeq m_{\rm c}(M_{\rm h})$ for $M_{\rm max}=M_{\rm h}$
(the black solid line and the red dashed line are very close), 
 at least for $M_{\rm h}>10^{10.5}\,{\rm M}_\odot$ and $m_*>10^7\,{\rm M}_\odot$.
Thus, we are justified to replace our original assumption $m_{\rm entry}=m_{\rm c}(M_{\rm h},z_{\rm entry})$ with $m_{\rm entry}=m_*(M_{\rm max},z_{\rm entry})$,
from which we can compute entry masses at redshifts much larger than $z=0.2$.
 
\begin{figure}
\includegraphics[width=1.\hsize,angle=0]{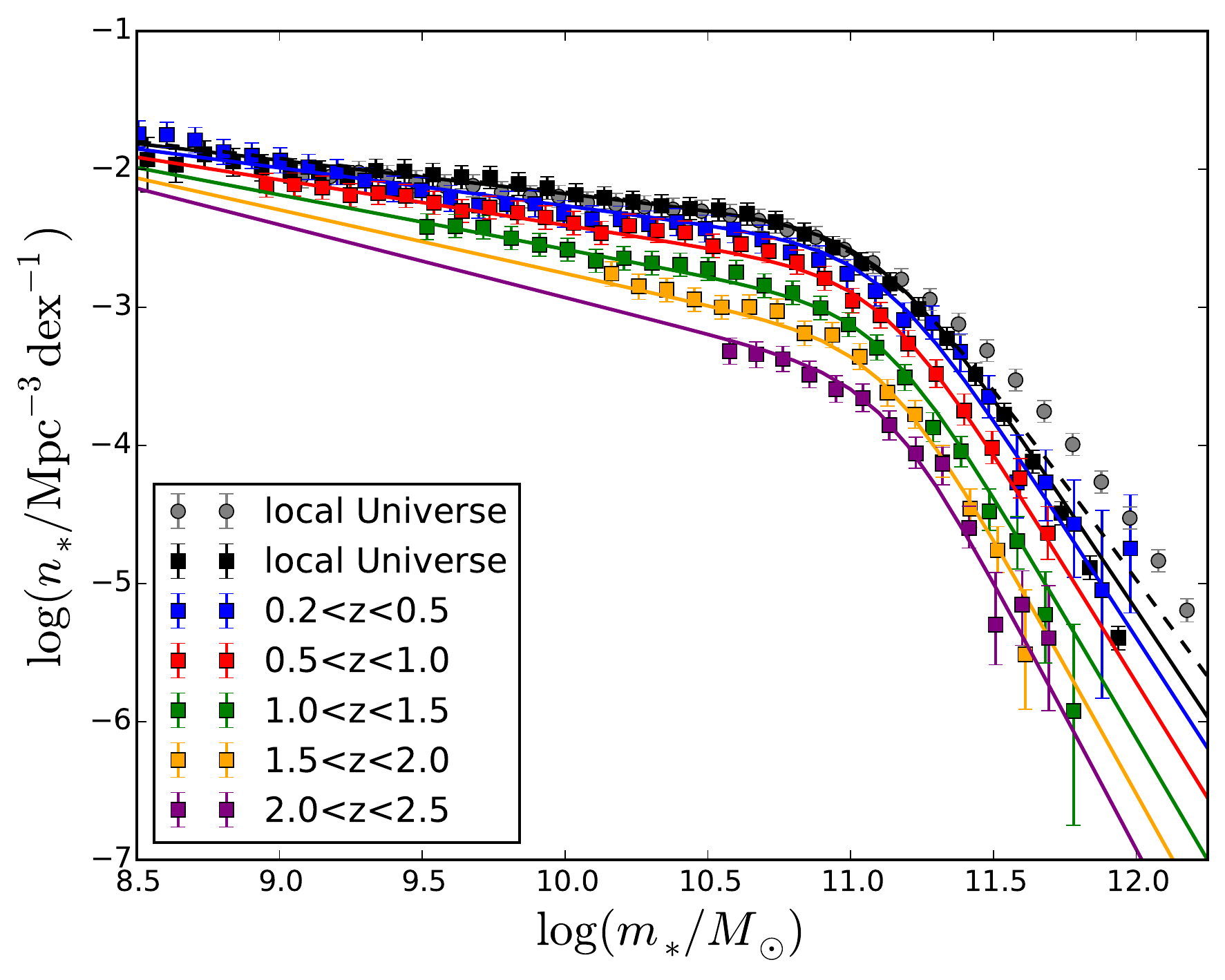} 
\caption{Observed stellar mass functions used to
compute the stellar - halo mass relation in Figs.~\ref{fig_AM_2ways} and \ref{fig_msmh}.
The local data ($z<0.2$, \emph{black squares}) are from \citet{yang_etal12}. 
The data at $0.2<z<2.5$ (\emph{blue, red, green, yellow} and \emph{purple squares})
are from \citet{muzzin_etal13}. 
The SMFs in the six redshift bins were fitted with a double power-law model
(\emph{curves}), the parameters of which were assumed to vary linearly with
redshift. See Appendix~\ref{appendix_fit} for more details about the fit.
The \emph{black, blue, red, green, yellow} and \emph{purple curves} show our
fits 
at $z=0.1,\,0.35,\,0.75,\,1.25,\,1.75,\,2.25$, respectively.
The \emph{gray circles} are data from \citet{bernardi_etal13} ($z<0.2$). They have not been used to fit the evolution of the SMF but are shown
for comparison.
The \emph{dashed black curve} is the fit we would have obtain by fitting
\citet{muzzin_etal13}'s data only and extrapolating them at $z\sim 0.1$. The
quality of the fit to the black symbols and the small difference between the
solid and dashed black curves proves the overall consistency of
\citet{muzzin_etal13}'s and \citet{yang_etal12}'s data. 
}
\label{fig_massfun}
\end{figure}

The observational data that we use for the AM are the SMFs of
\citet{yang_etal12} at $z<0.2$ and of \citet{muzzin_etal13} at $0.2<z<2.5$.
To avoid noise, we do the AM using four-parameter double-power-law fits to the observed SMFs rather than the data points themselves.
Furthermore, to compute $n_*(m_*,z)$, we do not use the best-fit parameters at redshift $z$.
We determine the parameter value at redshift $z$ by fitting a straight line to the evolution with redshift of the best-fit parameters
over six redshift bins covering the range $0<z<2.5$.
The exact fitting formula and the values of the parameters used to fit the SMF are presented in Appendix~\ref{appendix_fit}.
Fig.~\ref{fig_massfun} shows that this model for the galaxy SMF is in good agreement with the data points of both \citet{muzzin_etal13}
and \citet{yang_etal12}. 

To assess the consistency of the two data sets,
we have computed the local SMF by extrapolating the data of \citet{muzzin_etal13} to $z\sim 0.1$, without including the data of \citet{yang_etal12} in the fit.
The result (the black dashed line in Fig.~\ref{fig_massfun}) is intermediate between the
SMFs of \citet{yang_etal12} and \citet{bernardi_etal13}, but much closer to
the former than to the latter.

\begin{figure}
\includegraphics[width=1.\hsize,angle=0]{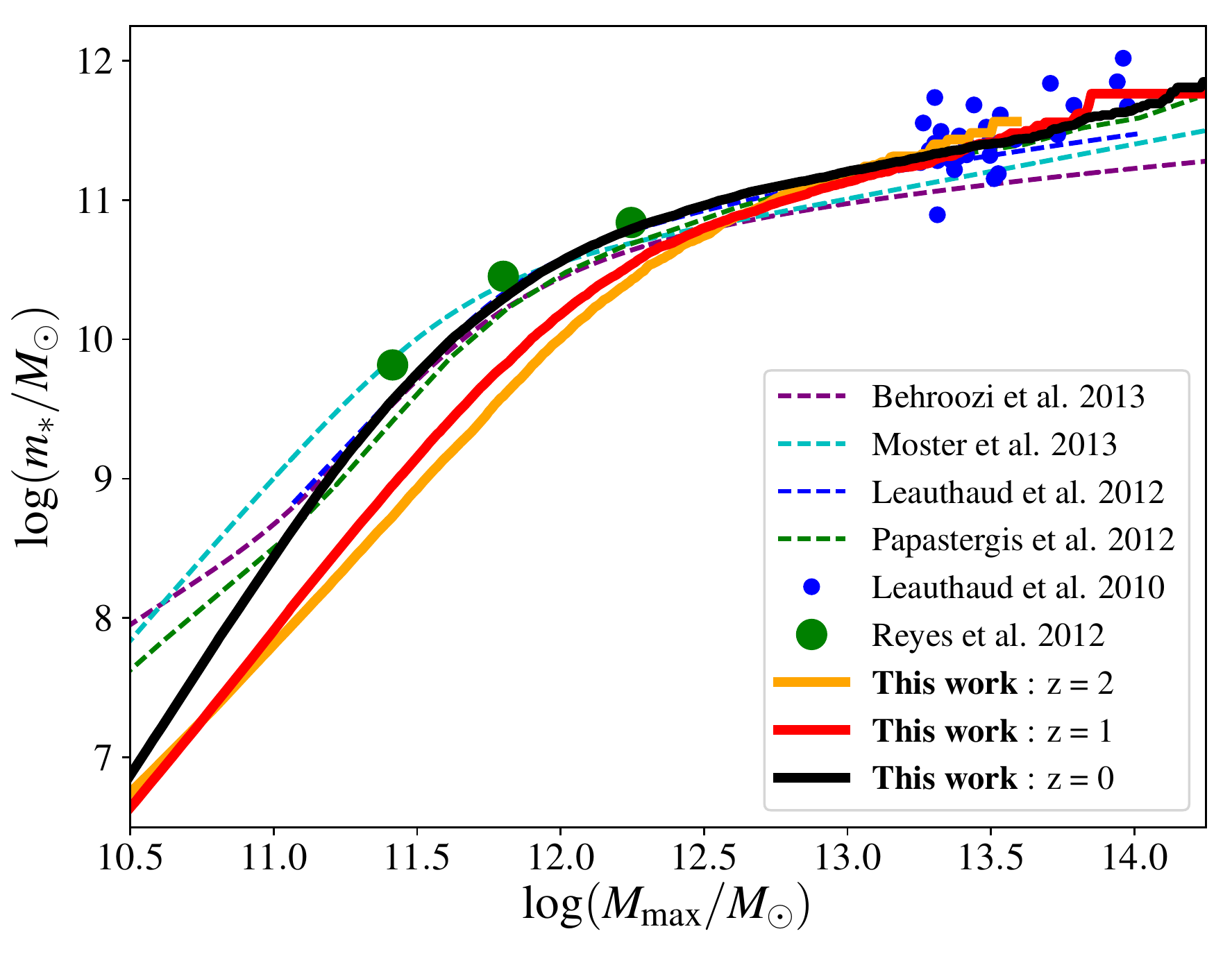} 
\caption{Stellar mass ($m_*$) - halo mass ($M_{\rm max}$) relation used to compute $m_{\rm entry}$. 
The relation is computed from AM at all timesteps in the merger tree.
For clarity, we show it only at $z=0$, $z=1$ and $z=2$ (thick solid black, red and orange curves, respectively).
The thin dashed curves show the results of previous studies by Behroozi et
al. (2013, purple), Leauthaud et al. (2012, blue), Papastergis et al. (2012, green) and Moster et al. (2013, cyan).
The circles are lensing data. Each large green circle is the result of stacking $\sim 10,000$ spiral galaxies \citep{reyes_etal12}, while the small blue circles are data points 
for individual galaxies \citep{leauthaud_etal10}.}
\label{fig_msmh}
\end{figure}

Fig.~\ref{fig_msmh} summarises the results of this section by showing the $m_*$ - $M_{\rm max}$ relation from AM at three different redshifts ($z=0,\,1,\,2$).
The black curve ($z=0$) is smooth up to $M_{\rm max}\simeq 10^{13.3}\,{\rm M}_\odot$, where the effects of low-number statistics in the N-body simulations begins to be important
(there are not many clusters in a volume of $10^6{\rm\,Mpc}^3$).
We also note that, for a fixed halo mass (e.g., $M_{\rm max}=10^{11.5}\,{\rm M}_\odot$), $m_*$ is higher at lower $z$, 
most likely because there has been more time to convert gas into stars.

In Fig.~\ref{fig_msmh}, we also compare our SMHM relation at $z=0$ with lensing data (\citealp{leauthaud_etal10}, \citealp{reyes_etal12}; circles) and previous AM/HOD models 
(\citealp{behroozi_etal13}, \citealp{leauthaud_etal12}, \citealp{papastergis_etal12}, \citealp{moster_etal13}; curves).
The agreement with lensing data is very good considering that lensing observations measure $M_{\rm h}\le M_{\rm max}$.
The agreement with previous AM/HOD models is also quite good (particularly in the mass range $10^{11}\,M_\odot<M_{\rm max}<10^{12.5}\,M_\odot$),
although models differ from one another at the level of  $0.1$ - $0.2\,$dex.
The impact that this disagreement may have on our results is discussed in Sect.~7.1.5.
To ease the comparison with future AM work we provide in Appendix~\ref{appendix_fit} a fit of our SMHM relation.

\section{The stellar masses of satellite galaxies today}

\subsection{Evolution without tidal stripping}
\label{secAM_no_strip}

We have described the AM method that we used to place a galaxy at the centre of each halo in the volume of our N-body simulations.
This procedure determines the stellar masses that satellite galaxies have when they enter a group or cluster environment.
We now consider how the masses of these galaxies evolve after their haloes have become subhaloes.
Here, we focus on the evolution without tidal stripping of stars, the effects of which will be discussed in detail in the next section.

In standard SAMs, a galaxy is composed of stars and cold gas, and is surrounded by a halo of hot gas. The halo of hot gas accretes mass from the intergalactic medium when the DM halo grows.
The hot gas cools and accretes onto the galaxy. The cold gas within the galaxy forms stars. When the galaxy enters a larger system and becomes a satellite,
the hot component associated with the galaxy can no longer grow and is depleted by ram-pressure stripping, tidal stripping or cooling onto the galaxy
(see \citealp{mccarthy_etal08} for a simple analytical model of how a subhalo is stripped of its hot gas). Only after the halo of hot gas has been stripped down to the size of the galaxy do ram-pressure and tidal stripping begin to remove
the cold gas within the galaxy \citep{bekki09}. Stars are the last to go because they are impervious to ram-pressure and can be stripped only by tides.
 
Our model is focussed on the stellar component and does not follow the presence of gas in either the hot or the cold component.
The stellar mass of a central galaxy is determined by an empirical relation that depends only on halo mass $M_{\rm max}$ and redshift, that is, cosmic time.
Its growth is the sum of two terms, the growth of stellar mass with halo mass at constant cosmic time and the growth of stellar mass with time at constant halo mass:
\begin{equation}
{{\rm d}\over{\rm d}t}m_*(M_{\rm max},t)=\left({\partial m_*\over\partial M_{\rm max}}\right)_t\dot{M}_{\rm max}+\left({\partial m_*\over\partial t}\right)_{M_{\rm max}}.
\label{SFR}
\end{equation}
The first term is directly related to the accretion of baryons onto the halo. Thus, it is natural to interpret the second term as the depletion of an existing gas reservoir by star formation
(keeping in mind that star formation is not the only process that may remove gas from galaxies).
In central or isolated galaxies, the first term dominates. We say that these galaxies are in an `accretion mode'.
In satellite galaxies, the only contribution to the star formation rate comes
from the second term. We say that these galaxies are in a `starvation mode'.
We note that our definition of starvation does not exclude accretion from a residual reservoir of hot gas.
This definition may not coincide with that of other authors, who define starvation as a complete shutdown of gas accretion onto the galaxy.

The transition from an accretion mode to a starvation mode at $z_{\rm entry}$ is not an assumption of our model.
It is a consequence of the fact that subhaloes do not gain mass, they lose it. For most subhaloes,
$\dot{M}_{\rm max}=0$ at $z<z_{\rm entry}$

In a more extreme scenario, the entire reservoir potentially available for star formation (hot gas and cold gas) is removed from satellite galaxies upon entry into the host halo.
In this `shutdown' scenario, $\dot{m}_*=0$ for all satellite galaxies at $z<z_{\rm entry}$.
We call this scenario the shutdown model because it corresponds to a complete shutdown of star formation in satellite galaxies.
The shutdown model and the starvation model set lower and upper limits, respectively, to the star formation that is possible in group and cluster environments.

\begin{figure}
\centering
\includegraphics[width=\hsize]{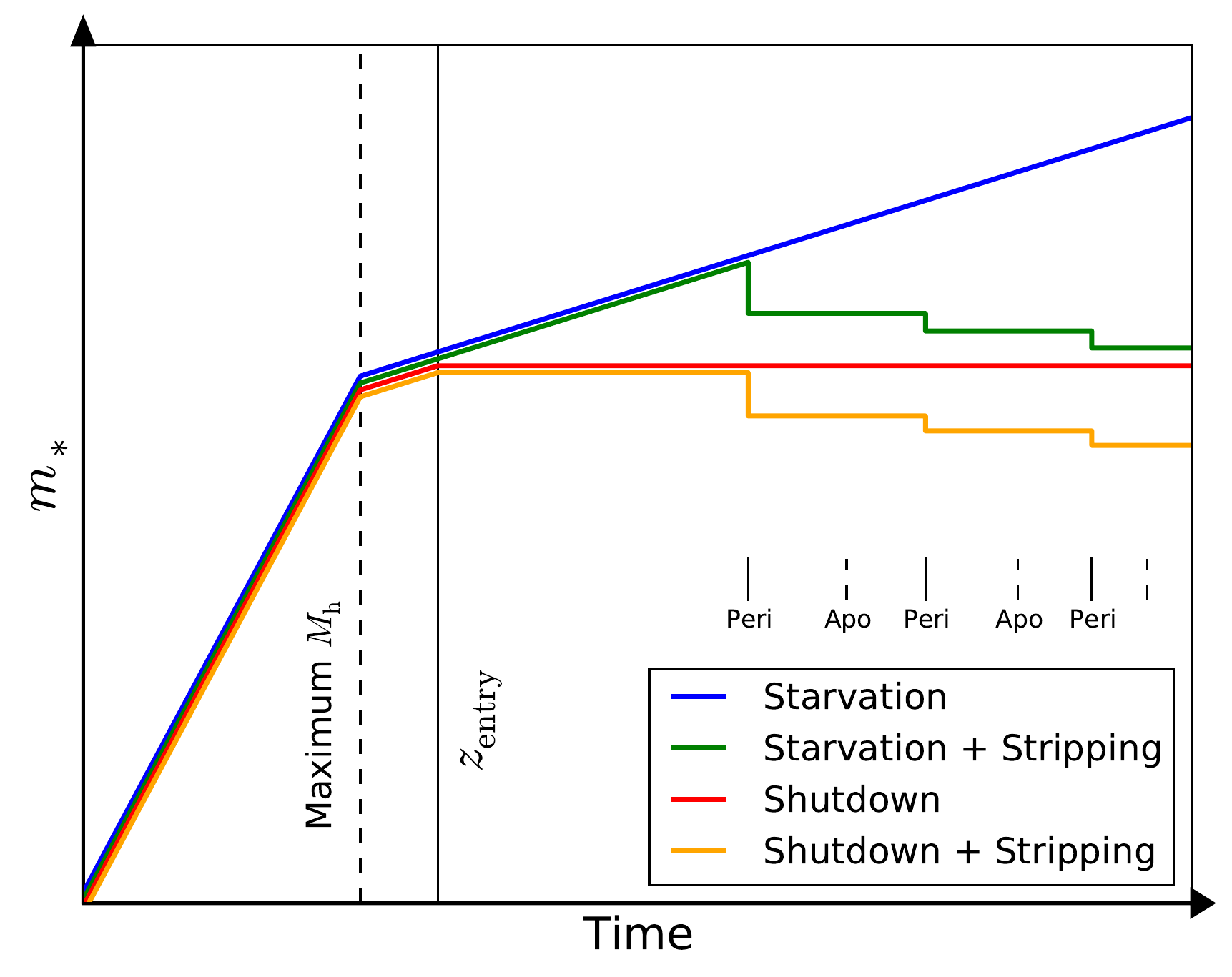} 
\caption{Illustration of our 4 scenarios for the evolution of stellar mass as a
  galaxy enters (at time $z_{\rm entry}$, \emph{black vertical bar}) and orbits (at times
  \emph{peri} and \emph{apo} for pericentres and 
  apocentres, respectively) in a group or cluster (and thus transitions from
  a central in a small group before entry to a satellite in a larger group or
  cluster after entry therein). 
In all 4 models, the stellar mass first grows as expected from abundance
matching with the current halo maximum mass,
once the (sub)halo mass reaches its maximum value (\emph{black dashed vertical bar}) the stellar mass grows more slowly until $z_{\rm entry}$ is reached. 
In the shutdown model (\emph{red}), the stellar mass after entry is maintained at the value at
entry.
In the starvation model (\emph{blue}), the stellar mass increases following the abundance matching prescription. 
Tidal stripping occurs at orbit pericentres (\emph{orange} and \emph{green} for shutdown
and starvation, respectively). After the first stars have been stripped no more star formation is allowed.
Satellite mergers are not considered in this illustration.
\label{fig_evolution}
}
\end{figure}

The blue and the red curve in Fig.~\ref{fig_evolution} illustrate the qualitative evolution of $m_*$ in the starvation and the model, respectively,
when we neglect the effects of tides.
Tidal stripping transforms the blue curve into the green one and the red curve into the yellow one, but here we focus on models without tidal stripping because we postpone its discussion to 
Sect.~\ref{tidesstars}.


In models without stripping, there is no mechanism that can remove stellar mass from galaxies, 
Hence at each timestep the stellar mass of a central galaxy is updated to the maximum between the sum of the stellar masses of its progenitors\footnote{By progenitors, we mean the galaxy main progenitor and all the satellites that have merged with it since the last timestep.}
and the mass predicted by the $m_*$ - $M_{\rm max}$ relation at the current $z$.
However, requiring that the mass of a central galaxy be the largest between the sums of the stellar masses of its progenitors and the mass
from AM method overestimates the masses of cD galaxies because the masses from AM already includes the effects of mergers.
This assumption applies to both the shutdown and the starvation model.

We deal with this problem by introducing a maximum halo mass $M_{\rm lim}$, above which we assume that dissipationless mergers
are the only opportunity for galaxies to grow. 
$M_{\rm lim}$ is a free parameter to be determined by fitting the SMF of galaxies (Sect.~6.1).
In haloes with $M_{\rm max}>M_{\rm lim}$, $m_*$ is the sum of the stellar masses of the progenitors
independently of what the AM relation prescribes.
The condition that $m_*(M_{\rm max})$ can never be lower than the value set by the AM relation is recovered in the limit $M_{\rm lim}\rightarrow\infty$.

\subsection{Tidal stripping of stellar mass}
\label{tidesstars}

We now need to estimate the effects of tidal stripping on the stellar mass of
the subhaloes.
As a gravitational dynamical process, tidal stripping does not distinguish
between a star and DM particle. Therefore, the tidal radius, $r_{\rm t}$ of the stellar
distribution should match the tidal radius, $R_{\rm t}$ of the subhalo, which
we computed in Sect.~\ref{tidal_circ} for the DM,
especially if we neglect the different dynamical structures of discs and haloes, which is beyond the scope of our analysis,
particularly since we do not distinguish between satellites of different morphological types.

However, while a subhalo loses an important fraction of its DM mass before its first pericentric passage 
(\citep{klimentowski_etal09}; although we cannot exclude that this may be due to incomplete relaxation of the subhalo),
the more concentrated stellar component is stripped almost entirely at pericentric passages 
(strong variations of the tidal accelaration along elongated orbits lead to a tidal shock at pericentre; \citealp{ostriker_etal72} and Fig.~3 of \citealp{klimentowski_etal09}).
Fig.~\ref{fig_evolution} illustrates the qualitative effect of adding tidal stripping on elongated orbits to our shutdown (now orange) and starvation (now green) models.
Since most satellite galaxies/subhaloes are on elongated orbits \citep{ghigna_etal98}, an impulsive model for tidal stripping of stars is justified.
In contrast, the circular-orbit approximation behind Eq.~(\ref{tidal_radius}) leads to errors that are too large.

Moreover, Eq.~(\ref{tidal_radius}) is based on the assumption that a particle is immediately stripped as soon as the tidal acceleration is larger than
the gravitational acceleration that keeps it bound to the satellite. In reality, it is also necessary that the impulse
\begin{equation}
\Delta{\bf v} = \int {\bf a}_{\rm t}{\rm\,d}t
\label{impulse}
\end{equation}
imparted by the tidal acceleration ${\bf a}_{\rm t}$ to the particle be sufficient for the unbinding condition
\begin{equation}
{1\over 2}({\bf v}_{\rm s}+\Delta{\bf v})^2+\Phi_{\rm s}\ge 0
\label{unbinding_condition} 
\end{equation}
to be satisfied, where ${\bf v}_{\rm s}$ is the velocity of the particle in the satellite before the tidal perturbation
and $\Phi_{\rm s}$ is the  gravitational potential of the satellite system.

The pericentric passage is where the tidal acceleration $a_{\rm t}$ is strongest but also where the passage is fastest.
The impulsive approximation consists of assuming that the integral in
Eq.~(\ref{impulse}) is dominated by the contribution around the pericentre.
This leads to
\begin{equation}
\Delta v \sim a_{\rm t} \,\Delta t_{\rm p} \sim  {|\alpha|}{{\rm
    G}M_{\rm h}(R_{\rm p})\over R_{\rm p}^3}\,
 \left({R_{\rm p}\over V_{\rm p}}\right)\,r \ ,
\label{impulsive_approximation}
\end{equation}
where  $\Delta t_{\rm p}\sim R_{\rm p}/V_{\rm p}$ is the duration of the pericentric passage
($R_{\rm p}$ and $V_{\rm p}$ are the pericentric radius and speed of the satellite in the host's reference frame),
$M_{\rm h}(R_{\rm p})$ is the host-halo mass within $R_{\rm p}$, $r$ is the distance of the particle from the centre of mass of the satellite,
$\alpha$ is the exponent of the mean density profile,
and
$a_{\rm t}$ has been evaluated with Eq.~(\ref{at_max}) from Appendix~A.
(Similar equations were derived by \citealp{spitzer58} for a point mass
perturber, and generalised to extended perturbers by
\citealp{gonzalez-casado_etal94} and \citealp{mamon00}.)

Let us assume that $\langle {\bf v}_{\rm s}\cdot{\bf a}_{\rm t}\rangle =0$, either
because ${\bf v}_{\rm s}$ has a random orientation or because the particle is
assumed to lie on a circular orbit (as stars in the discs of spiral and S0 galaxies),
in which case the only component of ${\bf a}_{\rm t}$ parallel to  ${\bf
  v}_{\rm s}$ is the azimuthal one, which vanishes when averaged over the orbit (Appendix~A).
Then, as the term ${\bf v}_{\rm s}\cdot\Delta{\bf v}$ in Eq.~(\ref{unbinding_condition}) vanishes,
substituting Eq.~(\ref{impulsive_approximation}) into Eq.~(\ref{unbinding_condition}) gives:
\begin{equation}
{1\over 2}\left[{|\alpha|} {{\rm G}M_{\rm h}(R_{\rm p})\over R_{\rm p}^2}\, {r_{\rm t}\over V_{\rm p}}\right]^2=-\Phi_{\rm s}(r_{\rm t})-{1\over 2}v_{\rm s}^2.
\label{rt1}
\end{equation}

\begin{figure} 
\includegraphics[width=\hsize,angle=0]{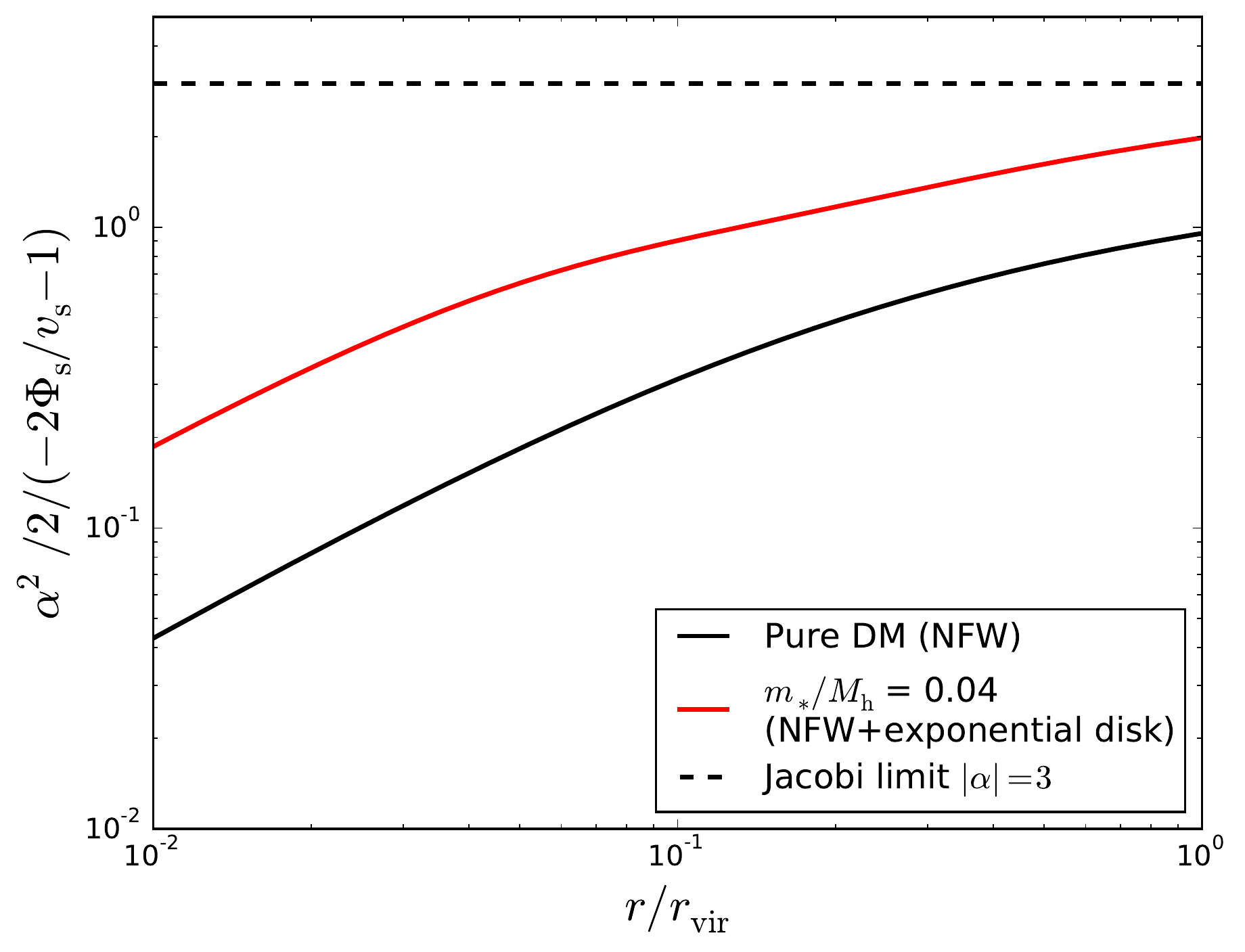} 
\caption{
Maximum efficiency of tidal stripping in the impulsive approximation as a function of the distance from the centre of the satellite
($r_{\rm vir}$ is the virial radius of the satellite and $\alpha=-3$;
the real efficiency is the maximum efficiency times $[V_{\rm c}(R_{\rm p})/V_{\rm p}]^2$).
The \emph{black curve} shows the radial dependence of the maximum efficiency for an NFW potential with $c=8$.
The \emph{red curve} shows the effect of embedding in the subhalo a disc of
mass $0.04\,M_{\rm h}$.
These values correspond to the maximum $m_{\rm stars}/M_{\rm h}$ ratio
allowed by AM and to $\lambda = 0.05$, respectively.
The horizontal \emph{dashed black line} correspond to a circular orbit in the Jacobi limit ($| \alpha |=-3$) }
\label{fig_tides}
\end{figure}

\begin{figure}
\includegraphics[width=\hsize,viewport=0 140 570 680]{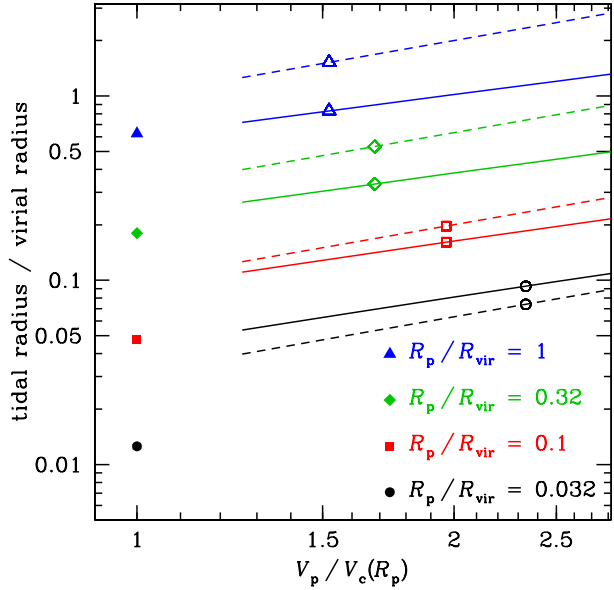} 
\caption{
Tidal radii of subhaloes for different orbital elongations (measured 
by the ratio of pericentre speed to circular velocity at pericentre) 
and different pericentric radii (colours correspond to different values of $R_{\rm p}/R_{\rm vir}$).
This figure compare the estimates from circular tidal theory
(Eq.~\ref{tidal_radius} with $\alpha=-3$, \emph{filled symbols}) and from 
  impulsive tidal theory using our full model
  (Eq.~\ref{rt3}, \emph{solid lines}) or the approximate formulation of Mamon
  (2000) (Eq.~\ref{rtmamon}, \emph{dashed lines}).
The lines stop at $V_{\rm p}/V_{\rm c}(R_{\rm p})=1.35$ because it makes no sense apply the impulsive theory to nearly-circular orbits.
The \emph{open symbols} on the solid and dashed lines show the mean $V_{\rm p}/V_{\rm c}(R_{\rm p})$
for a given $R_{\rm p}/R_{\rm vir}$, computed assuming an average apocentre-to-pericentre ratio of five \citep{ghigna_etal98}.
As the only purpose of this figure is to compare different approximations, we have assumed that both the host and the subhalo are 
described by $c=8$ NFW models and we have neglected the influence of the baryons on tidal radii.}
\label{fig_tidal_radii}
\end{figure}

With the substitution $V_{\rm c}^2(R_{\rm p})={\rm G}M_{\rm h}(R_{\rm p})/R_{\rm p}$, Eq.~(\ref{rt1}) becomes
\begin{equation}
{\alpha^2\over 2} {{\rm G}M_{\rm h}(R_{\rm p})\over R_{\rm p}^3}\,{V_{\rm c}^2(R_{\rm p})\over V_{\rm p}^2}\, r_{\rm t}^2=
-2\,\Phi_{\rm s}-v_{\rm s}^2 \ .
\label{rt2}
\end{equation}
If we make the further assumption that the test particle is on a circular
orbit, so that $v_{\rm s}^2(r_{\rm t})={\rm G}M_{\rm s}(r_{\rm t})/r_{\rm t}$
(as expected for a star in the disc a spiral galaxy), then Eq.~(\ref{rt2}) can be re-written in its final form:
\begin{equation}
\left[-2\,{\Phi_{\rm s}(r_{\rm t})\over v_{\rm s}^2(r_{\rm t})}-1\right] {M_{\rm s}(r_{\rm t})\over r_{\rm t}^3} =
{\alpha^2\over 2} \left[{V_{\rm c}(R_{\rm p})\over V_{\rm p}}\right]^2{M_{\rm h}(R_{\rm p})\over R_{\rm p}^3},
\label{rt3}
\end{equation}
where $R_{\rm p}$ and $V_{\rm p}$ are computed from the conservation of energy:
\begin{equation}
{1\over 2}\,V_{\rm p}^2+\Phi_{\rm h}(R_{\rm p}) = e
\label{e_con}
\end{equation}
and the conservation of angular momentum:
\begin{equation}
R_{\rm p}V_{\rm p}=j,
\label{j_con}
\end{equation}
where $e$ is the specific mechanical energy of the satellite, $j$ is the specific angular momentum and $\Phi_{\rm h}$ is the gravitational potential of the host system.
Although $e$ and $j$ are not really conserved because the satellite is
subject to the dynamical friction drag force (Eq.~\ref{eom}), their variations between snapshots are small 
and the values used to solve Eqs.~(\ref{e_con}) and~(\ref{j_con}) are those
measured at the snapshot just before the pericentric passage, where $\dot{R}$
changes sign.

Eq.~(\ref{rt3}) is identical to Eq.~(\ref{tidal_radius}), except that the term  $|\alpha|$ is now replaced by:
\begin{equation}
\epsilon_{\rm ts} ={\alpha^2/2\over -2\Phi_{\rm s}(r)/v_{\rm s}(r)^2-1} \left[{V_{\rm c}(R_{\rm p})\over V_{\rm p}}\right]^2.
\label{e_ts}
\end{equation}

To compute how $\epsilon_{\rm ts}$ depends on the distance $r$ from the centre of the satellite, 
we assume that all galaxies have an exponential scale-length \citep{mo_etal98}:
\begin{equation}
r_{\rm d} = \lambda R_{\rm vir}/2,
\label{exp_radius}
\end{equation}
the value of which is never allowed
to decrease. Here, $\lambda$ is the spin parameter measured in the N-body simulation. 
At the denominator of Eq.~(\ref{exp_radius}), there is $2$ rather than $\sqrt 2$ because $\lambda$ is defined as in \citet{bullock_etal01}.
The exponential profile is truncated at the radius $r_{\rm t}$ computed with Eq.~(\ref{rt3}). 
All stars outside $r_{\rm t}$ are removed from the galaxy and reassigned to the ICL
without modifying the satellite's profile inside $r_{\rm t}$.
The assumption that all satellite galaxies are discs is admittedly extreme but we do not expect that this assumption significantly affects any of our results\footnote{The satellite population is dominated by
disc morphologies: spirals, S0s (the spiral arms are no longer visible because all gas has been consumed or removed), dSphs (discs that puffed up because of
either stellar feedback or tidal interactions).}.

Although the ICL is, by definition, intra-cluster, i.e., it is not associated with individual galaxies,
we can imagine that, in the beginning, stars stripped from galaxies will form tidal tails, which can still be associated with the galaxies from which they originated 
(e.g., the Magellanic Stream and the Magellanic Clouds). Only later will phase mixing transform these tails into the extended envelopes of central galaxies.
We thus begin by storing the stars stripped from individual galaxies in an ICL component that is still associated with the satellites from which it came.
When the satellites merge, we transfer the stellar mass in this component to the ICL associated with the outer envelopes of the central galaxy.

Fig.~\ref{fig_tides} shows the radial dependence of the factor
$\alpha^2/2/(-2\Phi_{\rm s}/v_{\rm s}^2-1)$, which sets the maximum
efficiency of tidal stripping, for pure DM configuration (black solid curve) and when a disc is embedded in the subhalo (red curve).
The figure shows that stripping is less efficient for stars in the central parts of a satellite galaxy.
It also shows that $\epsilon_{\rm ts}\ll 3$ everywhere.

Besides the coefficient in Fig.~\ref{fig_tides}, the multiplicative factor 
$[V_{\rm c}(R_{\rm p})/V_{\rm p}]^2<1$ is the only significant difference between the results of the impulsive approximation  (Eq.~\ref{rt3}) and the instantaneous approximation
(Eq.~\ref{tidal_radius}). 

The dependence of the tidal radius on $R_{\rm p}$ and $V_{\rm p}$ was
first proposed by \citet{gonzalez-casado_etal94} and confirmed with N-body 
simulations by \citet{ghigna_etal98}).
\citet{mamon00} considered a similar model, in which he assumed $\Phi_{\rm s}\sim -{\rm G}M_{\rm s}(r_{\rm t})/r_{\rm t}$, and found a simpler version 
of Eq.~(\ref{e_ts}) without the term in Fig.~\ref{fig_tides} in front of the square bracket. He obtained the following formula, where $V_{\rm c}$ is the circular velocity of the host halo and $v_{\rm c}$ the circular velocity of the subhalo.

\begin{equation}
r_{\rm t} \approx {V_{\rm p}\over V_{\rm c}(R_{\rm p})}{v_{\rm c}(r_{\rm t})\over V_{\rm c}(R_{\rm p})} R_{\rm p}
\label{rtmamon}
\end{equation}

Fig.~\ref{fig_tidal_radii} compares the tidal radii for our impulsive model (Eq.~\ref{e_ts}; solid curves), the model of Mamon (dashed curves) and
the instantaneous-tides circular-orbit approximation (filled symbols).
By definition, the circular tidal theory is only valid for circular orbits,
whereas the impulsive theory is not valid for circular orbits. 
Hence, we expect the tidal radii computed from the circular theory to be more accurate for orbits with $V_{\rm p}=V_{\rm c}(R_{\rm p})$ and the tidal radii computed from the impulsive theory
to be more accurate for orbits with $V_{\rm p}\gg V_{\rm c}(R_{\rm p})$.
However, most satellites have elongated orbits.
The open symbols in Fig.~\ref{fig_tidal_radii} show the mean elongations, measured by $V_{\rm p}\gg V_{\rm c}(R_{\rm p})$, for orbits with different pericentric radii $R_{\rm p}/R_{\rm vir}$,
according to \citet{ghigna_etal98}.
Smaller pericentric radii correspond to higher orbital elongations.
Comparing the ordinates of the open symbols to those of the filled ones shows
that the circular theory underestimates the average tidal radius, particularly for satellites with small pericentric radii, and thus overestimates tidal stripping.
Fig.~\ref{fig_tidal_radii} also shows that the simple theory of \citet{mamon00} provides a
good estimate of the tidal radius for satellites with $R_{\rm p}\sim 0.1R_{\rm vir}$, although it 
overestimates the effects of tides for satellites on highly elongated orbits.

We therefore compute the tidal radius $r_{\rm t}$ for the stellar component
with Eq.~(\ref{rt3}), which is more accurate. 
The implications for our results of using Eq.~(\ref{rt3}) rather than Eq.~(\ref{tidal_radius}) will be discussed in Sect.~\ref{discus_tides}.

The best way to test the validity of our model for tidal stripping is to compare it with idealised (i.e., non-cosmological) numerical simulations, which follow the dynamics of stars more accurately than our analytic calculations
while retaining full control of the gravitational potential and the orbital configuration.
\citet{kazantzidis_etal13} used idealised simulations to study how tidal stirring can transform a dwarf irregular (a discy dwarf) into a dwarf spheroidal. They assumed a host with the size of the Milky Way and explored three orbital 
configurations: $(R_{\rm a},\,R_{\rm p}) = (125,\,25)\,$kpc, $(R_{\rm a},\,R_{\rm p}) = (125,\,50)\,$kpc and $(R_{\rm a},\,R_{\rm p}) = (250,\,50)\,$kpc, where $R_{\rm a}$ and $R_{\rm p}$ are the apocentric radius and
the pericentric radius, respectively. In the simulations with $R_{\rm p}=50\,$kpc, tidal stripping was negligeable. 
For similar orbital configurations, stripping is negligeable in our model, too.
In the simulation with $R_{\rm p}=25\,$kpc, \citeauthor{kazantzidis_etal13}
found the same qualitative behaviour that is shown by the models with stripping in Fig.~\ref{fig_evolution}. Quantitatively, the stellar mass of the satellite decreased by $5-10\%$ at each pericentric passage
if the stripping potential corresponds to that of an NFW profile, as it does in our model.
This figure is broadly consistent with our results, in which a satellite galaxy loses $\sim 20-25\%$ of its stellar mass over two pericentric passages on average (Sect.~6).

A more careful examination shows that the agreement is not so straightforward. If we apply our model of tidal stripping to the simulations of \citet{kazantzidis_etal13},
assuming the same stellar mass and radius as Kazantzidis et al., our model predicts the stellar mass stripped at the first pericentric passage is $\sim 1\%$ rather than the $5\%$ value found by Kazantzidis et al.
in the simulation with $(R_{\rm a},\,R_{\rm p}) = (125,\,25)\,$kpc. The stellar mass stripped at each pericentric passage is highly sensitive to both $R_{\rm p}$ and the radius of the satellite galaxy.
The reason why tidal stripping is not negligeable in our model despite being much weaker than suggested by Kazantzidis et al. is that the stellar component is less concentrated in our galaxies than in the dwarf of Kazantzidis et al.
However, the comparison should keep in mind that
\citeauthor{kazantzidis_etal13} defined $m_*$ as the stellar mass within $0.7\,$kpc, corresponding to $1.7$ exponential radii of the disc of the satellite galaxy.
Observationally, disc galaxies in the central regions ($r<0.1r_{200}$) of clusters tend to have surface brightness profiles with residuals above an exponential fit at large radii 
(they have a type III `antitruncated' profile; \citealp{pranger_etal17}). Pranger et al. interpreted this observation as a tidal effect. Instead of truncating discs, tides cause them to be more extended by pulling their outer regions.
A lot of the stellar mass that Kazantzidis et al. considered as lost because it moved out of the central $0.7\,$kpc may still be in the disc at slightly larger radii.
Our condition for stripping is much stronger because it requires gravitational unbinding (Eq.~\ref{unbinding_condition}).
Thus, it is not surprising that we find less stripping in our model.
The interesting question is what stellar mass loss Kazantzidis et al. would have found if they had defined $m_*$ as the total mass of the stars that are gravitationally bound to the satellite.
We have no access to their simulations to answer this question.

\section{Results}

\subsection{Total SMF} 
\label{totsmf}

\begin{figure*}
\begin{center}$
\begin{array}{cc}
\includegraphics[width=0.5\hsize]{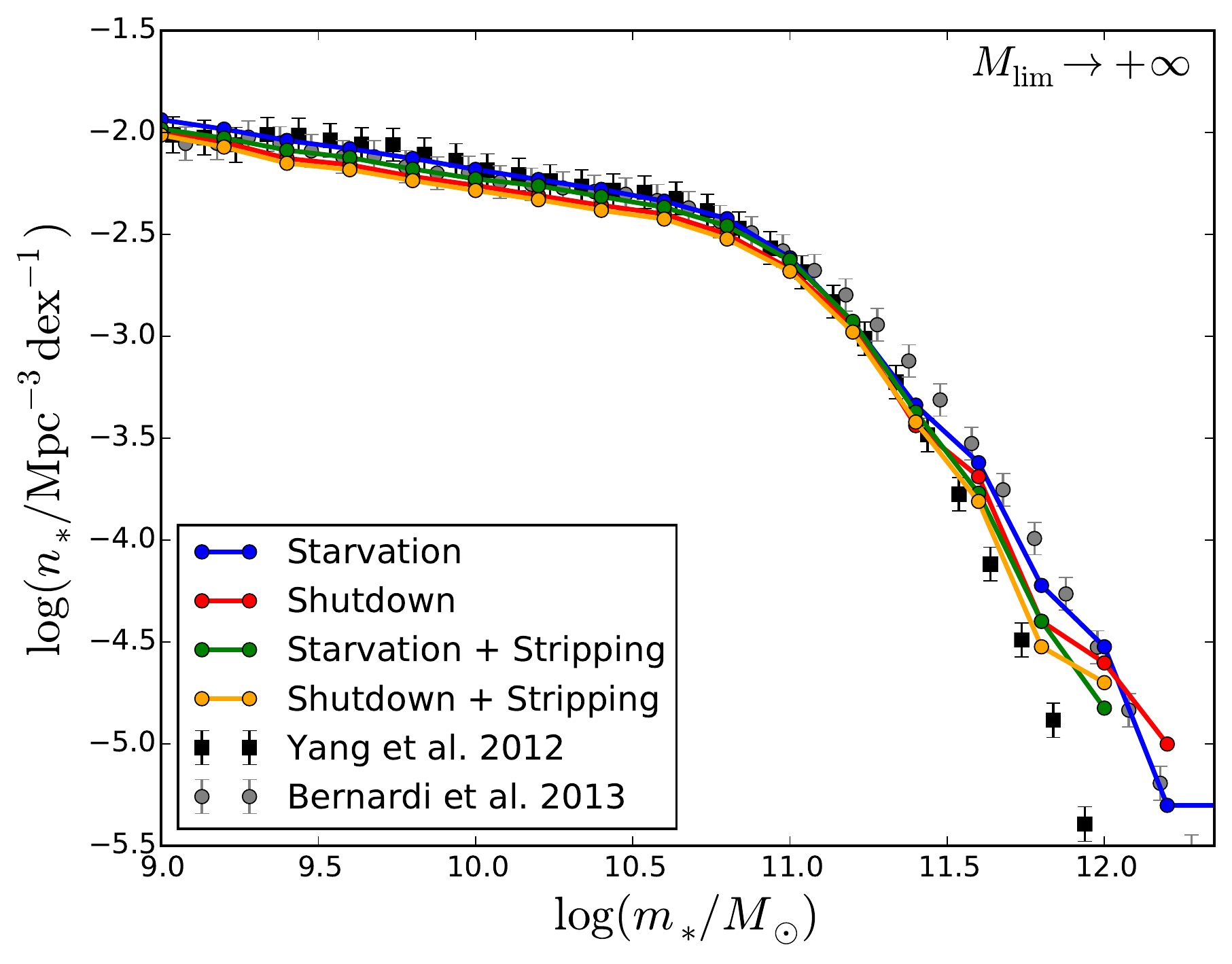} &
\includegraphics[width=0.5\hsize]{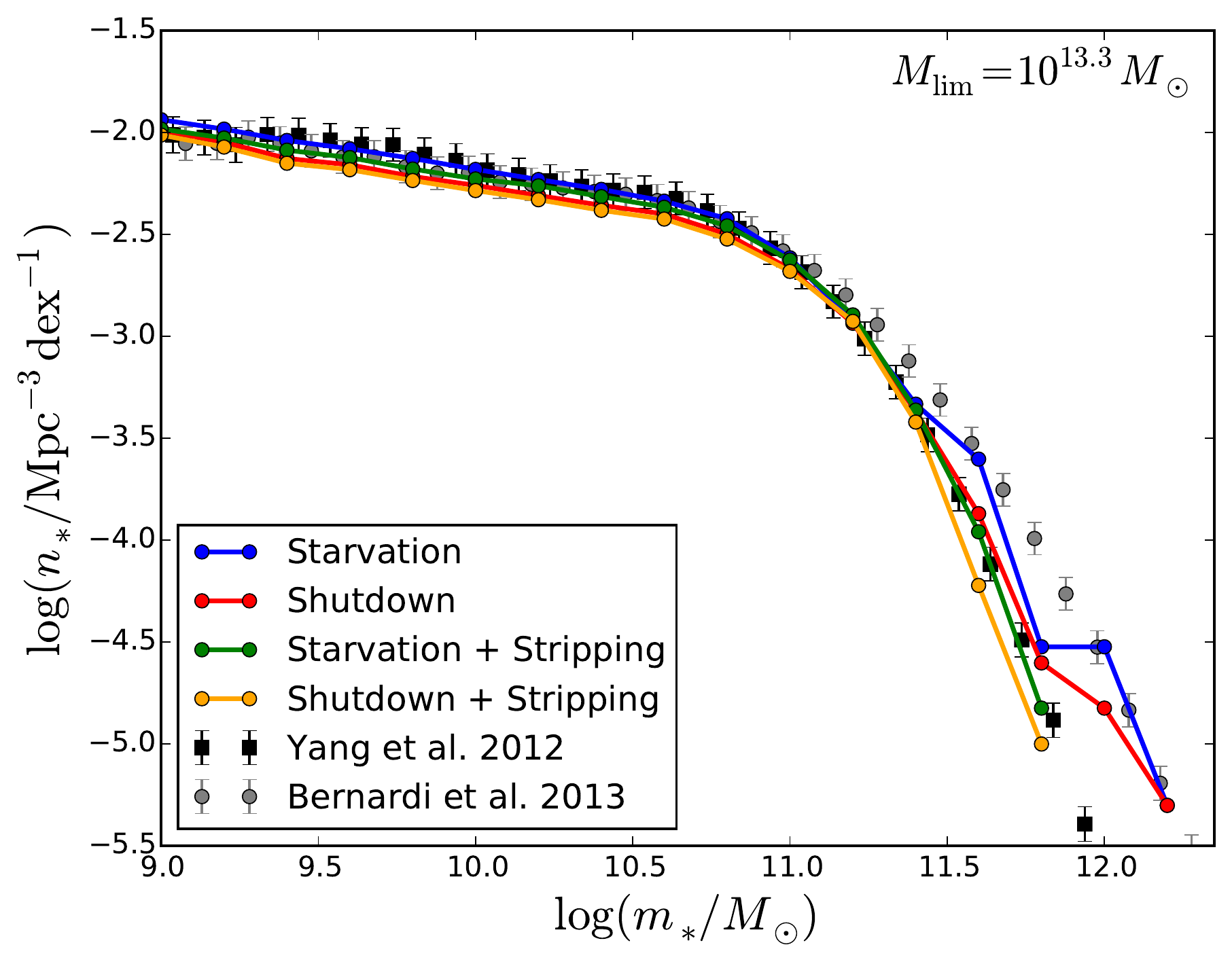} 
\end{array}$
\end{center}
\caption{Stellar mass functions at $z=0$ predicted by our four models
  (\emph{curves}, colour-coded as in the legends and in
  Fig~\ref{fig_evolution}) compared with the observations of  
Yang et al. (2012; \emph{black squares} with error bars) and Bernardi et
al. (2013; \emph{grey circles} with error bars). 
{\bf Left}: the stellar mass is always the maximum between the stellar mass
from abundance matching and the sum of the stellar masses of the progenitors.
{\bf Right}: when the halo mass is $M_{\rm max}>M_{\rm
  lim}=10^{13.3}\,{\rm M}_\odot$, the stellar mass is the sum of the masses of the
progenitors even when this mass is lower than the value obtained from
abundance matching.
}
\label{fig_smf}
\end{figure*}

In Sect.~5, we have described two models, the shutdown model and the starvation model, each in a version without and a version with tidal stripping.
We now compare these four models with the local galaxy SMFs by \citet{yang_etal12} and Bernardi et al. (2013; Fig.~\ref{fig_smf}).
 
At $m_*\lsim 3\times 10^{11}\,{\rm M}_\odot$, the four models are indistinguishable from one another and they are all in excellent agreement with the SMF by \citet{yang_etal12}.
However, at $m_*\gsim 3\times 10^{11}\,{\rm M}_\odot$, all models tend to be above the SMF of Yang et al. The tendency is stronger for the starvation model
without stripping than for the other three models.
This finding may seem surprising because the starvation model without stripping applies to all haloes and subhaloes an AM procedure that should
reproduce the SMF of Yang et al., by construction.
The discrepancy arises because, if the stellar mass returned by this procedure is smaller than the sums of the stellar masses of the progenitors of a galaxy,
it is this sum and not the value returned by the AM procedure that is used to assign a stellar mass to this galaxy. 
If we assign to all galaxies a mass $m_*$ such that the SMF of Yang et
al. is reproduced, by construction, and
we increase the masses of some of these galaxies (typically, the most massive ones, which have greatest number of progenitors),
logically our SMF will contain more massive galaxies than the one of Yang et al.
In other words, if all our haloes had a single progenitor, the blue curve in Fig.~\ref{fig_smf} would fit the square symbols by construction.
The discrepancy between the blue curve and square symbols is linked to the merging histories of galaxies.
The stellar mass above which the blue curve begins to differ from the SMF of  \citet{yang_etal12} is the one above which
dry (dissipationless) mergers become  the dominant
growth mechanism and star formation is negligible \citep{cattaneo_etal11,bernardi_etal11,bernardi_etal11_let}. 

By enforcing the AM relation $m_*=m_*(M_{\rm max})$ even when $m_*$ is larger than the sum of the stellar masses of the progenitors,
we effectively allow star formation in massive galaxies, which we know to be red from observations \citep{kauffmann_etal03,baldry_etal04}
We can deal with this problem by introducing a halo mass limit $M_{\rm lim}$, above which $m_*$ is simply the sum of the stellar masses of the progenitors,
independently of the AM relation (Sect.~\ref{secAM}).
This is equivalent to assuming that there is a limit mass, above which dry mergers are the only growth mechanism.

Observations \citep{kauffmann_etal03,baldry_etal04}, physical models
\citep{dekel_birnboim06},  and SAMs
\citep{bower_etal06,cattaneo_etal06,croton_etal06} 
suggest a limit mass of order $10^{12}\,M_\odot$
(\citealp{cattaneo_etal06} fitted the colour-magnitide distribution in the SDSS with $M_{\rm lim}\sim 10^{12.4}\,{\rm M}_\odot$).
We were therefore surprised to discover that the starvation plus stripping model fits the \citet{yang_etal12} data  for 
$M_{\rm lim}=10^{13.3}\,{\rm M}_\odot$ (Fig.~\ref{fig_smf}b, green curve).
Our explanation for this finding is that AM is not a physical model for the baryonic mass that
is able to condense to the centre in a halo of given mass: $m_*$ is the net result of star formation, dry mergers and stripping.
Tidal stripping is responsible for the extended envelopes of giant
ellipticals, the masses of which are underestimated by studies based on magnitudes from the SDSS pipeline \citep{bernardi_etal17}.
Had we calibrated the $m_*$ - $M_{\rm max}$ relation on the SMF of \citet{bernardi_etal13},
which includes the light from the outer regions and therefore the debris of tidally disrupted satellites,
 the green
curve in Fig.~\ref{fig_smf}b would have shifted to higher masses by an amount
comparable to the difference between the SMFs of \citet{yang_etal12} and
\citet{bernardi_etal13}.
To bring the green curve back on the data points of \citet{yang_etal12} would have then required $M_{\rm lim}\lsim 10^{12.6}-10^{12.7}\,{\rm M}_\odot$
(i.e., the mass limit that shifts the blue curve on the black squares in our calibration), in better agreement with previous studies.

\subsection{Conditional SMF}

The conditional SMF $N(m_*|M_{\rm h})$ is defined so that $N(m_*|M_{\rm h}){\rm\,d}m_*$ is the average number of galaxies with mass between $m_*$ and $m_*+{\rm d}m_*$ in a host halo of mass $M_{\rm h}$
(we have omitted the dependence on $z$ because, in this section, we are only interested in the local Universe).
It can be split into the contributions of central and satellite galaxies, in which case the former integrates to unity (there is only one central galaxy per halo).

In this section, we compare the four models in Fig.~\ref{fig_smf}b to the conditional SMF measured by \citet{yang_etal12}.
A meaningful comparison requires: i) that we apply their same definition of host-halo mass 
and ii) that we apply their same criterion to decide which galaxies belong to a group or cluster.

\begin{figure}
\includegraphics[width=1.\hsize,angle=0]{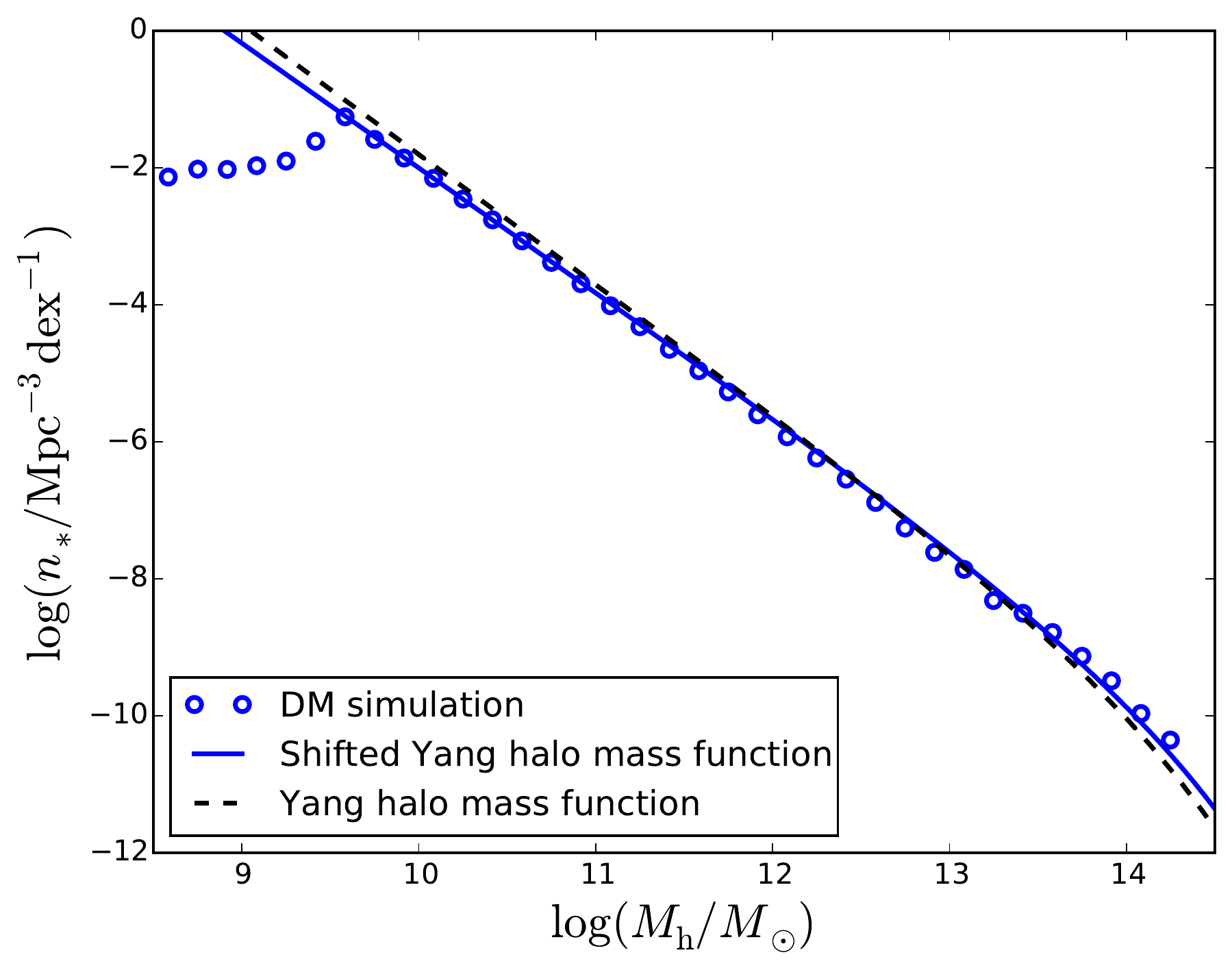} 
\caption{Halo mass function assumed by Yang et al. (2012; \emph{black dashed curve}) compared to the one that we extract from our N-body simulation (\emph{blue circles}). 
The Yang et al. halo mass function (HMF) is mapped into our extracted HMF 
with a linear transformation ${\rm log\,}M_{\rm h}\mapsto a{\rm log\,}M_{\rm h}+b$,
where $a$ and $b$ are fit to our HMF, and the best-fit HMF is shown as the
\emph{blue curve}.
The blue circles show clearly the resolution of our N-body simulations, which contains $1024^3$ particles in a comoving volume of $(100{\rm\,Mpc})^3$.}
\label{fig_shift_Yang}
\end{figure}

\begin{figure*}
\includegraphics[width=1.0\hsize ,height=1.22\hsize ,angle=0]{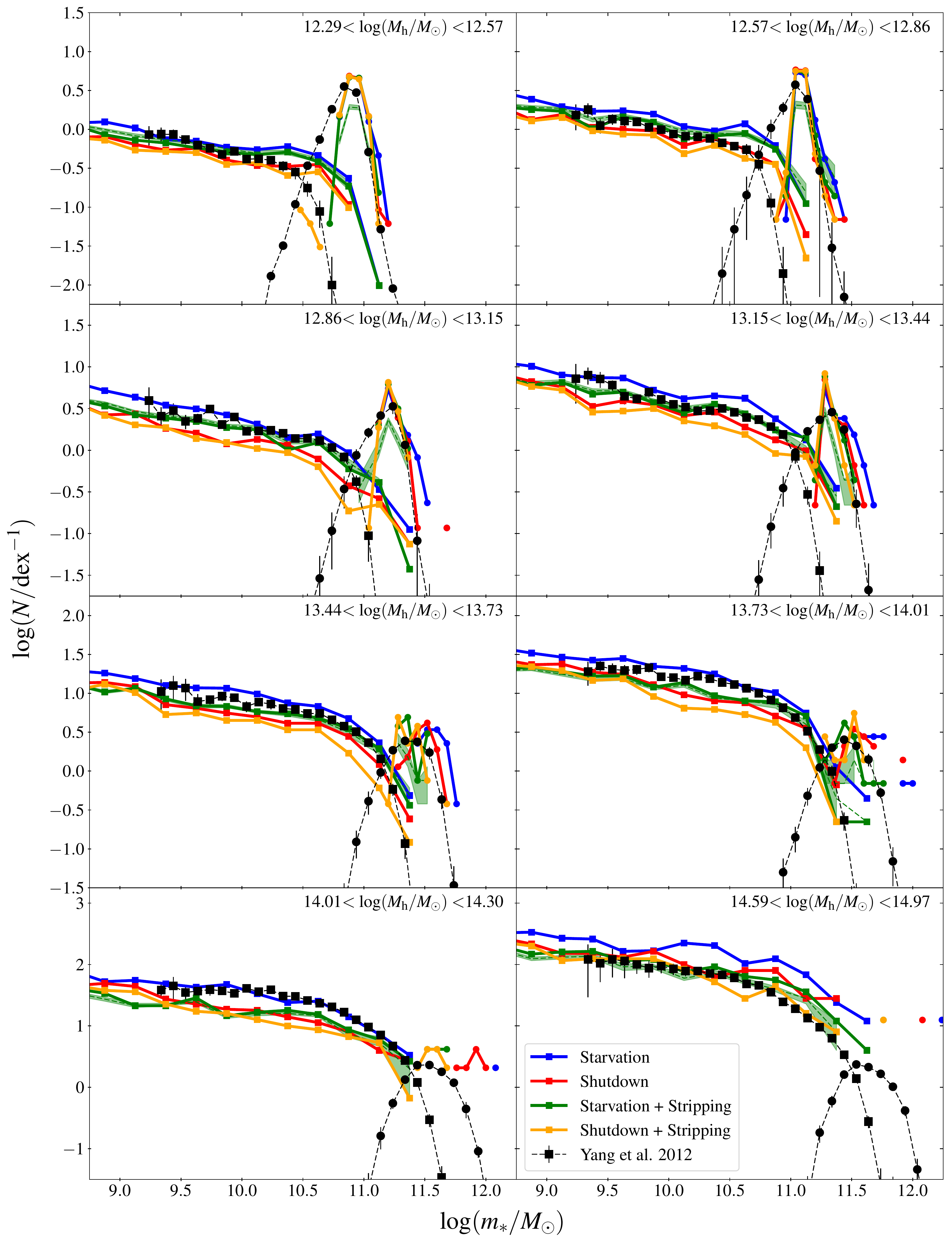} 
\caption{Conditional SMF for our different models (\emph{thick solid curves}, see legends and
    Fig.~\ref{fig_evolution}) compared to the observations of Yang et
  al. (2012; \emph{black points} with error bars). 
The panels correspond to 
bins of group mass.
\emph{Squares}) and \emph{circles}) show Yang et al.'s decomposition of the data in central and satellite galaxies. 
The same decomposition has been applied to the models.
In models with stripping (\emph{green} and \emph{orange curves}), the tidal
radius $r_{\rm t}$ has been computed with Eq.~(\ref{rt3}). 
The thin green dashed curves show how the green thick curves vary when we introduce a scatter of $0.2\,$dex in SMHM relation.
They are medians over a hundred realisations. The upper and lower envelopes of the green shaded areas around them correspond to upper and lower quartiles, respectively.
In all panels, $M_{\rm lim}=10^{13.3}\,{\rm M}_\odot$ (see
Sect.~\ref{totsmf}).
}
\label{fig_group}
\end{figure*}

We start with point (i). \citet{yang_etal12} did not measure $M_{\rm h}$ dynamically. They inferred the group mass $M_h$ from the group total luminosity $L$ by using the AM relation:
\begin{equation}
\int_{L}^\infty n_{\rm gr}(L'){\rm\,d}L'=\int_{M_{\rm h}}^\infty \tilde{n}_h(M_{\rm h}'){\rm\,d}M_{\rm h}',
\label{am2}
\end{equation}
where $n_{\rm gr}$ and $\tilde{n}_{\rm h}$ are \citet{yang_etal12}'s group luminosity function and halo mass function, respectively.
The problem is that the halo mass function $\tilde{n}_{\rm h}$ that they
computed with \citet{sheth_tormen02}'s formula (the black dashes in
Fig.~\ref{fig_shift_Yang} )
is different from the halo mass function $n_{\rm h}$ that we measure in our N-body simulation (the blue open circles), also because the cosmology is not identical.
To overcome this problem, we have fitted a linear transformation that maps  $\tilde{n}_{\rm h}$ into $n_{\rm h}$ (i.e., the black dashes into the blue solid line).
We have applied this transformation to the intervals of $M_{\rm h}$ within which \citet{yang_etal12} determined the conditional SMF and  we have used the transformed intervals to 
select host haloes of corresponding mass in our N-body simulation.

For point (ii), we have reanalysed the density profiles of \citet{yang_etal12}'s groups and verified that they are truncated at $R_{180}$, 
the radius within which the mean density equals $180$ times the mean density of the Universe. $R_{180}/R_{\rm vir}$ depends on concentration.
We have computed $R_{180}$ for all  haloes in the N-body simulations and used this radius to decide which satellite galaxies should be assigned to a host
when computing the conditional SMF.

Fig.~\ref{fig_group} compares the conditional SMF in our four models with \citet{yang_etal12}'s data after taking points (i) and (ii) into account.
A number of conclusions can be drawn from this comparison.

First, the difference between the blue curve and the green one (or the red curve and the orange one) is usually smaller than the difference between the blue curve and the red one.
In other words, the effects of stripping are smaller than the uncertainty from our ignorance of the stellar mass $\Delta m_*$ formed after $t_{\rm entry}$.

Second, all our models predict an excess of massive satellites in low mass groups ($M_{\rm h}<10^{13.44}\,{\rm M}_\odot$), though, at $m_*<10^{10.5}-10^{11}\,{\rm M}_\odot$,
data points for satellites tend to lie in the range allowed by our models (between the blue and the orange curve).

Third, 
the starvation model with tidal stripping (green curves) is the one that,
despite this problem, is overall in best agreement with the conditional SMF of \citet{yang_etal12}
(our N-body simulation contains very few clusters; therefore, the last two panels in Fig.~\ref{fig_group} are affected by poor statistics).
Fig.~\ref{fig_group} was plotted for $M_{\rm lim} = 10^{13.3}\,{\rm M}_\odot$, but
these conclusions are based on the conditional SMF of \emph{satellite} galaxies, the masses of which are insensitive to the value of $M_{\rm lim}$.

\begin{figure}
\includegraphics[width=1.\hsize,angle=0]{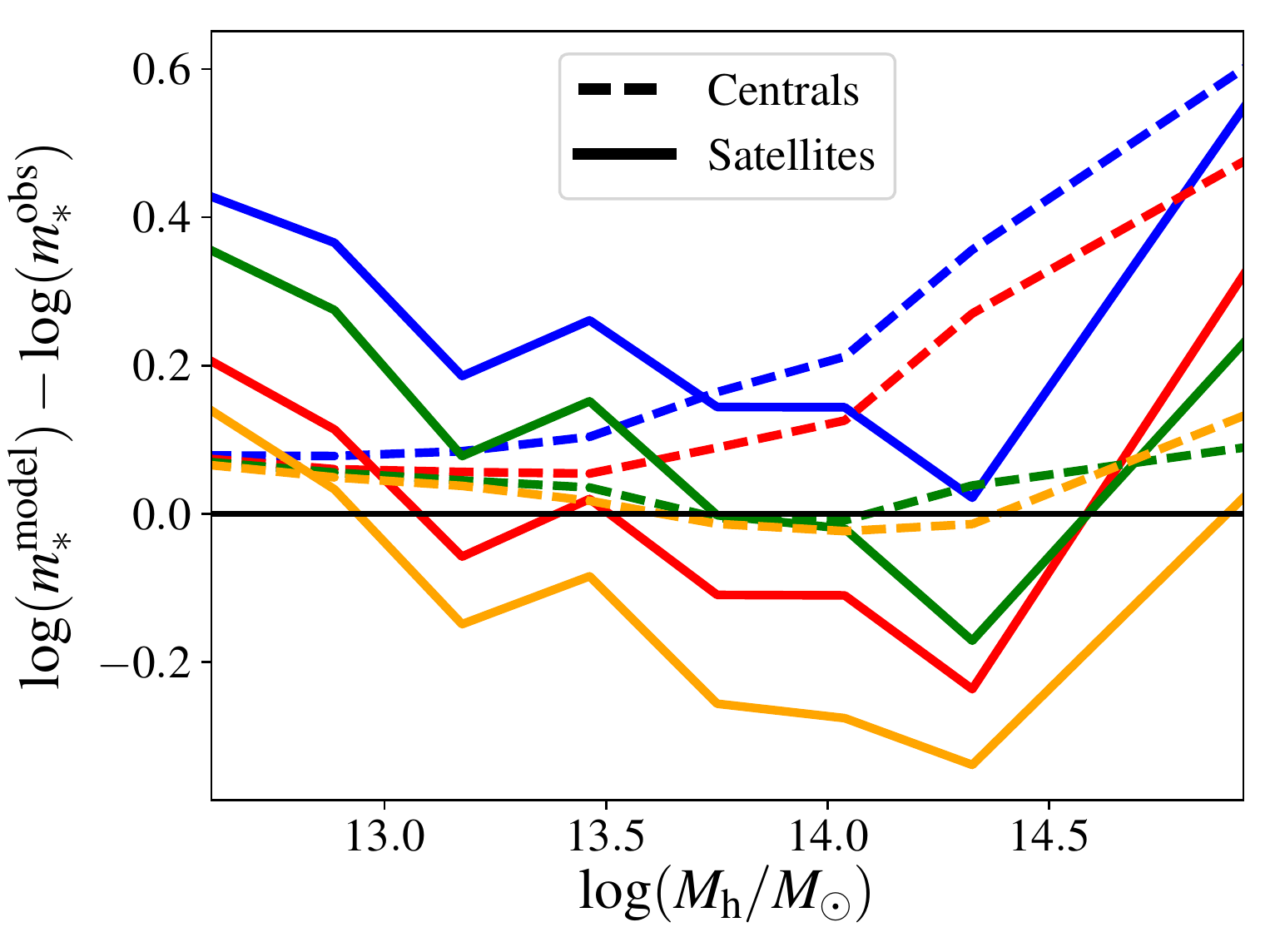} 
\caption{Differences between model predicted total stellar masses and observed
  total stellar masses as a function of host halo (group) mass, for central (\emph{dashed curves}) and
  satellite (\emph{solid curves}) galaxies. 
Models are colour-coded as in Figs.~\ref{fig_evolution}, \ref{fig_smf} and Fig.~\ref{fig_group}.}
\label{fig_diff_log_mass}
\end{figure}

Fig.~\ref{fig_diff_log_mass} compares predictions and observations for the 
conditional SMFs displayed  in Fig.~\ref{fig_group} in a more quantitative
manner. 
For each bin of group mass $M_{\rm h}$, we compute the mean stellar mass of the central galaxy and the mean total mass of all satellite galaxies, 
and compare these masses to observations by taking their logarithmic differences.
A model in perfect agreement with the observations would coincide with the black horizontal line ${\rm log\,}m_*^{\rm model}-{\rm log\,}m_*^{\rm obs}=0$ everywhere.
$M_{\rm lim}$ has been tuned so that the models with tidal stripping match
well the observations at all stellar masses and group masses, but results for central galaxies (\emph{dashed curves}) at $M_{\rm h}< 10^{13}-10^{13.5}\,{\rm M}_\odot$ 
and satellite galaxies (\emph{solid curves}) at all masses are insensitive to the value of $M_{\rm lim}$.
The solid curves should be interpreted with caution because they are the result of a sum over all $m_*$.
For instance, the shutdown+stripping model (orange curve) 
is above the data at $M_{\rm h}\sim 10^{12.5}\,{\rm M}_\odot$ because of a few massive satellites,
while the conditional SMF for the corresponding model is below most data
points in the mass bin 
$12.29< \log(M_{\rm h}/{\rm M}_\odot)<12.57$
(Fig.~\ref{fig_group}).
However, there are two considerations that we can make from the solid curves in Fig.~\ref{fig_diff_log_mass}:
\begin{itemize}
\item{Tidal stripping reduces the total stellar mass of satellite
galaxies by typically $0.1\,$dex ($0.2\,$dex at most, green vs. blue and orange
vs. red curves).}
\item{The starvation model with tidal stripping (green curve) and the shutdown model without tidal stripping (red curve)
provide a comparably good fit to the total stellar mass of satellites in a group.}
\end{itemize}
Therefore, if we were to draw a conclusion based on the total stellar mass of satellites alone, 
we should concede that there is a degeneracy between the gas mass that accretes onto galaxies and the stellar mass that is stripped from them, and
that  a model with stellar mass loss
 intermediate between the predictions of the shutdown model and the starvation model
 (between the red and the blue curves)
could fit the observed conditional SMF without the need for any tidal stripping.
Nevertheless, tidal stripping is expected to occur on physical grounds.
Furthermore, there is observational evidence outside this work that
the shutdown of star formation in satellite galaxies is not instantaneous (see the discussion in Sect.~\ref{gasacc} and references therein).
Hence, it is reassuring that the most astrophysically plausible model is the one that returns a comparatively best fit to the data in Figs.~10-11.

\subsection{ICL}
\label{icl}
\begin{figure}
\includegraphics[width=1.\hsize,angle=0]{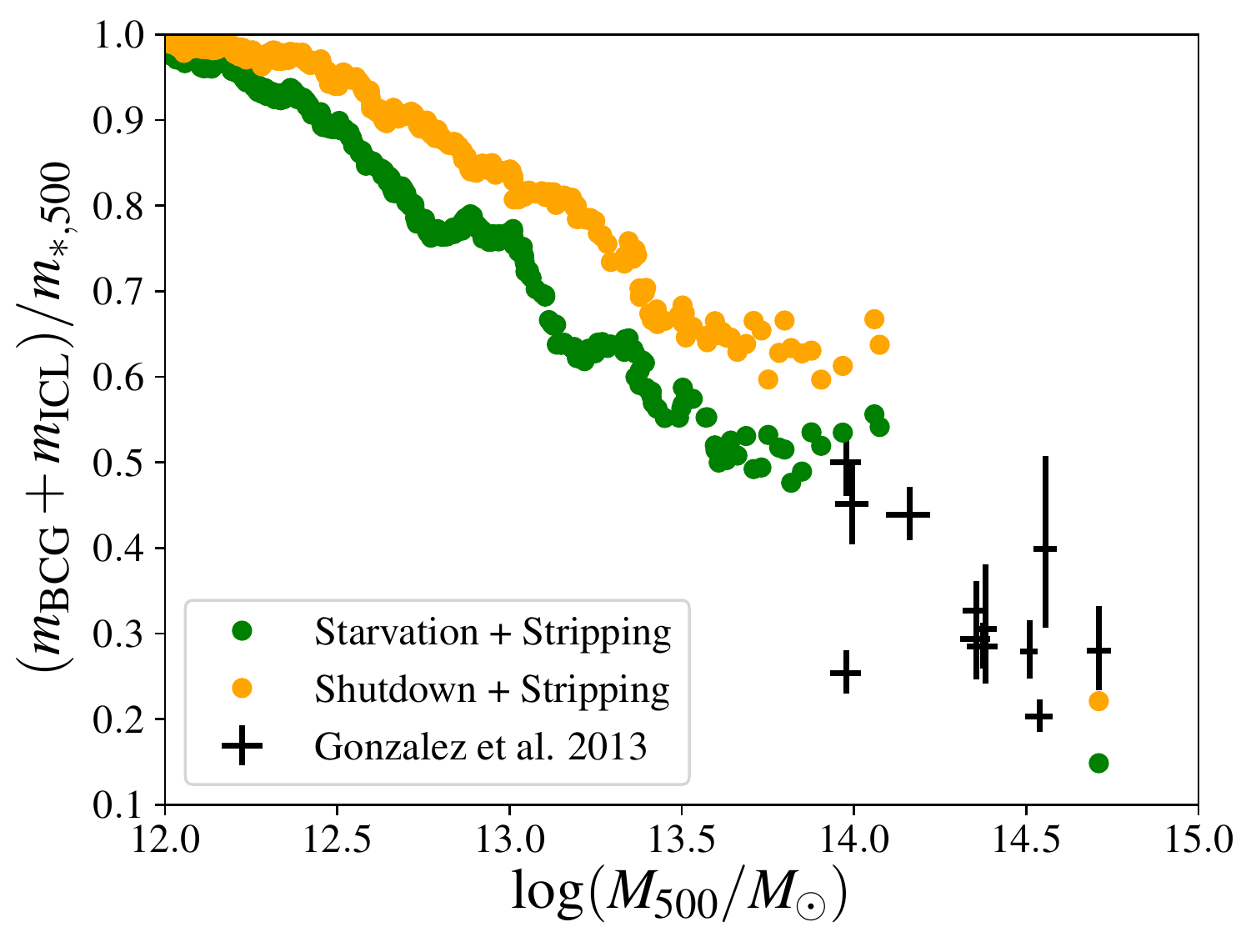} 
\caption{Fractional contribution of cD galaxy inclusive of its ICL to the total stellar mass of a group or a cluster.
Only models with tidal stripping display an ICL component.
Green and orange circles refer to the starvation model and the shutdown model, respectively.
The crosses are the error bars for the observations of
\citet{gonzalez_etal13}.
Model predictions have been shown as a function of $M_{500}$ (the total mass enclosed in 
 sphere of radius $R_{500}$, whithin which the average density  equals 500 times the
critical density of the Universe) to match the definition of cluster mass used by \citet{gonzalez_etal13}.
}
\label{fig_cd}
\end{figure}

As an independent test of our model, we have compared
 our predictions for the contribution of cD galaxies (inclusive of the ICL) to the total stellar masses of clusters with 
the observations of \citet{gonzalez_etal13}.
Fig.~\ref{fig_cd} shows that, although there are very few clusters in our N-body simulations, our starvation
plus tidal stripping model (green circles) matches the observed trend (crosses)
for the ratio of BCG+ICL stellar mass over total stellar mass.

\begin{figure}
\includegraphics[width=1.\hsize,angle=0]{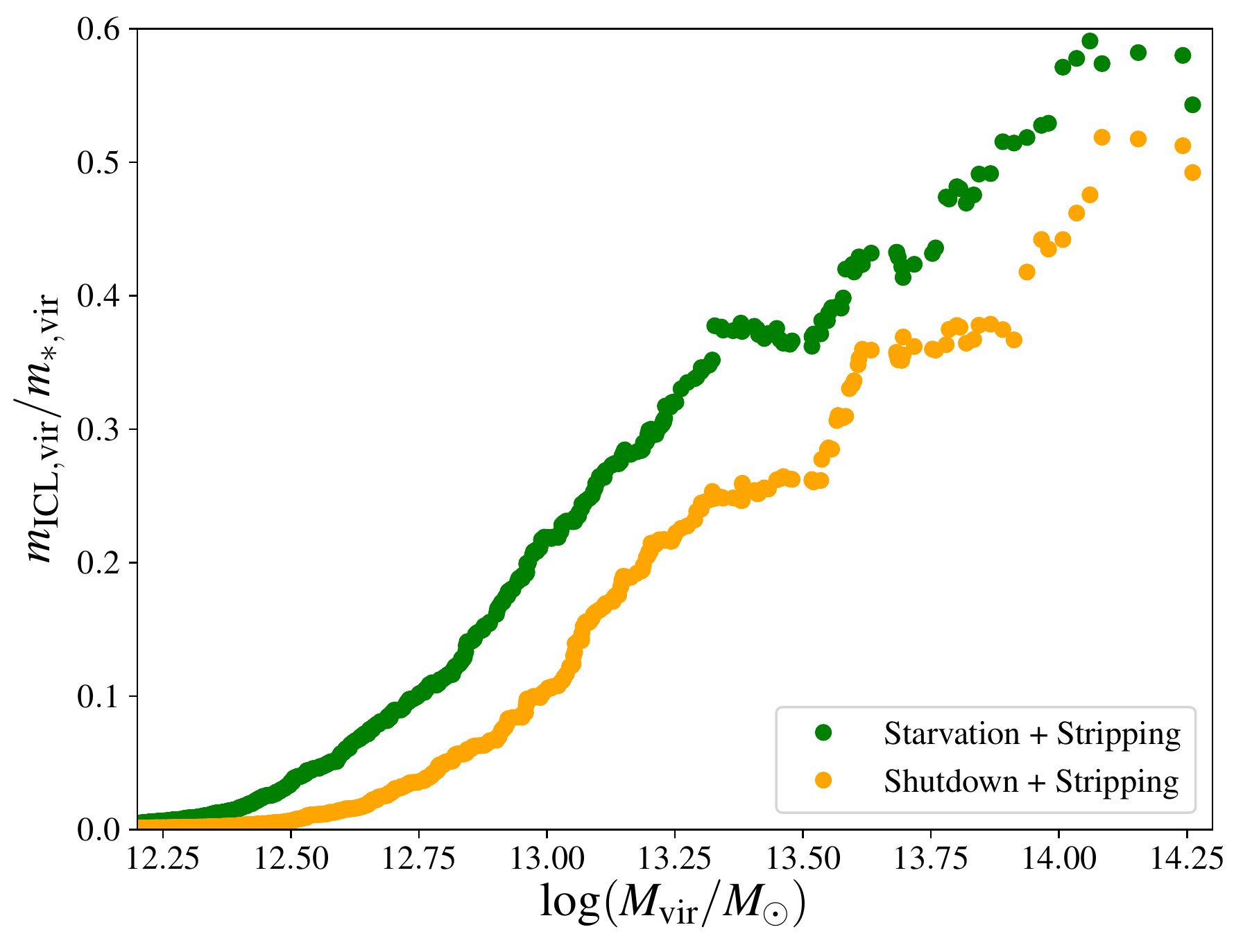} 
\caption{Fractional contribution of the outer envelopes to the total stellar masses within $R_{500}$.
This represent the total diffuse light fraction within $R_{\rm vir}$.
The model with gas accretion (green) allows for more stripping than the model in which accretion
shuts down immediately when a galaxy becomes a satellite (orange).
}
\label{fig_outerenv}
\end{figure}

More interesting (but more difficult to compare with observations) is the
contribution of the ICL 
to the total stellar mass within a cluster.
Fig.~\ref{fig_outerenv} shows this contribution when we consider not only the extended envelops of cD galaxies but all stars stripped from galaxies over the entire cluster out to $R_{\rm vir}$.
It shows that stellar mass fraction in the ICL increases with halo mass.
 
In Fig.~\ref{fig_cd}, we had shown the BCG+ICL mass fraction within $R_{500}$ (the radius of a sphere within which the mean density equals $500\,$times the critical density of the Universe)
for consistency with \citet{gonzalez_etal13}. In our cosmology, the virial radius corresponds to $\Delta_{\rm c}=102$ at $z=0$, but we find that the ratio of the ICL mass $m_{\rm ICL}$ to the total stellar mass $m_*$
is very similar within $R_{\rm vir}$ and $R_{500}$ for haloes up to $\sim 10^{13.5}\,M_\odot$.
In clusters, $m_{\rm ICL}/m_*$ is $\sim 20\%$ smaller within $R_{500}$ than it is within $R_{\rm vir}$.
The implication is that the ICL is more concentrated than the total light of the cluster\footnote{Stripped stars are stored first in an ICL component associated with it parent satellite, when the satellites merge, we transfer the stellar mass in this component to the ICL associated with the central galaxy. Therefore we define $m_{\rm ICL}$ within $R_{\rm vir}$ (respectively $R_{500}$) as the sum of the mass of the ICL component from galaxies within $R_{\rm vir}$.}.
Fig.~\ref{fig_outerenv} shows that, in clusters, the ICL, defined as the total stellar mass stripped from galaxies, whether it still forms a tidal stream around the galaxies themselves or whether it has merged into the extended envelope of a cD galaxy,
may amount to nearly half of the total stellar mass with $R_{\rm vir}$.

In an article that appeared when ours was about to be submitted, \citet{bernardi_etal17} argued against the interpretation that difference between the SMFs of \cite{baldry_etal12} and \cite{bernardi_etal13} is due to the ICL. Their claim is that the difference is entirely due to the different way the photometry is done.
The magnitude provided by the SDSS are based on fitting an exponential and a de Vaucouleurs surface-brightness profile separately and retaining the value for the profile that fits better. The Python image-morphology software PyMorph fits the surface-brightness profiles of galaxies much more accurately because it allows for the presence of both an exponential and a Sersic component.
Bernardi et al. (2017) correctly argued that the difference is more than semantic because there is no doubt that a model with five free parameters can fit the surface-brightness profiles of galaxies more accurately than a model with two and thus return more accurate photometry.

As a Sersic-exponential profile provides an excellent fit to the surface-brightness profiles of luminous red galaxies out to eight effective radii (about 100 kpc),
\citet{bernardi_etal17} concluded that the difference between PyMorph and SDSS magnitudes cannot be due to the ICL. 
This conclusion is based on the fact that they define the ICL as any residual luminosity above the Sersic-exponential fit.
This definition is entirely reasonable from an observer’s standpoint. 
However, the Sersic-exponential profile is nothing more than a useful fitting formula. Another functional form with more free parameters may fit the surface-brightness profile far beyond eight effective radii, eliminating the need for the ICL altogether.
We do not question the claim by Bernardi et al. (2017) that PyMorph returns objectively more accurate magnitudes than the SDSS pipeline.
We enquire about the physical reason why giant ellipticals have extended light profiles, be they or not above a Sersic-exponential fit.
Following \cite{gallagher_ostriker72}, we pursue the hypothesis that the extended envelopes of giant ellipticals are the debris of tidally disrupted galaxies,
and define the ICL as the light from stars that have been tidally stripped from galaxies.
This definition is of no assistance to an observer who wishes to measure the ICL.
However, it is significant that when we compute the ICL mass according to our definition, we recover a lot of the difference between the SMFs of \cite{baldry_etal12} and \cite{bernardi_etal13} (see Fig.~\ref{fig_smf}, the gap between the blue and the green curve).

Bernardi et al. (2017) have also argued that the ICL should be centred on the centres of the clusters and should thus affect the magnitudes of central
galaxies more than it affects those of satellites, while their work shows that, for a same luminosity, the difference between PyMorph and SDSS magnitudes is about the same for both central and satellite galaxies. However, this is not a problem if one adopts our definition of the ICL because tidal stripping is expected
to affect satellite galaxies, too. In fact, satellite galaxies will first develop long tidal tails and then these tails will coalesce into the extended envelopes of the central systems. This is how minor mergers have plausibly contributed to the considerable size evolution of elliptical galaxies from $z = 2$ to the present
(e.g., Naab et al. 2009, van Dokkum et al. 2010, Tal and van Dokkum 2011, Cooper et al. 2012, Shankar et al. 2013).


\section{Discussion}
\label{discus}
In this section, we discuss the uncertainties that affect our results. They come from 
i) the resolution of the N-body simulation and the model for orphan galaxies that we have introduced to overcome the effects of limited resolution;
ii) uncertainties regarding the amount of star formation a satellite galaxy experiences post accretion onto the host halo;
iii) uncertainties regarding the amount of stellar mass loss experienced by satellite galaxies as a consequence of tidal stripping and heating;
iv) scatter in the SMHM relation; and v) the AM method itself.
We also discuss the excess of massive satellites in groups with $M_{\rm h}<10^{13.44}\,{\rm M}_\odot$ predicted by all our models.

\subsection{Model uncertainties}

\subsubsection{N-body resolution and orphan galaxies}
\label{discuss_strip}

\begin{figure}
\includegraphics[width=1.\hsize,angle=0]{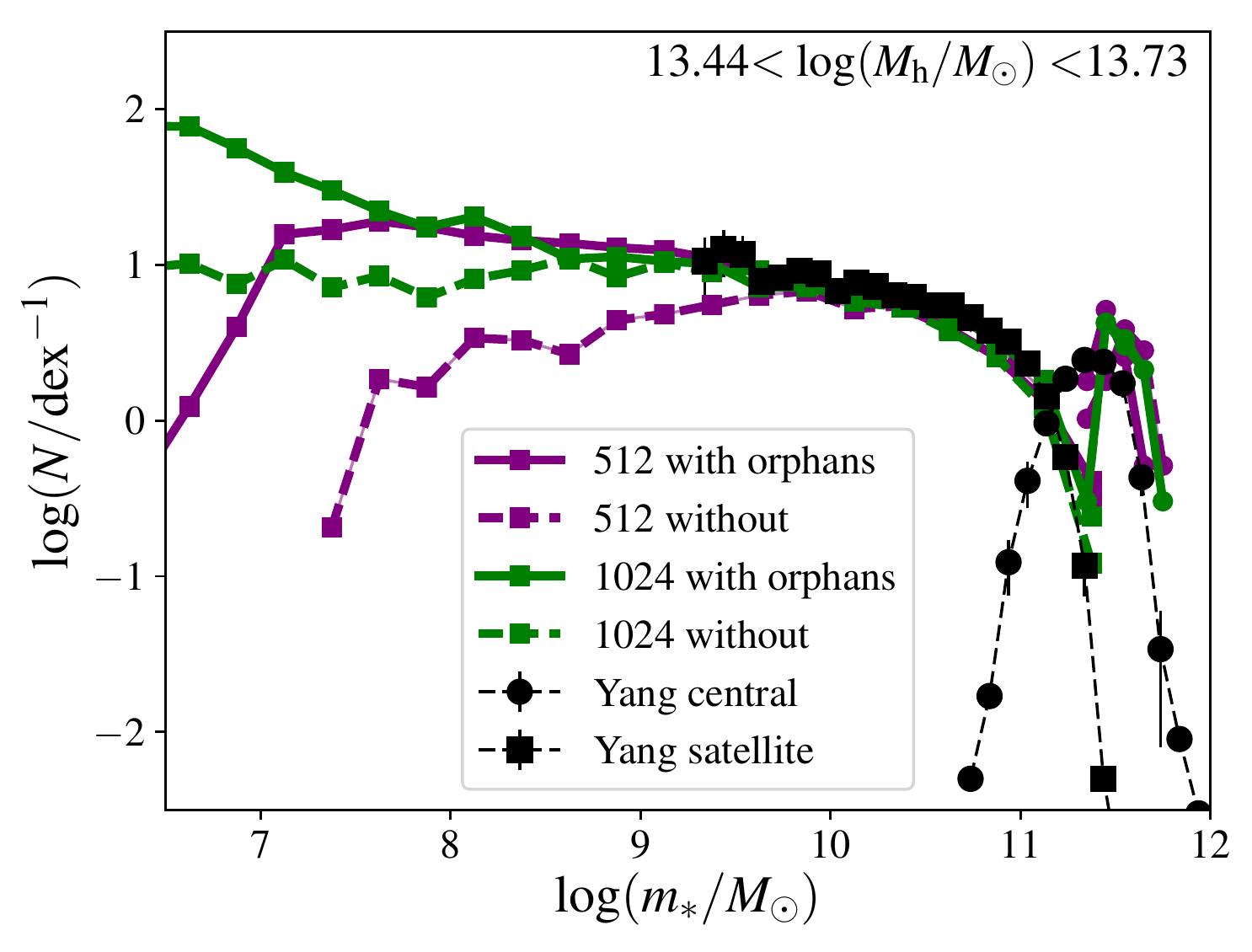} 
\caption{Sensitivity of the conditional stellar mass function predicted by the starvation+stripping
  model 
to the resolution of the N-body simulation, degraded from $1024^3$ (\emph{green}) to
$512^3$ particles (\emph{purple lines}), and to the merger time of orphan
galaxies with the central galaxy: immediately (\emph{dashed lines}) or only at the
first pericentre following the expected time of orbital decay by dynamical
friction (\emph{solid lines}). 
The figure shows the case for haloes with $13.44 <
  \log(M_{\rm h}/{\rm M}_\odot) < 13.73$.
In the mass range probed by the observations (\citealt{yang_etal12}; \emph{black symbols} with error bars), 
the model with orphan galaxies has converged because the simulations with
$512^3$ and $1024^3$ particles give very similar conditional SMFs.
}
\label{fig_comp512-1024}
\end{figure}

In Sect.~3.2, we have treated ghost subhaloes as systems with well-defined orbits in a static spherical potential. Cosmological haloes 
contain substructures that contribute to their gravitational masses and perturb the orbital motions of subhaloes.
Our work does not consider the contribution of substructures to the gravitational potential of their host because our calculations are based on exclusive masses
(our halo masses do not include the masses of substructures. We made this choice because the NFW model fits the density profiles of DM haloes more accurately when substructures are removed.
If the mass distribution of substructures followed the NFW profile of the host halo, 
their merging timescales would be shorter  by typically $10\%$.
In reality, it is entirely possible that the interaction with other substructures may scatter a subhalo on an orbit with a longer merging timescale.
However, \citet{hayashi_etal07} have shown that the isopotential surfaces inside a halo are much smoother than 
the density distribution and relatively insensitive to the presence of
substructure.

The assumption that haloes are spherical is another simplification. Real haloes are triaxial. At $z=0$, the typical minor-to-major axis ratio of the virial ellipsoid ranges from $0.75$ at $M_{\rm h}\sim 10^{11}\,M_\odot$
to $0.5$ at $M_{\rm h}\sim 3\times 10^{14}\,M_\odot$ \citep{despali_etal17}. Triaxiality increases at small radii but
dissipation makes DM haloes substantially rounder at small radii than suggested by dissipationless simulations \citep{springel_etal04}.
Furthermore, as expected from Poisson's equation,
the gravitational potential tends to be much more spherical than the mass distribution.
Indeed,
\citet{hayashi_etal07} find that a flattening (minor-to-major axis ratio) of
$\sim 0.4$ in the mass distribution corresponds to a flattening of only $\sim
0.75$ for the isopotential contours the minor to major axis ratios of the isopotential contours are $\approx
0.75$, hence much greater than the corresponding ratios for the density contours ($\approx 0.4$).

In this work, the approximation of a static spherical potential applies to ghost subhaloes only. At a given stellar mass, the fraction of satellite galaxies with unresolved (ghost) subhaloes depends on the resolution of the
N-body simulation. If all subhaloes of satellite galaxies with $m_*>10^9\,M_\odot$ were resolved, our results would be independent of this approximation.
Hence, while it is difficult to estimate, {\it a priori}, the errors introduced by treating ghost subhaloes as systems with well-defined orbits in a static spherical potential,
it is easy to do it, {\it a posteriori}, by performing resolution studies.

To test the sensitivity of our results to N-body resolution and to our
modeling of orphan galaxies, we have repeated our entire analysis on a simulation with the same cosmology, the same volume, the same initial conditions,
but only $512^3$ particles instead of $1024^3$, and we allow ourselves to immediately merge
orphan galaxies with the central galaxy of their host halo (i.e. to use the original merger tree without the addition of ghosts and orphan galaxies).
We focus the comparison on our best fit model (starvation plus stripping) and on the mass range $10^{13.44}\,{\rm M}_\odot<M_{\rm h}<10^{13.73}\,{\rm M}_\odot$,
but the results for this case also apply to the other models and mass bins.

Fig.~\ref{fig_comp512-1024} compares the conditional SMF for this model and
mass range varying the resolution of the simulation and the treatment of
orphans.
With $512^3$ particles, the conditional SMFs with (\emph{solid purple curve}) and without (\emph{dashed purple curve}) orphans 
differ at $m_*\lsim 10^{10}\,{\rm M}_\odot$.
However, with $1024^3$ particles, the resolution is so good that delaying the mergers
of orphans with central galaxies (\emph{solid green curve}) or
not (\emph{dashed green curve}) makes little difference above $m_* = 10^{8.5}\,\rm M_\odot$.
The treatment of orphans is a small correction and therefore a negligible source of uncertainty in relation to our conclusions.

Above $m_*\sim 10^{10}\,{\rm M}_\odot$, the conditional SMFs for the $512^3$ simulation without orphans (\emph{dashed purple curve}) and the $1024^3$ simulation without orphans (\emph{dashed green curve})
are very similar, suggesting that numerical convergence has been reached.
Most interesting, however, is the agreement of the $512^3$ and $1024^3$
simulations 
when orphans are included, as we see convergence (solid green and purple
lines) 
in the conditional SMF down
to $10^{7.7}\,\rm M_\odot$.
This proves that the inclusion of orphans aids in achieving convergence to correct solution (also see \citealt{guo_etal11}).
At $m_*\gsim 3\times 10^8\,{\rm M}_\odot$, the $512^3$ simulation with orphans is at least as good as the $1024^3$ simulation without orphans.

\begin{figure}
\includegraphics[width=1.\hsize,angle=0]{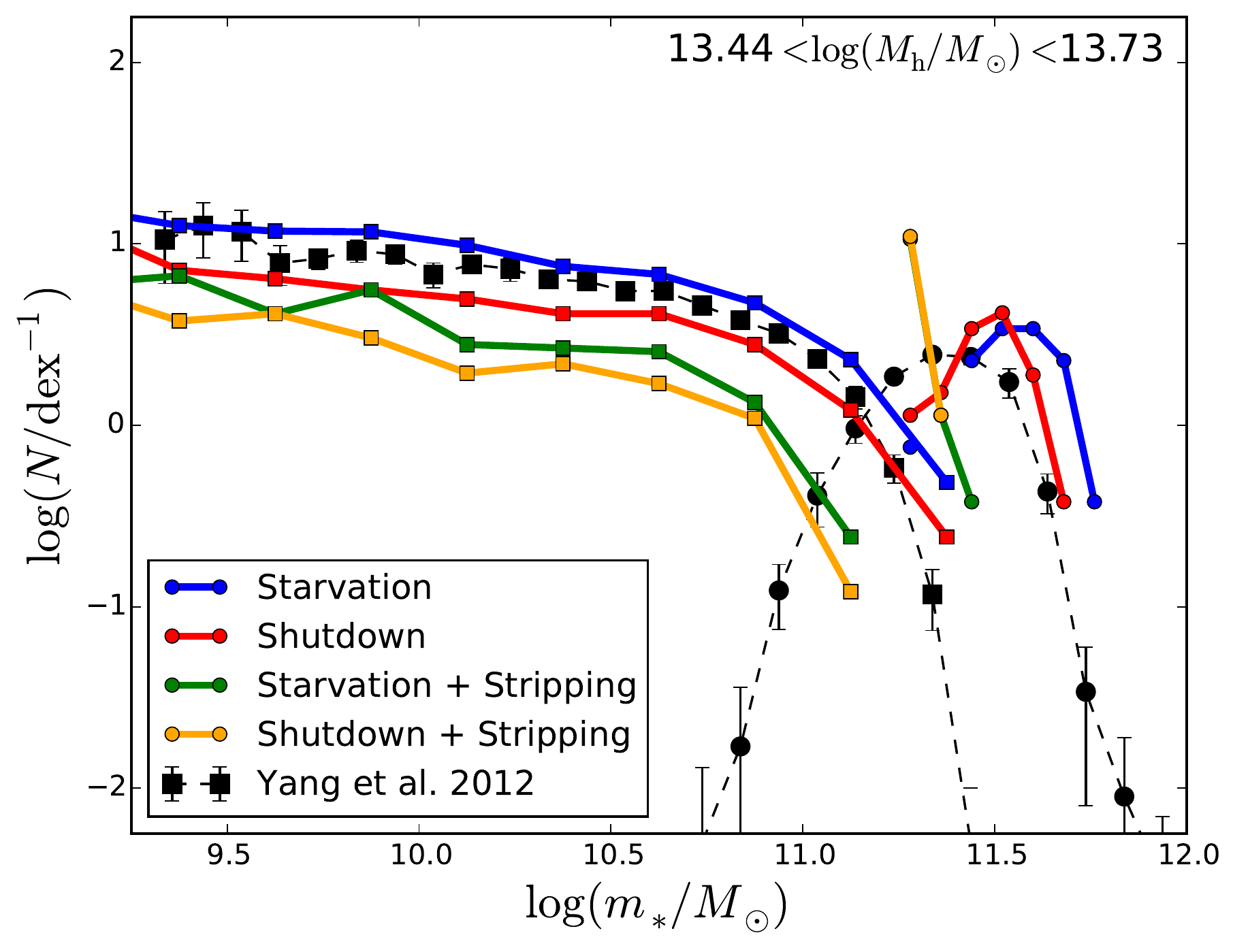} 
\caption{Conditional SMF for our different models (curves) compared to the observations of Yang et al. (2012; black data points with error bars).
This figure is identical to the $10^{13.44}\,{\rm M}_\odot<M_{\rm h}<10^{13.73}\,{\rm M}_\odot$ panel of Fig.~\ref{fig_group} except for the calculation of tidal stripping (the green and orange curves).
Here, tidal stripping is computed in the circular-orbit instantaneous-tide approximation, using
Eq.~(\ref{tidal_radius}) with $\alpha=-3$.}
\label{fig_csmf}
\end{figure}

\subsubsection{Gas accretion onto satellites}
\label{gasacc}
Estimating how much gas accretes onto satellite galaxies after entering a group or cluster environment is less straightforward than testing for resolution effects,
but the simple assumption that star formation shuts down immediately cannot be correct.
\citet{weinmann_etal06} used a version of the Munich semianalytic model in which there was no accretion onto satellite galaxies \citep{croton_etal06}.
They found that the fraction of faint satellites with red colours was overestimated by a factor of $\sim 2-3$.
All semianalytic models published in those years shared the same problem (e.g., \citealt{fontanot_etal09}). Indeed,
\citet{cattaneo_etal07} ran the GalICS semianalytic model on merger trees from a cosmological hydrodynamic simulation.
GalICS assumed no gas accretion on satellites and predicted a much higher fraction of quenched galaxies than the hydrodynamic simulation.

A delayed quenching scenario can be parameterised by two timescales: the time $t_{\rm delay}$ during which a galaxy
keeps forming stars after entering a group or a cluster and the $t_{\rm quench}$, over which the star formation
rate rapidly decays after $t_{\rm delay}$ has elapsed.
Several authors have investigated these timescales.
\citet{mahajan_etal11} split galaxies between infalling, backsplash and
virialised, and combined the fraction of star forming galaxies
observed in the SDSS with cosmological $N$-body simulations to quantify
projection effects. Their analysis suggests that quenching is delayed until
galaxies reach the virial radius on their way out of the cluster  after the
first pericentric passage.
As shown in Fig.~\ref{fig_tff} in Appendix~\ref{orbtimes},
galaxies that are at the virial radius today, on their way out after their
first pericentric passage, and have typical 1st apocenters close to the
turnaround radius at that time (3 to 4 virial radii at that time),
 entered the group/cluster environment $\sim 3\,$Gyr ago
 (Fig.~\ref{fig_tff}
 in Appendix~\ref{orbtimes}), and passed through the pericenter $\sim
 1.6\,$Gyr ago. 
Therefore,
according to the modeling of \citeauthor{mahajan_etal11}, star formation is
quenched $\sim 3$ Gyr after cluster entry and $\sim 1.6\,$Gyr after the first pericentric passage.
\citet{wetzel_etal13} used an N-body simulation to measure the characteristic time since $t_{\rm entry}$ of a galaxy at a given $R/R_{\rm vir}$,
and constrained $t_{\rm delay}$ and $t_{\rm quench}$ by measuring the fraction of red galaxies in SDSS groups/clusters
as a function of the distance from the centre.
A slow progressive fading of star formation since $t_{\rm entry}$ would blur the bimodal distribution of galaxy colour.
In contrast, the observations are consistent with a long delay ($t_{\rm
  delay}= 2-4\,$Gyr) followed by rapid quenching ($t_{\rm
  quench}=0.2-0.8\,$Gyr).
\citet{haines_etal15} performed a similar study to match the observed  distribution of
the fraction of star-forming galaxies with the predictions of times since entry as a function of
position in projected phase-space (PPS) from cosmological $N$-body simulations. They
conclude that star formation declines exponentially after entering the virial
radius on a timescale of $1.7\,Gyr$.
A similar analysis by \citet{oman_hudson16} suggests that star formation in cluster satellite galaxies is rapidly quenched 
within $\sim 1-2\,$Gyr from the first pericentric passage.
Another recent similar study based on both SDSS and higher-redshift data leads to delay times of 2 to
$5\,Gyr$ \citep{fossati_etal17}.

Further evidence in support of delayed quenching comes from chemical abundances.
When a galaxy ceases to accrete pristine gas but keeps forming stars, its metal content relative to hydrogen increases.
From the metal abundances of red galaxies with $m_*\lsim 10^{10.5}\,{\rm M}_\odot$, 
\citet{peng_etal15} inferred that they must have behaved as closed boxes for
$\sim 4\,\rm Gyr$ before they eventually run out of gas.
The higher metallicities of satellite galaxies were interpreted as evidence that this is due to starvation by the environment.
While the complete starvation of gas accretion in \citeauthor{peng_etal15}'s picture seems to conflict with semianalytic models\footnote{The difference is largely due to the assumed star formation efficiencies.
Semianalytic models usually assume
shorter star formation timescales than those of \citet{peng_etal15}, at least at low stellar mass.
Hence, they need sustained accretion to keep star formation going for several gigayears.},
there is consensus that star formation cannot have been quenched instantaneously at $t_{\rm entry}$.

In conclusion, while it is not straightforward to determine what fraction of
the gas associated with a subhalo will accrete onto the satellite galaxy it
contains and what fraction will be stripped (mainly by ram pressure, which is
more important than tidal stripping for gas\footnote{In field galaxies, {\sc Hi} discs are more extended than stellar discs. In satellite galaxies, it is often the contrary
because ram pressure has stripped their outer parts. Were tidal stripping the dominant phenomenon, the {\sc Hi} disc would be truncated at the same
radius as the stellar disc because tides do not differentiate between gas and stars.}),
and while the precise value will also depend on the feedback one assumes
\citep{tomozeiu_etal16},
there appears to be observational consensus that star formation is quenched
$1 - 2\,$Gyr after the first pericentric passage.
Therefore, the starvation model, which prevents further accretion from the environment but allows star formation to continue until the first pericentric passage,
seems a much more plausible assumption than to assume a complete shutdown of star formation at the entry time.

\subsubsection{Tidal stripping}
\label{discus_tides}

Tidal stripping is an inevitable dynamical process, but its analytic modelling is not straightforward and requires simplifying assumptions.
The most common assumption is instantaneous tides applied to satellites are on circular orbits. However, for a fixed pericentric radius, this assumption gives an upper limit rather than a realistic estimate for the stellar mass
that is tidally stripped from galaxies because most orbits are highly elongated.
The circular-orbit approximation underpredicts the conditional SMF even when it is applied to the starvation model, which corresponds to the maximum
possible star formation in satellite galaxies (\emph{green curve} in Fig.~\ref{fig_csmf}) because it results
in stellar mass loss of satellite galaxies that can be as large as 0.5 dex.
On the contrary, the difference between the starvation model without stripping (\emph{blue curve}) and the observations (\emph{horizontal black line}) in Fig.~\ref{fig_diff_log_mass}
shows that  the stellar mass that can plausibly be stripped from galaxies is $\lsim 0.15-0.2\,$dex ($\lsim 0.1-0.15\,$dex if we allow for $\gsim 10\%$ stellar mass loss through stellar evolution between $z_{\rm entry}$ and $z=0$).
This upper limit is obtained by comparing a model without stripping to the observations. It is therefore totally independent of any physical model of tidal stripping.

Reassuringly, the more sophisticated model in Sect.~\ref{tidesstars} predicts tidal stripping by $\sim 0.1\,$dex ($\sim 25\%$) on average, in agreement with the upper limit above.
This value (based on the green curve in Fig.~\ref{fig_diff_log_mass}) has been computed assuming the maximum tidal acceleration but also assuming that the tidal acceleration acts only 
for a very short time around the pericentric passage (corresponding to the part of the orbit shown as a \emph{thick} solid red line in Fig.~\ref{fig_traj}).
Using the average acceleration for a test particle (equal to half the maximum acceleration; Appendix~A) while retaining the second assumption will most likely underestimate the tides.
However, this results in an average tidal stripping of $0.07-0.08\,$dex on average, so the quantitative difference is small.

The assumption of circular orbits is as incorrect for the DM as it is for the stars, but we kept using it to compute the DM lost by ghost subhaloes because
DM is stripped all the way down to the centre, not just at pericentre \citep{klimentowski_etal09}, so the impulsive approximation is not necessarily much more accurate.
The question is the extent to which its inaccuracy affects our conclusions.
Tidal radii have no consequences on the survival times of ghost subhaloes, which are by \citet{jiang_etal08}'s formula.
Their only effect on the tidal stripping of stars is through the value of $R_{\rm p}$.
If a ghost subhalo is stripped too heavily (the most likely outcome of our approximation), it will suffer less dynamical friction.
Less dynamical friction implies less orbital decay. The pericentric radius will be overestimated and the tidal stripping of stars will be less efficient than for the correct value of $R_{\rm p}$.
Therefore, we can be confident that $0.07-0.08\,$dex is a plausible lower limit for the stellar mass lost by galaxies owing to dynamical friction.

In practice, ghost subhaloes (and thus this calculation) were introduced as a way to beat the resolution limit of our N-body simulation.
A posteriori, our resolution is so good that conditional SMFs are very similar with or without them (Fig.~\ref{fig_comp512-1024}).
Thus, any error in our calculation of the pericentric radii of ghost subhaloes is bound to have a limited impact on the conclusions of this article.

\subsubsection{Scatter in the SMHM relation}

Our AM procedure (Sect.~1.2) assumes that the $m_*$ - $M_{\rm max}$ relation does not contain any scatter. In this section, we discuss how scatter can affect our conclusions.

Observationally, to determine the scatter in stellar mass at constant halo mass, one needs a method to measure $M_{\rm h}$.
 \citet{yang_etal09} estimated the masses of groups from their luminosities (Sect.~6.2) and found a scatter in ${\rm log}\,m_*$ of $\sigma_{{\rm log}\,m_*}\simeq 0.17\,$dex in $m_*$ at constant $M_{\rm h}$.
\citet{more_etal09} found a similar result ($\sigma_{{\rm log}\,m_*}\simeq 0.16\,$dex) using halo masses from satellite kinematics.
\citet{leauthaud_etal12} performed a more sophisticated analysis by fitting simultaneously the galaxy SMF, clustering data (correlation functions) and halo masses from galaxy-galaxy lensing.
They found an intrinsic scatter of about $0.2\,$dex after subtracting errors
from photometry, photometric redshifts and spectral-energy-distribution (SED)
fitting.
\cite{behroozi_etal13} used AM to infer a scatter of $0.22\pm0.02$ dex.
\citet{coupon_etal15} repeated the same analysis with more recent data and confirmed their results.

Implementing scatter in our models requires a more sophisticated approach than simply applying random errors to the stellar masses determined from the AM relation.
If we simply perturbed the AM relation, our models would no longer reproduce the galaxy SMF because of the Eddington bias.
We overcome this problem by splitting the galaxy population into pairs. One galaxy has logarithmic stellar mass ${\rm log}\,m_*$. The other has logarithmic stellar mass ${\rm log}\,m_*+{\rm log}\, \Delta{m_*}$, where ${\rm log}\, \Delta{m_*}$
is a random number from a Gaussian distribution with standard deviation $\sigma_{{\rm log}\,m_*}$ and zero mean.
Scatter is implemented by swapping the haloes of the two galaxies. This swapping introduces the
requires scatter in the stellar mass-halo mass relation without changing
the actual stellar mass function of the galaxies.

This procedure means that the conditional SMFs computed by our models are now dependent on the random way in which galaxy population has been splitted into pairs
but we can obtain robust results by averaging over many different realisations. The thin green dashed curves in Fig.~12 show the median conditional SMF for the starvation plus stripping model
over a hundred realisations with $\sigma_{{\rm log}\,m_*}=0.2$, while the green shaded areas show upper and lower quartiles for the same hundred realisations.
The absence of systematic differences between the thin green curves and the thick green curves (the model without scatter) proves that scatter adds noise but will not bias our conclusion.
Thin lines and shaded areas have been shown for the starvation plus stripping model only not to overcrowd the figure.

\subsubsection{Uncertainties in the SMHM relation}

Fig.~\ref{fig_msmh} shows that the SMHM relations derived with different AM/HOD models differ at the level of $0.1$ - $0.2\,$dex in stellar mass.
One could interpret these differences as a measure of the intrinsic uncertainty of the SMHM relation from AM.
To understand the implications that such an uncertainty may have for our results,
we begin by discussing the origin of these differences. 

The first source of difference is the SMF used to constrain the SMHM relation.
The SMHM relation of \citet{moster_etal13} differs from those of 
other authors at low masses because they used the SMF of \citealp{li_white09}, which contains a higher a number density of galaxies with $10^9\,M_\odot<m_*<10^{10.2}\,M_\odot$ than the SMFs
of \citet{baldry_etal08,baldry_etal12},  \citet{leauthaud_etal12},
\citet{yang_etal12}, \citet{papastergis_etal12} , and
\citet{moustakas_etal13}.

Secondly, the halo mass functions assumed by different authors can come either from N-body simulations or from the \citet{sheth_tormen02} formula, which is calibrated on N-body simulations.
Even if the cosmologies assumed by different authors were completely identical, there would still be an uncertainty of about $10\%$ in the halo mass function from the halo finder \citep{knebe_etal13}.

Finally, one can use the AM method, as we have done, or one can assumed a parametric SMHM relation and constrain its parameters so that it fits the SMF.
The results obtained with the two methods will be very similar but not necessarily identical.
One can also consider or not consider the presence of scatter (Sect.~7.1.4).
Models with scatter find lower stellar masses for a given halo mass at high masses to compensate for the Eddington bias.

We argue that these systematic uncertainties are not important for our conclusions because our analysis focusses on the differences in stellar mass between field and satellite galaxies.
We apply the same N-body simulation, the same halo finder and the same AM procedure to both. 
Therefore, these uncertainties cancel out in the relative comparison, as would any systematic error in the photometry or the initial mass function.

The only real question is whether our SMF $n_*(m_*,z)$, which is constructed from data at different redshifts (\citealp{yang_etal12} at $z<0.2$, \citealp{muzzin_etal13} at $0.2<z<2.5$),
is fully consistent with the conditional SMF of \citet{yang_etal12}, to which we compare our results.
Fig.~\ref{fig_massfun} shows that, for  $10^9\,M_\odot<m_*<10^{11.5}\,M_\odot$
the local SMF assumed for this work (black solid curve) is fully consistent with both the SMF of Yang et al. (2012; black symbols with error bars)
and the SMF of \citet{muzzin_etal13} extrapolated to $z=0.1$ (black dashed curve).

\subsection{The massive satellite excess in low-mass groups}

The most noticeable discrepancy between our models and the observations of \citet{yang_etal12} is the excess of massive satellites with $m_* \gta 10^{11}\,M_\odot$ in low-mass groups (Fig.~\ref{fig_group}).
This excess cannot be due to star formation after $z_{\rm entry}$ or to underestimated stripping because, at $M_{\rm h}\lta 10^{13}\,M_\odot$ it is present even in the shutdown model when tidal stripping
is computed with the instantaneous-tide circular-orbit approximation (which largely overestimates the magnitude of the phenomenon; Sect.~\ref{discuss_strip}).

There are two possible explanations for this discreapancy.
First, \citet{yang_etal12} may have classified as central systems that, in the N-body simulation, our halo finder classifies as satellites, see \cite{skibba_etal11} and \cite{lange_etal17} for a discussion of this phenomena.
\citet{bernardi_etal17} have analysed SDSS groups with a group finder called {\sc redMaPPer}, which differ from the one used by \citet{yang_etal12}.
They have remarked that: ``Many of the objects which Yang et al. classify as being centrals in groups less massive than $10^{14}\,M_\odot$ are called satellites by {\sc redMaPPer}''.
Second, a satellite with $m_*$ comparable to the central galaxy in a small group of often $2-3$ objects is very different system from a satellite in a cluster,
whose total mass is much larger than that of any satellite.
The \citet{jiang_etal08} formula (combined with our prescription, which requires one galaxy out of four to complete an additional orbit after a time $t_{\rm df}$ has elapsed since entry)
may fail when applied to nearly-equal-mass binary systems.
If their merging time is systematically overestimated, this could explain the excess of massive satellites in low-mass groups.

\section{Conclusions}

The original goal of this work was to estimate the stellar mass lost by galaxies in groups and clustering due to tidal stripping
by comparing the distribution of entry masses to observations of the conditional SMF.
As our work progressed, we realised that this original approach was too simplistic because the distribution of entry masses (shown by the red curves in Fig.~\ref{fig_group}, except for satellites that have merged)
is below the data points for most values of $m_*$. If we look at the mass variation from $z_{\rm entry}$ to $z=0$, then galaxies have gained stellar mass, not lost it.

The simplest refinement of this analysis is to assume that $\Delta m_{\rm strip}= m_{\rm entry}+\Delta m_*-m_*$, where $m_{\rm entry}$ is the stellar mass at $z_{\rm entry}$, $\Delta m_*$ is the mass of the
stars formed between $z_{\rm entry}$ and $z=0$, and $m_*$ is the stellar mass at $z=0$.
The problem is that $\Delta m_*$ is considerably uncertain. 
We can obtain an upper limit for $\Delta m_{\rm strip}$ by assuming that, in satellite galaxies, stellar mass grows with halo mass following the same relation that holds for central galaxies.
This assumption maximises $\Delta m_*$. If we follow this approach, we find that the stellar mass loss from satellite galaxies is $\lsim 0.15-0.2\,$dex.
However, this would imply that central and satellite galaxies have
similar SFRs. This is inconsistent with observations, which indicate
that satellite galaxies have, on average, lower SFRs than centrals of
the same stellar mass (i.e., \cite{weinmann_etal06}, \cite{wetzel_etal13} and references therein). Hence, this model clearly has to
be regarded an extreme upper limit for $\Delta m_*$.

The upper limit for $\Delta m_{\rm strip}$ includes both the stellar mass lost due to tidal
stripping and the decrease in stellar mass that results from stellar evolution.
A typical satellite galaxy is accreted into its host halo at a median
redshift of $z_{\rm entry}\sim 0.5$, which corresponds to a look-back time of
$\sim 5\,$Gyr. Assuming passive evolution, a typical quiescent galaxy will
lose betweeen 10 and 20 percent of its mass over a period of $\sim 5\,$Gyr
(e.g. \citealp{fioc97}). Accounting for this passive
evolution, gives a more stringent upper limit $\Delta m_{\rm strip}\lsim 0.1-0.15\,$dex.

This is an upper limit because of the uncertainty on $\Delta m_*$. Since $m_{\rm entry}-m_*<0$ and $m_{\rm entry}+\Delta m_*^{\rm max}-m_*>0$,
it is possible to find a plausible value of $\Delta m_*$ for which $\Delta m_{\rm strip}= m_{\rm entry}+\Delta m_*-m_*=0$,
that is, the analysis above cannot rule out a model without stripping.

We have compared this indirect result with direct analytic estimates of the stellar masses that galaxies lose due to tidal stripping.
The simplest estimates based on instantaneous tides and circular orbits are highly inaccurate because most satellites are on highly elongated orbits.
These estimates predict much more stripping that is allowed by the upper limit derived in this article.
More sophisticated estimates assume impulsive stripping on elongated orbits (satellite galaxies lose stars at each pericentric passage).
In this article, we have improved previous analytic models of impulsive tides \citep{spitzer58,gonzalez-casado_etal94,mamon00} and have used our results (Eq.~\ref{rt3})
to predict stripping by $0.07-0.1\,$dex on average, which is consistent with our upper limit $\lsim 0.1-0.15\,$dex.

We consider $0.07\,$dex ($17\%$) to be a reasonable lower limit for the stellar mass lost by galaxies owing to tidal stripping because:
i) its calculation is based on the average rather than the maximum tidal acceleration; 
ii) we assumed that the tides acts only for a very short time interval around the pericentre, and
iii) pericentric radii that may be overestimated (in the case of ghost subhaloes) but not underestimated (tides are stronger for closer pericentric passages).

Our best estimate for the stellar mass lost owing to tidal stripping, $\sim 0.07-0.1\,$dex ($\sim 17-25\%$),
is consistent with a picture in which $\Delta m_*$ is close to $\Delta m_*^{\rm max}$, i.e., one in which satellite galaxies are quenched several gigayears
after entering a group or cluster environment and, in any case, after the first pericentric passage
\citet{mahajan_etal11,wetzel_etal13, haines_etal15,peng_etal15,oman_hudson16,fossati_etal17}.

Our model predicts that the fraction of stars that contribute the ICL increase with the mass of the host system.
In clusters, stars tidally stripped from galaxies are predicted to contribute to half of the total light within $R_{500}$.

\section*{Acknowledgments}
We thank Julien Devriendt for taking care of running the dark matter
simulations used for this work and for computing and providing us the
associated merger trees.
We also thank the anonymous referee and Marina Trevisan for useful comments.
Frank van den Bosch is supported by the Klaus Tschira Foundation and by the US
National Science Foundation through grant AST 1516962.

\bibliographystyle{mn2e}

\bibliography{ref_av}

\appendix

\section{Theory of tides in the circular orbit approximation}
\label{appendix_A}

In this appendix, we compute the tidal acceleration ${\bf a}_{\rm t}$ that a host system exerts on a test particle of a satellite.
We also compute the tidal radius $r_{\rm t}$ of the satellite in the approximations that: 
\emph{i}) the satellite is on a circular orbit, so that the gravitational potential is static in a co-rotating frame, and
\emph{ii}) the test particle is instantaneously stripped as soon as ${\bf a}_{\rm t}$ exceeds the gravitational acceleration that keeps the
particle bound to the satellite.
The latter is an approximation because a net outward acceleration is a necessary but not sufficient condition for tidal stripping.
For it to be sufficient, the speed $\int{\bf a}_{\rm t}{\rm\,d}t$ imparted by the acceleration to the particle must be large enough to unbind it.
In Sect.~\ref{tidesstars}, we build on these results and generalise them to
non-circular orbits and non-instantaneous tides. 

Let $M_{\rm h}$ and $M_{\rm s}$ be the masses of the host (of centre of mass
H) and the satellite (of centre of mass S). Let H and S be the respective
centres of mass of the halo and subhalo, call O the center of mass of the
halo+subhalo system and P the position of a particle in the subhalo.
Finally, denote ${\bf r}\equiv \overrightarrow{\rm SP}$
and ${\bf R}\equiv \overrightarrow{\rm HS}$.
Because of assumption (\emph{i}), H and S rotate around O
 with angular velocity ${\bf\Omega}$, such that
\begin{equation}
\Omega = \sqrt{ G\,(M_{\rm h}+M_{\rm s}) \over R^3 } \ .
\label{Omega}
\end{equation}
In a co-rotating reference frame, the satellite is subject to two
accelerations that cancel one another:
the gravitational attraction of the central system and the centrifugal acceleration.
The particle P is subject to four accelerations:
the gravitational attraction of the satellite, the gravitational attraction of the host, the centrifugal acceleration and the Coriolis acceleration.
The sum of the last three accelerations defines the tidal acceleration ${\bf
  a}_{\rm t}$, which can be written
\begin{eqnarray} 
\dot{\bf v}&=&{\bf a}_{\rm t}-{\bf\nabla}\Phi_{\rm s} \ , \\
\label{net_acceleration}
{\bf a}_{\rm t} &=& -{\bf\nabla}\Phi_{\rm h}
-{\bf\Omega}\times({\bf\Omega}\times\overrightarrow{\rm
  OP})-2{\bf\Omega}\times{\bf v}\ ,
\label{tidal_acc}
\end{eqnarray}
where ${\bf v}$ is the velocity of the particle in the co-rotating frame,
$\Phi_{\rm h}$ is the gravitational potential of the host and $\Phi_{\rm s}$ is the gravitational potential of the satellite.
Assumption (\emph{ii}) applied to equation~(\ref{net_acceleration}) implies
that the particle will be  tidally stripped when $|{\bf a}_{\rm t}| > \nabla
\Phi_{\rm s}$ in
Eq.~(\ref{net_acceleration}).

To proceed further, we must make additional assumptions. Here we assume that:
\emph{iii}) P is in the orbital plane of the binary, so that
$-{\bf\Omega}\times({\bf\Omega}\times\rm\overrightarrow{\rm
  OP})=\Omega^2\overrightarrow{\rm OP}$, and
\emph{iv}) P turns around S on a circular orbit with angular velocity ${\bf\Omega}$, so that ${\bf v}=0$ in the co-rotating frame
(as it is the case for the Moon, which co-rotates in phase-locking with the Earth). 
The impact of these assumptions on the value of $r_{\rm t}$ will be explored at the end of this appendix with a numerical experiment.
By using (\emph{iii}) and (\emph{iv}), 
Eq.~(\ref{net_acceleration}) becomes $\dot{\bf v}=-{\bf\nabla}\Phi_{\rm eff}$, where
\begin{equation}
\Phi_{\rm eff}=\Phi_{\rm h}+\Phi_{\rm s}-{1\over 2}\Omega^2\,\overline{\rm OP}^2.
\label{phi_eff}
\end{equation}

\begin{figure} 
\includegraphics[width=1.\hsize,angle=0]{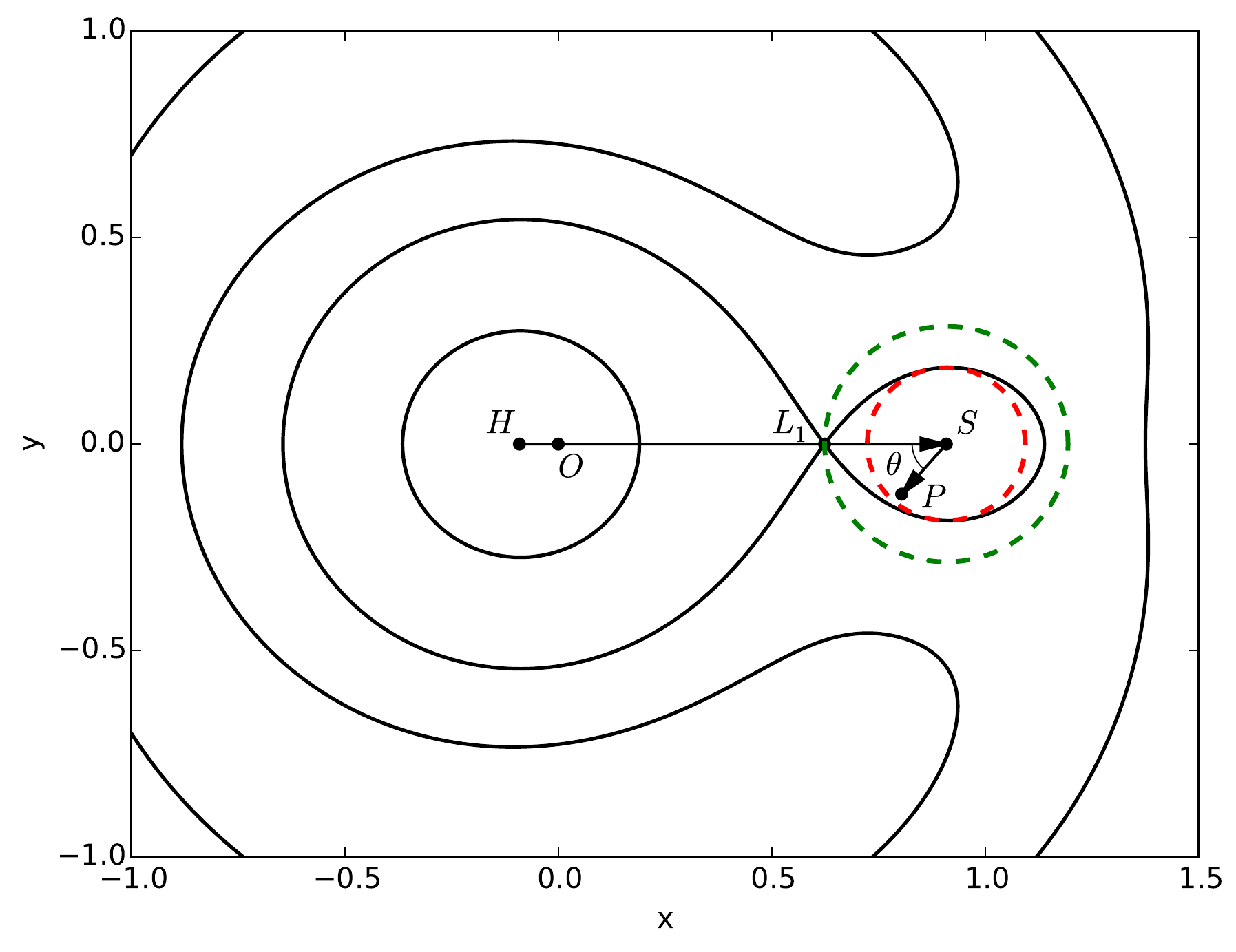} 
\caption{Equipotential curves for the effective potential $\Phi_{\rm eff}$ (Eq.~\ref{phi_eff}) of a two-body system composed of a host halo and a subhalo, centred on H and S,
respectively (\emph{contours}). H and S are on circular orbits around the centre of mass O of the two-body system.
The First Lagrangian Point $\rm L_1$ separates the Roche lobes of the host and the satellite. P is a test particle within the subhalo. 
The figure is for a satellite-to-host mass ratio of $M_{\rm s}/M_{\rm h}=0.1$.
It assumes that both the host halo and the subhalo are described by an NFW
profile with $c=8$ and that the subhalo lies at the virial radius of the
host halo, which is used to scale the coordinates (so that 
$\overline{\rm HS}=1$).
The figure is shown in a reference frame centred in O and co-rotating with the two-body system.
A circular orbit around S through $\rm L_1$ (\emph{larger green dashed circle}) lies outside the Roche lobe of the satellite.
A particle on this orbit is tidally stripped.
The real value of $r_{\rm t}$ corresponds to the radius of the \emph{smaller red dashed circle}, i.e., the largest circle centred on S
to be entirely contained in the Roche lobe of the satellite. }
\label{fig_phi_eff}
\end{figure}

\begin{figure} 
\includegraphics[width=1.\hsize,angle=0]{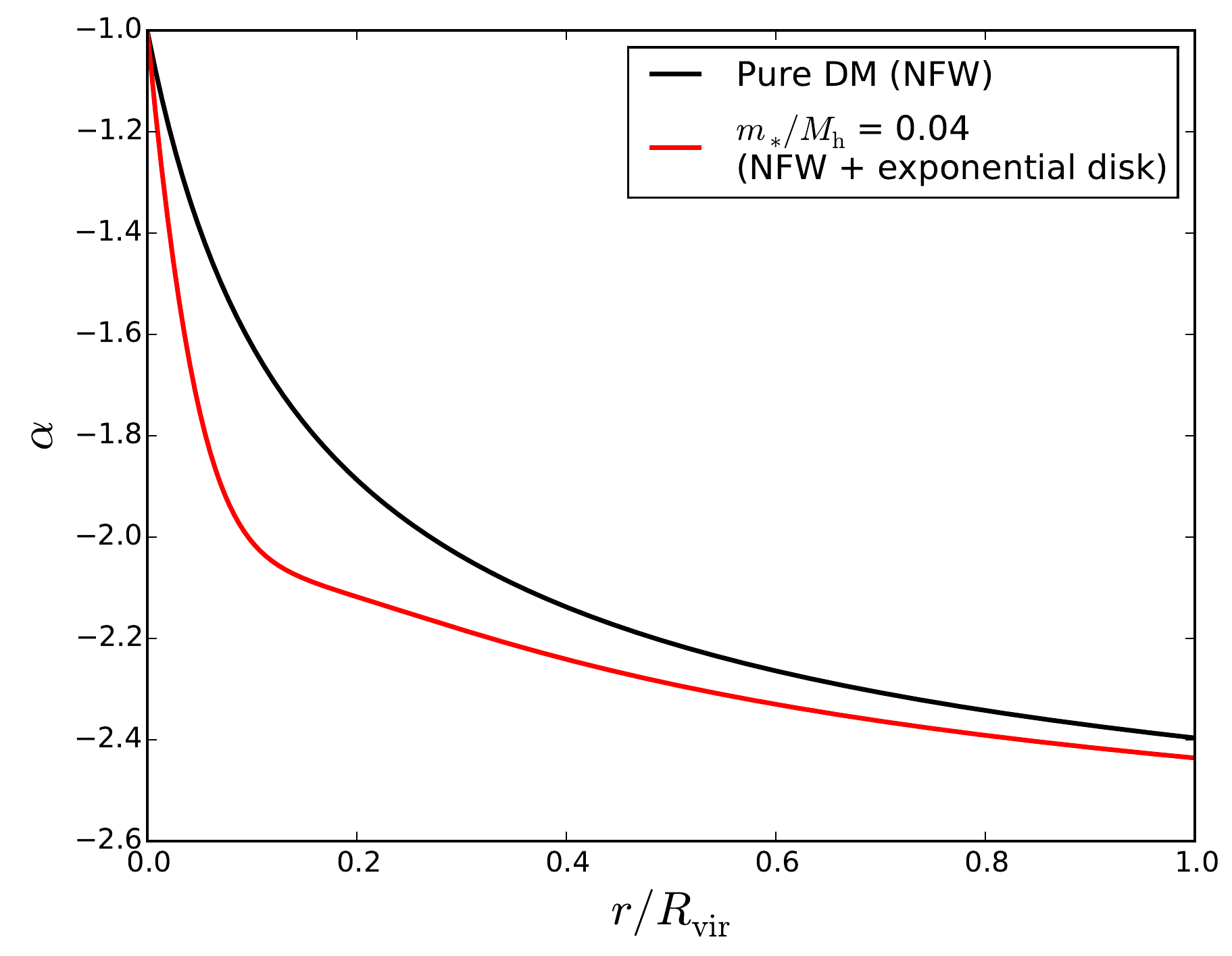} 
\caption{The logarithmic slope $\alpha$  of the mean density profile
  (Eq.~\ref{alpha}) as a function of the spherically averaged radial
  coordinate 
in units of the virial radius,
for an NFW halo with $c=8$ (\emph{black solid line}) and the same halo with
an additional exponential disc of mass equal to $4\%$ the total mass of the
system (\emph{red solid line}).
Note that $m_{\rm *}/M_{\rm h}=0.04$ is the maximum stellar-to-halo mass
ratio allowed by abundance matching.
The exponential disc is assumed to have a scale-length $\lambda R_{\rm
  vir}/2\sim 0.025R_{\rm vir}$, where $R_{\rm vir}$ is the halo's virial
radius. }
\label{fig_alpha}
\end{figure}

Fig.~\ref{fig_phi_eff} shows the equipotential contours for $\Phi_{\rm eff}$ in a particular case used for illustrative purposes.
In this particular case, $M_{\rm s}/M_{\rm h}=0.1$,
$\overline{\rm HS}=R_{\rm vir}$, and both the host and the satellite are described by an NFW profile with concentration $c=8$
($R_{\rm vir}$ is the host virial radius).
The equipotential contours show a saddle point $\rm L_1$ on the segment HS,
which is the First Lagrangian Point $\rm L_1$, which separates the Roche lobes of the host and the satellite.
The particle P is tidally stripped if its orbit spills outside the Roche lobe of the satellite.
In the next five paragraphs, we shall calculate $r_{\rm t}\sim \bar{\rm
  SL_1}$ in the limit that: 
\emph{v}) $M_{\rm s}\ll M_{\rm h}$,
so that $\rm O\rightarrow H$, and \emph{vi}) $\overline{\rm SP}\ll
\overline{\rm HS}$,
so that we can expand ${\bf a}_{\rm t}$ in powers of $r/R$
(Jacobi limit). However, it is important to understand that, even if
$\bar{\rm SL_1}$ could be computed exactly,
it would still provide an approximate estimate for $r_{\rm t}$ because a
circle of radius  $\bar{\rm SL_1}$ and centre S lies outside the Roche lobe of the satellite.
The implication is that the real tidal radius is $r_{\rm t}<\bar{\rm SL_1}$.

We can write
$\overrightarrow{\rm OP}={\bf R}+{\bf r}$ in the limit that $\rm O\rightarrow H$ (assumption \emph{v})\footnote{Throughout this appendix, lowercase letters refer to distances from S, uppercase letters to distances from H.}.
If the mass profile $M_{\rm h}(R)$ of the host is spherically symmetric (assumption \emph{vii}),
Eq.~(\ref{tidal_acc}) can be re-written as \citep{gonzalez-casado_etal94}
\begin{equation}
{\bf a}_{\rm t} = \left[-{G\,M_{\rm h}(|{\bf R}+{\bf r}|)\over |{\bf R}+{\bf r}|^3}+{ G\,M_{\rm h}(R)\over R^3}\right]({\bf R}+{\bf r}),
\label{at1}
\end{equation}
where we have used Eq.~(\ref{Omega}) in the limit $M_{\rm s}\ll M_{\rm h}$  to eliminate $\Omega^2$ and we have neglected
the Coriolis term because of assumption (\emph{iv}).

Expanding ${\bf a}_{\rm t}$ in a Taylor series to first order, we derive 
\begin{equation}
{\bf a}_{\rm t} = {G\,M_{\rm h}(R)\over R^3} \left[-3+{R\over M_{\rm h}(R)}{{\rm d}M_{\rm h}(R)\over{\rm d}R}\right]{r\over R}\cos\theta{\bf\,R}
\label{at2}
\end{equation}
By introducing the slope of the mean density profile of the host
\begin{equation}
\alpha = {{\rm
    d}\ln\overline \rho \over {\rm d}\ln r}  =
{R\over M_{\rm h}(R)}{{\rm d}M_{\rm h}(R)\over{\rm d}R}-3 \ ,
\label{alpha}
\end{equation}
Eq.~(\ref{at2}) can be re-written in the the simpler form
\begin{equation}
{\bf a}_{\rm t} = \alpha \, {G\, M_{\rm h}(R)\over R^3} r\cos\theta\,{{\bf
    R}\over R} \ .
\label{at3}
\end{equation}

The term $\alpha\cos\theta{\bf\,R}$ in Eq.~(\ref{at3}) is directed as $-\cos\theta{\,\bf R}$ ($\alpha<0$ because the density decreases with radius).
In a system of polar coordinates $(r,\theta)$ centred on S,
\begin{equation}
\left(\alpha\cos\theta\,{{\bf R}\over R}\right)_r = |\alpha|\cos^2\theta,
\label{r_term}
\end{equation}
\begin{equation}
\left(\alpha\cos\theta\,{{\bf R}\over R}\right)_\theta = |\alpha|\cos\theta\sin\theta,
\label{theta_term}
\end{equation}
where $\alpha$ appears in absolute value on the right-hand side because ${\bf R}$ points from the central system to the satellite
(and thus $-{\bf R}$ points like ${\bf r}$ when $\theta=0$).
Eq.~(\ref{r_term}) shows that the radial component of  ${\bf a}_{\rm t}$ is always positive, while
Eq.~(\ref{theta_term}) shows that the azimuthal component has zero average.
Therefore, the maximum acceleration is:
\begin{equation}
\langle{a}_{\rm t,max}\rangle={|\alpha|}{{\rm G}M_{\rm h}(R)\over R^3}r.
\label{at_max}
\end{equation}

And averaging over all $\theta$ gives a mean outward acceleration of:
\begin{equation}
\langle{a}_{\rm t}\rangle={|\alpha|\over 2}{{\rm G}M_{\rm h}(R)\over R^3}r.
\label{at_average}
\end{equation}

\begin{figure*}
\begin{center}$
\begin{array}{cc}
\includegraphics[width=0.5\hsize]{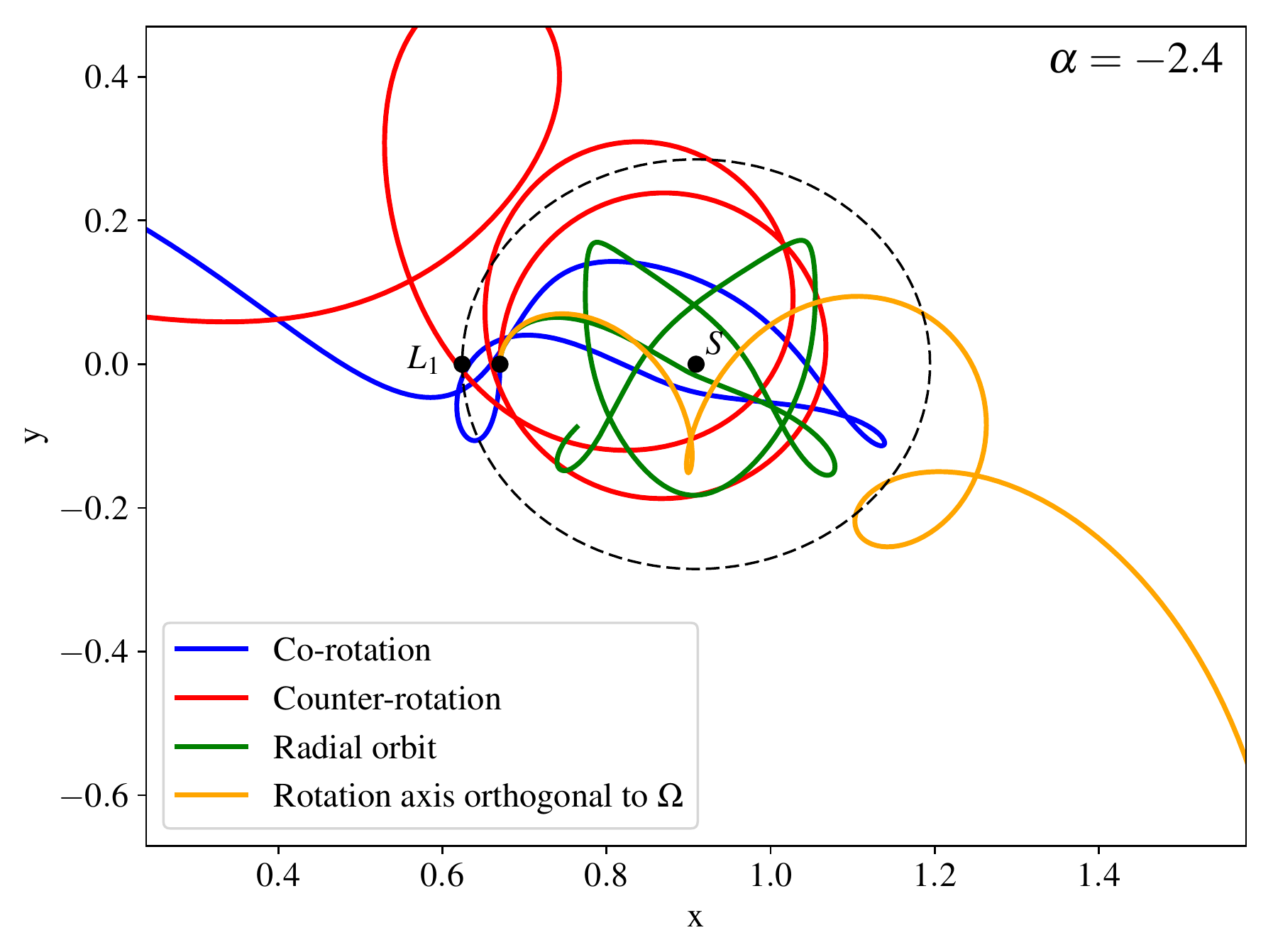} &
\includegraphics[width=0.5\hsize]{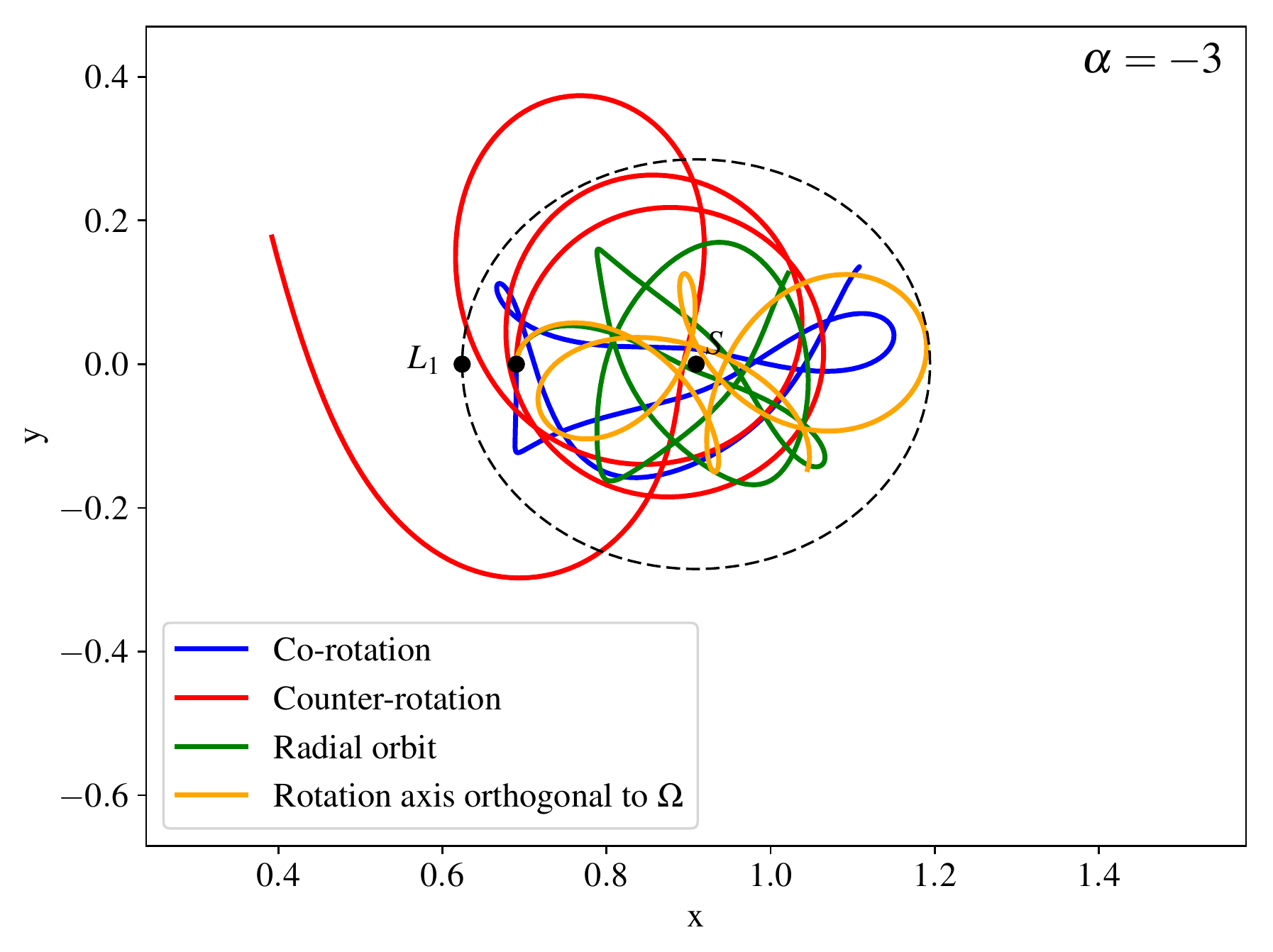} 
\end{array}$
\end{center}
\caption{Orbits of four test particles during one revolution of S around O. All four test particles start at the position P of First Lagrangian Point computed with Eq.~(\ref{jacobi_limit}), as opposed
to the true First Lagrangian Point $\rm L_1$, but their initial conditions for ${\bf v}$ are not the same: a particle co-rotates with $\Omega$, another counter-rotates with respect to $\Omega$,
a third one rotates on a plane  orthogonal to the orbital plane of the two-body system and the fourth one starts on a purely radial orbit.
The left and the right panel correspond to $\alpha=-2.4$, that is, $\alpha(R_{\rm vir})$ for $c=8$, and $\alpha=-3$, respectively.
In the first case, all the particles on circular orbits are stripped from the satellite.
Only the particle on a radial orbit (\emph{green curve}) is retained.
In the second case, only the counter-rotating particle (\emph{red curve})
escapes from the satellite. All others are retained. The units on the $x$ and the $y$ axes are host-halo virial radii.
}
\label{fig_orbits}
\end{figure*}

We compute $r_{\rm t}$ by finding the point of the segment HS for which $a_{\rm t} = -\nabla\Phi_{\rm s}=GM_{\rm s}(r)/r^2$,
where $M_{\rm s}(s)$ is the mass profile of the satellite, for which we assume spherical symmetry (assumption \emph{vii}).
Eq.~(\ref{at3}) gives \citep{dekel_etal03}
\begin{equation}
{G\,M_{\rm s}(r_{\rm t})\over r_{\rm t}^3}=|\alpha|{G\,M_{\rm h}(R)\over R^3},
\label{jacobi_limit}
\end{equation}
from which $r_{\rm t}$ can be computed numerically. There is no factor of two dividing $|\alpha|$ in Eq.~(\ref{jacobi_limit})
because this equation is for $\theta=0$  and not the result of an average
($\widehat{\rm HSL_1}=0$).

The $\alpha$ parameter determines the strength of tidal stripping. The higher its absolute value, the lower the tidal radius $r_{\rm t}$.
For a DM halo described by the NFW profile, 
$\alpha$ decreases from $\alpha=-1$ at $R=0$ to $-2.5<\alpha<-2.2$ at $R=R\sim R_{\rm vir}$ (Fig.~\ref{fig_alpha}, black curve).
The upper and lower limits for $\alpha(R_{\rm vir})$ corresponds to $c=4$ and $c=12$, respectively.
The limit for $R\rightarrow\infty$, $\alpha=-3$, corresponds to the classical Jacobi limit for a point mass, whose average density over a sphere of radius $R$ decreases as $R^{-3}$.
The presence of a luminous galaxy at the centre of the halo causes the decrease of $\alpha$ to be much more rapid at first but then much slower because the limit at infinity has to be the same (Fig.~\ref{fig_alpha}, red curve).
The value of $\alpha$ in the nearly flat part of the curve depends not only
on $c$ but also on the baryon-to-DM mass ratio and the baryon scale length with respect to that of the DM.
Despite these uncertainties, a value $-3<\alpha<-2$ was to be expected, because the flatness of the rotation curves of spiral galaxies in their outer parts imply $\alpha\sim -2$ on the scale of the optical radius.

To test the accuracy of Eq.~(\ref{jacobi_limit}) in recovering the correct value of $r_{\rm t}$, we have performed a  numerical experiment,
in which we start from the pure DM configuration in Fig.~\ref{fig_phi_eff}, we set up a range of initial conditions $({\bf r},{\bf v})$ for the test particle 
and we integrate their orbits to find under what conditions the particles escape from the satellite.
The experiment retains the assumptions that the satellite is on a circular
orbit (\emph{i}), and that the mass distributions of the host and the
satellite are spherically symmetric (\emph{vii}), but allow us to relax the
other five assumptions
(the non-sphericity of discs has a small effect on the total gravitational potential of DM plus baryons).

We compute the position of $\rm L_1$ by solving Eq. ~(\ref{jacobi_limit}) for $\alpha(R_{\rm vir};c=8) = -2.4$ and compare this result, derived from assumptions (\emph{v}) and (\emph{vi}), to the real position
of $\rm L_1$ in our configuration. 
Then, we consider four test particles with the same initial position
(they all start at $\rm L_1$ computed with Eq.~\ref{jacobi_limit}), but with different initial velocities.
Three of the four particles start on a circular orbit with $v_{\rm orb}^2=GM_{\rm s}(r_{\rm t})/r_{\rm t}$, where $v_{\rm orb}$ is the orbital speed of P around S in an inertial frame.
One co-rotates with ${\bf \Omega}$, another counter-rotates and the orbital plane of the third one is orthogonal to that of the binary (the third case allows us to relax assumption \emph{iii}).
The fourth particle starts with zero speed at the apocentre of a purely radial orbit. All four are stripped from the satellite in less than an orbital time ($2\pi/\Omega$), the duration of the numerical experiment.

We then progressively lower the value of $\alpha$ until the particles start so close to S that they are all able to remain within the satellite. 
Both $\alpha=-2.4$ (the value obtained from Eq.~\ref{alpha}) and $\alpha=-3$ (the classical Jacobi limit) correspond to initial conditions for P
in between the smaller and the larger dashed circle in Fig.~\ref{fig_phi_eff}.
The $\alpha$ below which stripping is prevented depends on the initial condition for ${\bf v}$ but the dependence is not strong.
Hence, assumptions (\emph{iii}) and (\emph{iv}) are likely to have a minor effect on the value of $r_{\rm t}$.
Assuming $\alpha=-2.4$ overestimates $r_{\rm t}$ even in a pure DM configuration.
So does $\alpha=-3$, but only slightly.
The results of a numerical experiment in which P is positioned on the inner dashed circle of Fig.~\ref{fig_phi_eff}. are qualitatively similar to those shown in the right panel of Fig.~\ref{fig_orbits}.
The only difference is that the red curve makes many more orbits around S before it escapes from the satellite.

The numerical test in Fig.~\ref{fig_orbits} illustrates the limitations of our analytic approach because it shows
Eq.~(\ref{jacobi_limit}) with $\alpha=-2.4$ from Eq.~(\ref{alpha}) overestimates the real tidal radius (particles on circular orbits tend to be stripped even if they are at $r<r_{\rm t}$).
However, the right panel of Fig.~\ref{fig_orbits} shows that $r_{\rm t}$ is approximately recovered for a higher value of $|\alpha|$ corresponding to $\alpha=-3$.
In principle, this value could depend on the distance between the satellite and the central, but Fig.~\ref{fig_alpha} shows that, for $r/R_{\rm vir}>0.1$, $\alpha$ should be fairly independent of radius, especially in presence of 
baryons (red curve).

\section{Orbital times} 
\label{orbtimes}
Orbital times cannot simply be computed with energy conservation, because, as
the halo increases in mass, its gravitational potential is not stationary.
We have integrated the equation of motion $\ddot {\bf r} = - \Phi'(r)\,{\bf
  r}/r$ in a non-stationary NFW potential
\begin{equation}
\Phi(r,t) = -V_{\rm vir}^2\,{\ln(1+c\,r/r_{\rm vir})\over f(c)\,r/r_{\rm vir}} \ ,
\end{equation}
where $f(c) = \ln(1+c)-c/(1+c)$
\citep{cole_etal96}.
We computed orbits by assuming that the potential varies in time as the
median evolution of cosmological haloes.
We ran the PWGH code of \cite{vandenbosch_etal14} with the same cosmological
parameters as used in our cosmological $N$-body simulation, saving the median
halo mass and concentration as a function of redshift and lookback time.
Our method is similar to that of \cite{vandenbosch_etal16}.
We considered a variety of initial apocenters, and both radial orbits, and
orbits with initial apocentric velocity equal to 0.45 times the circular
velocity at the apocentric radius.
We measured the time from halo entry (1st passage inside the evolving halo
virial radius) to pericenter, to virial radius on the way out, and to the 2nd
apocenter.
We fit the initial redshifts to reach these 3 radii at $z=0$.

\begin{figure}
\centering
\includegraphics[width=\hsize,viewport=0 140 610 666]{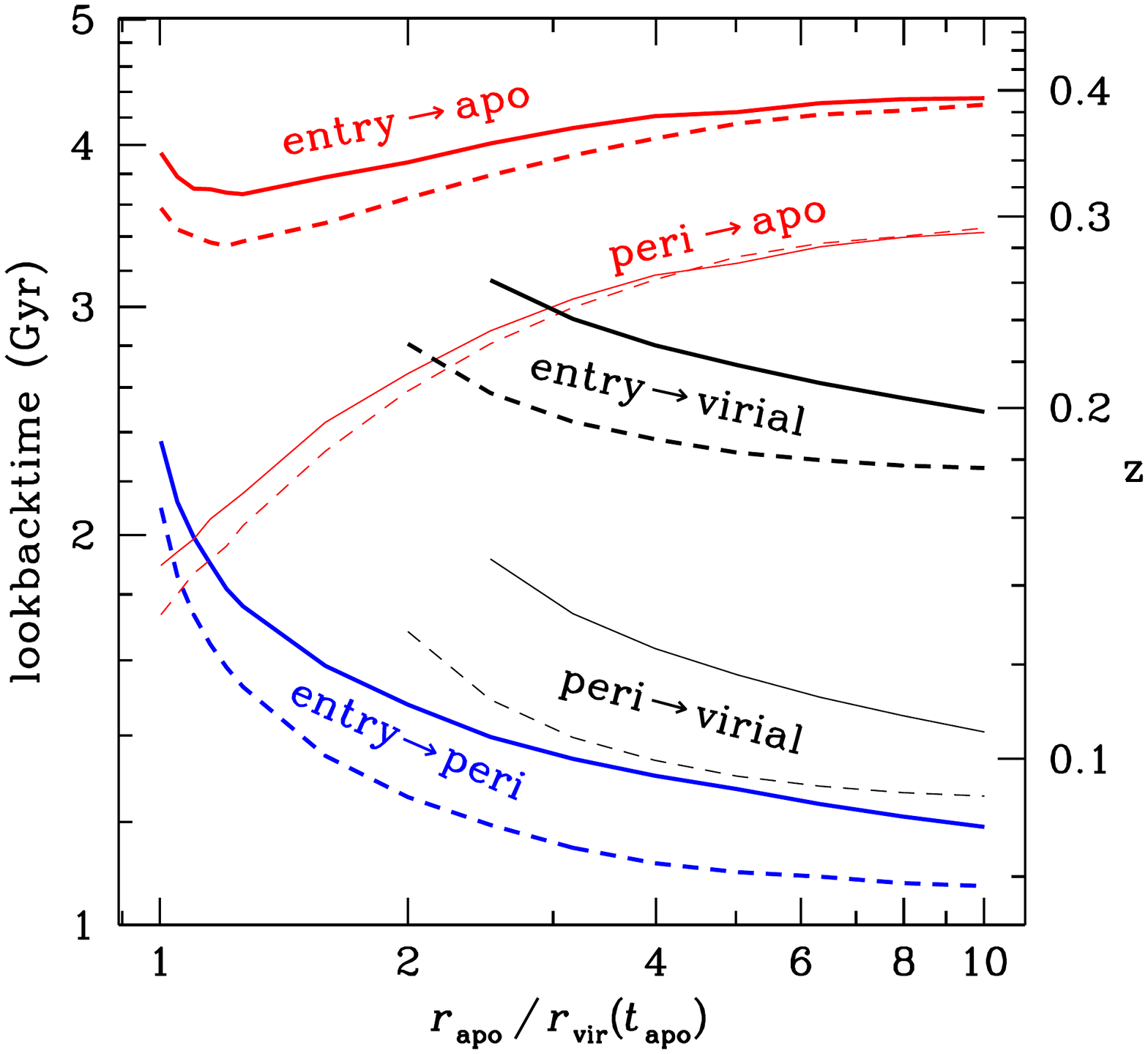} 
\caption{Lookback times and corresponding redshifts for galaxies to be at
  different locations at $z=0$: pericenter (\emph{blue}), virial radius on the way
  outwards (\emph{black}) and apocenter (\emph{red}), for radial orbits (dashed) and
   orbits of typical elongations (\emph{solid}), starting the clock at cluster
   entry (\emph{thick}) or at pericenter (\emph{thin}).
\label{fig_tff}}
\end{figure}

Figure~\ref{fig_tff} displays the lookback times and corresponding
redshifts for a galaxy to reach the pericenter (blue), virial radius on the
way out (black) and apocenter (red) at $z=0$.
While the dashed lines indicate the orbital times for radial orbits, the
solid lines represent orbits of typical apocenter/pericenter ratios of 5
\citep{ghigna_etal98}, where we used an apocentric (tangential) velocity of
0.45 times the circular velocity at that radius.
The $z$=0 halo mass is $10^{14} \rm M_\odot$, but the results very fairly
little with the halo mass (except that the virial radius is harder to reach
on the way out for more massive haloes).
The times to reach the pericenter and the virial radius decrease for
increasing initial apocentric radius (in units of the initial $R_{\rm vir}$),
because galaxies starting at large radii travel faster through the
halo.
On the other hand, the time to reach the 2nd apocenter increases for increasing 1st
apocentric radius (in units of the initial virial radius), because there is
more distance to travel outwards (e.g. from 1st pericenter to 2nd apocenter),
unless the initial apocenter is small, where the speed effect overcomes the
distance effect.

\section{Fitting parameters for the SMF and the SMHM relation} 
\label{appendix_fit}

In this appendix, we explain how we use the data points of \citet{yang_etal12} for the local Universe and \citet{muzzin_etal13} at $z>0.2$ to construct the SMF $n_*(m_*,z)$ that we use for AM.

We fit the data points in each redshift bin (centred on $z_i$) with a double-power-law function of the form: 
\begin{equation}
n_i\!\left( m_*\right) = {1\over N_i}\left[\left({m_*\over m_i}\right)^{\alpha_i}+
\left({m_*\over m_i}\right)^{\beta_i}\right]^{-1},
\label{eq_fitSMF}
\end{equation}
where $N_i$, $m_i$, $\alpha_i$ and $\beta_i$ are free parameters.

After having determined $N_i$, $m_i$, $\alpha_i$ and $\beta_i$ for each $z_i$, we fit the evolution of $N$, $m_*^0$, $\alpha$ and $\beta$ with $z$ by assuming the linear dependences:

\begin{equation}
\begin{aligned}
\log N\!\left( z\right) = n_1z + n_0 \\
\log m_*^0\!\left( z\right) = x_1z + x_0 \\
\alpha\!\left( z\right) = \alpha_1z + \alpha_0\\
\beta\!\left( z\right) = \beta_1z + \beta_0
\end{aligned}
\label{eq_fit_param_z}
\end{equation}
Table~\ref{tab_fit}1 gives the best-fit values for the fitting parameters $n_0$, $x_0$, $\alpha_0$, $\beta_0$ and $n_1$, $x_1$, $\alpha_1$, $\beta_1$.
The quality of the fit is shown on Fig.~\ref{fig_massfun}.
The parameters in Table~\ref{tab_fit}1 specify the SMF that we use for the AM.

\begin{table}
\label{tab_fit}
\caption{Best-fit parameters characterising the SMF at all redshift smaller than $2.5$. The SMF is fitted at a given $z$ by equation Eq.~\ref{eq_fitSMF} with four parameters which are assumed to be linear function of $z$ (see Eq.\ref{eq_fit_param_z}).}
\begin{center}
\begin{tabular}{cccccccc}
\hline
\hline
$n_0$ & $x_0$ & $\alpha_0$ & $\beta_0$ \\
\hline
2.39 & 11.12 & 0.228 & 3.07 \\
\hline
\hline
$n_1$ & $x_1$ & $\alpha_1$ & $\beta_1$ \\
\hline
0.501 & $4.6*10^{-3}$ & 0.133 & 0.367 \\
\hline
\end{tabular}
\end{center}
\end{table}

\begin{figure}
\includegraphics[width=1.\hsize,angle=0]{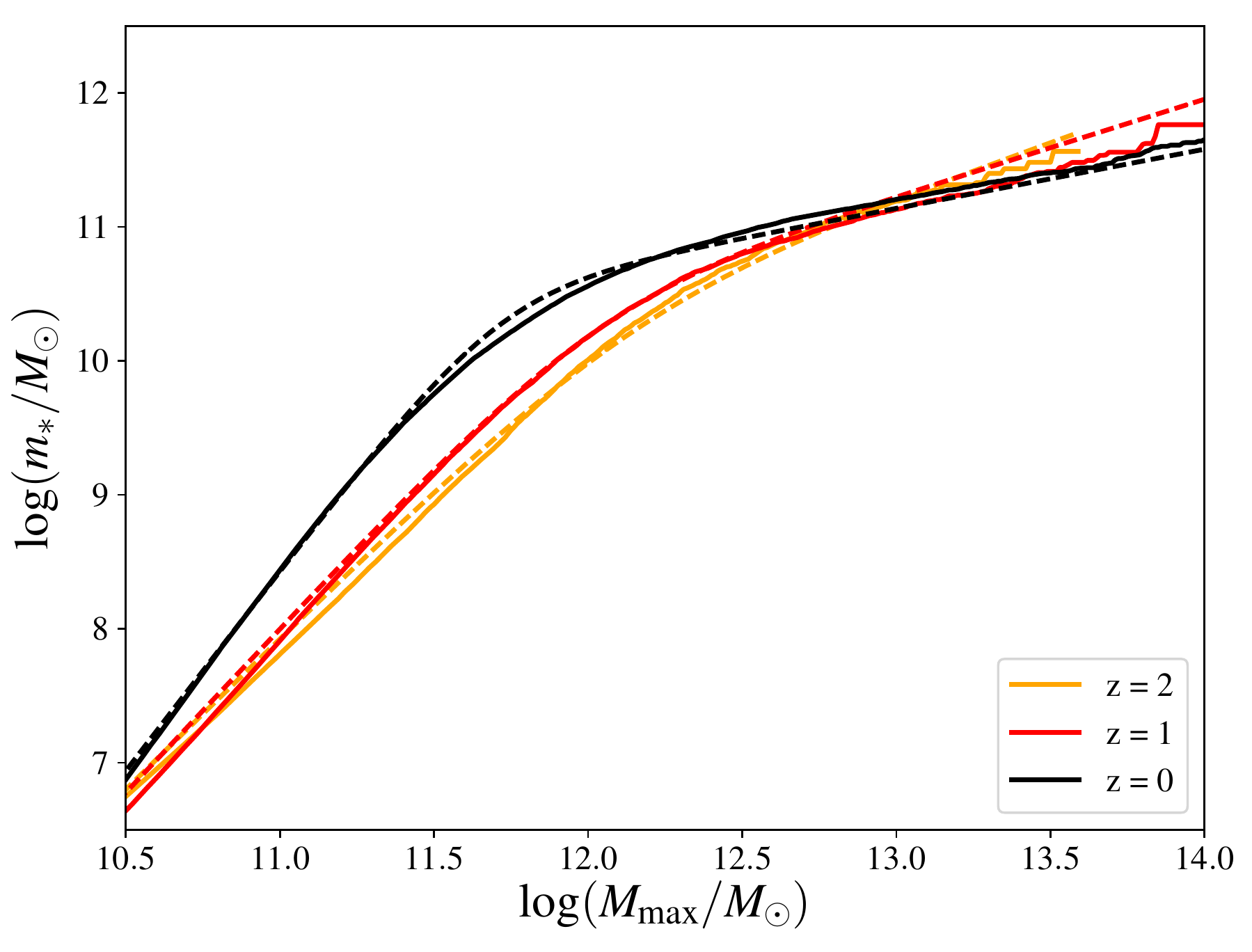} 
\caption{Comparison between the fitted stellar mass ($m_*$) - halo mass ($M_{\rm max}$) relation (dotted lines) and the original relation
used in this work (solid line) at $z=0$, $z=1$ and $z=2$.
The only purpose of this fitting is to facilitate the comparison with future AM work. The best-fit parameters are given in table~\ref{tab_fit2}2.
}
\label{fig_msmh_fit}
\end{figure}

To ease the comparison with future work, we apply the same fitting procedure to the SMHM relation.
Following \citet{moster_etal13},
we obtain a good fit to the relation from AM (Fig.~\ref{fig_msmh_fit})
for a double-power-law function of the form
\begin{equation}
m_*\!\left( m_h,z\right) = M\!\left( z\right)\left[\left(\frac{m_h}{m_h^0\!\left( z\right)}\right) ^{-\gamma\left( z\right)}+
\left(\frac{m_h}{m_h^0\!\left( z\right)}\right) ^{-\eta\left( z\right)}\right]^{-1},
\label{eq_fitSMHM}
\end{equation}
where:
\begin{equation}
\begin{aligned}
\log M\!\left( z\right) = m_1\frac{z}{1+z} + m_0, \\
\log m_h^0\!\left( z\right) = y_1\frac{z}{1+z} + y_0, \\
\gamma\!\left( z\right) = \gamma_1\frac{z}{1+z} + \gamma_0,\\
\eta\!\left( z\right) = \eta_1\frac{z}{1+z} + \eta_0.
\end{aligned}
\label{eq_fit_param_z2}
\end{equation}
Table~\ref{tab_fit2}2 gives the best-fit value for the fit parameters
$m_0$, $y_0$, $\gamma_0$, $\eta_0$ and $m_1$, $y_1$, $\gamma_1$, $\eta_1$.
We stress that the fitting formula in Eq.~\ref{eq_fitSMHM} and the parameters in Table~\ref{tab_fit2}2 are used nowhere in our analysis.
They have been inserted purely to ease comparison with our work.

\begin{table}
\label{tab_fit2}
\caption{Best-fit parameters characterising the SMHM relation at all redshift smaller than $2.5$. The SMHM relation is fitted at a given $z$ by equation Eq.~\ref{eq_fitSMHM} with four parameters which are assumed to be linear function of $z/(1+z)$ (see Eq.\ref{eq_fit_param_z2}).}
\begin{center}
\begin{tabular}{cccccccc}
\hline
\hline
$m_0$ & $y_0$ & $\gamma_0$ & $\eta_0$ \\
\hline
10.57 & 11.69 & 3.04 & 0.417 \\
\hline
\hline
$m_1$ & $y_1$ & $\gamma_1$ & $\eta_1$ \\
\hline
-0.085 & 0.685 & -1.16 & 0.607 \\
\hline
\end{tabular}
\end{center}
\end{table}

\label{lastpage}
\end{document}